\documentclass[aps,prl,shownopacs,onecolumn,superscriptaddress]{revtex4}
\usepackage{bm,color,amsmath,amssymb,mathrsfs,latexsym,graphicx,psfrag}

\newcommand{\braket}[2]{\left\langle #1 | #2 \right\rangle}
\newcommand{\bra}[1]{\left\langle#1\right|}
\newcommand{\ket}[1]{\left|#1\right\rangle}

\newcommand{\beq}{\begin{equation}}
\newcommand{\eneq}{\end{equation}}

\begin{document}
\title{Emergent Many-Body Translational Symmetries of Abelian and Non-Abelian Fractionally Filled Topological Insulators}
\author{B. Andrei Bernevig} 
\affiliation{Department of Physics, Princeton University, Princeton,
NJ 08544}
\author{N. Regnault}
\affiliation{Laboratoire Pierre Aigrain, ENS and CNRS, 24 Rue Lhomond, 75005 Paris, France}


\begin{abstract}
The energy and entanglement spectrum of fractionally filled interacting topological insulators exhibit a peculiar manifold of low energy states separated by a gap from a high energy set of spurious states. In the current manuscript, we show  that in the case of fractionally filled Chern insulators, the topological information of the many-body state developing in the system resides in this low-energy manifold. We identify an emergent many-body translational symmetry which allows us to separate the states in quasi-degenerate center of mass momentum sectors. Within one center of mass sector, the states can be further classified as eigenstates of an emergent (in the thermodynamic limit) set of many-body relative translation operators. We analytically establish a mapping  between the two-dimensional Brillouin zone for the Fractional Quantum Hall effect on the torus and the one for the fractional Chern insulator. We show that the counting of quasi-degenerate levels below the gap for the Fractional Chern Insulator should arise from a folding of the states in the Fractional Quantum Hall system at identical filling factor. We show how to count and separate  the excitations of the Laughlin, Moore-Read and Read-Rezayi series in the Fractional Quantum Hall effect into two-dimensional Brillouin zone momentum sectors, and then how to map these into the momentum sectors of the Fractional Chern Insulator. We numerically check our results by showing the emergent symmetry at work for Laughlin, Moore-Read and Read-Rezayi states on the checkerboard model of a Chern insulator, thereby also showing, as a proof of principle, that non-Abelian Fractional Chern Insulators exist. 
 \end{abstract}
\date{\today}


\maketitle

\section{Introduction}

Band theory  formed  the foundation of solid state physics for the past century. The theory analyzes the one-body motion of electrons through a crystal and obtains their wavefunctions as Bloch states dependent on a crystal momentum which takes values in a space equivalent to a $d$-dimensional torus (in $d$ dimensions). Insulators hold a special place in band theory: at a first look, they are its most boring aspect, having a full gap between occupied and unoccupied bands, no low-energy excitations, and hence no interesting properties at zero temperature.  This viewpoint permeated the physics of much of the last century, but is, as we now know, untrue.  The space of the eigenvectors of the Bloch Hamiltonians  can be thought of as a unitary matrix, which in the case of insulators exhibits several gauge symmetries related to permutations of the occupied and un-occupied bands, thereby creating a complex manifold. The Hamiltonians can be regarded as maps from the Brillouin zone (BZ) to these manifolds, which, as mathematics teaches us, can be nontrivial. Anytime an insulator with a nontrivial map (Hamiltonian) is placed in the vicinity of the vacuum, gapless edge or surface states appear on the boundary and cross the energy space between the valence and conduction band. One of the first examples of such behavior was Haldane's Chern insulator model on the graphene lattice \cite{haldane-1988PhRvL..61.2015H}. This model exhibits the physics of the Integer Quantum Hall (IQH) effect but does not have an overall applied magnetic field, thereby preserving the translational symmetry of the initial lattice. The IQH does not require any discrete symmetry to exist. The field of topological insulators has further evolved to include other symmetries such as time-reversal \cite{kane-PhysRevLett.95.226801,Bernevig15122006,fu-PhysRevB.76.045302}, charge conjugation and point-group symmetries.

Most of the research on topological insulators has focused on one-body electron Bloch physics. While several interesting works on insulators with interactions exist \cite{pesin-naturephy2010,rachel-PhysRevB.82.075106,wen-PhysRevB.82.075125}, most of them focus on obtaining a mean-field state with topological properties through interactions. Recently, it has been proposed and substantiated \cite{regnault-PhysRevX.1.021014,neupert-PhysRevLett.106.236804,sheng-natcommun.2.389} that fractionally filled bands of Chern insulators at $1/3$ and $1/5$ filing can exhibit an incompressible state in the same universality class as the Abelian $1/3$ and $1/5$ Fractional Quantum Hall (FQH) states. This state was dubbed the Fractional Chern Insulator (FCI). Its appearance is quite surprising, as the FQH and FCI problems vary considerably in several important ways, in the lack of a constant Berry curvature in the FCI, its large lattice filling, the absence of  holomorphic and anti-holomorphic ($z, z^*$)  structures in the lowest band and the presence of Umklapp processes that favor density wave states.

In a previous paper\cite{regnault-PhysRevX.1.021014} we have shown that the groundstate of the interacting $1/3$ fractionally filled band with Chern number $1$ of the  checkerboard Hamiltonian supports a $3$-fold degenerate groundstate separated by a gap from the excited states and which is a featureless liquid with momentum orbital occupation number $1/3$. We showed that the spectrum of quasiholes (spectrum at lower filling) by a gap separated into two manifolds: a high energy uninteresting set of spurious states, and a low-energy set of quasiholes whose counting empirically matches that of the quasihole states in an Abelian $\nu=1/3$ filling FQH state. Similar results were obtained for the particle entanglement spectrum \cite{sterdyniak-PhysRevLett.106.100405}, a procedure used to obtain the topological part of the excitation spectrum directly from the groundstate  wavefunction \cite{li-08prl010504}. In the FCI, we have empirically found that the counting of states in the lower energy manifold, be they groundstates or quasiholes, exhibits a peculiar, yet un-determined  structure. On a finite size $N_x \times N_y$ lattice, a translationally invariant Hamiltonian for $N_e$ electrons has a spectrum classified by total many-body lattice momenta $\sum_{i=1}^{N_e} k_{i x}\; (\mod 2\pi)$ and $\sum_{i=1}^{N_e} k_{iy}\; (\mod 2 \pi)$, with each particle momentum $k_{ix,y}$ taking values  $2\pi j/N_x, 2 \pi l/N_y$, $j=1\ldots N_x$, $l=1\ldots N_y$. We have found that the counting of states in the low energy (or entanglement energy) manifold per momentum sector exhibits multiple degeneracies, although the states themselves are not degenerate - there are no other exact degeneracies besides the point-group enforced ones, such as inversion. In many cases, we have found that an empirical Pauli principle borrowed from the FQH physics can sometimes (but not always) explain the degeneracies observed. While the degeneracy of some model wavefunctions has been worked out for the FQHE on a lattice \cite{kol-1993PhysRevB.48.8890}, these results cannot be applied to the FCI.

Our prior analysis strongly suggests the existence of an emergent symmetry for FCI's at rational filling factors. In the present manuscript we show that such a symmetry is a many-body translational symmetry of the type exactly present in the FQH effect. We show analytically that, if the FCI state is in the same phase as the FQH state (including in the presence of quasiholes), the spectrum separates into  quasi-degenerate  center of mass sectors which exhibit identical counting. Within each center of mass sector, the Hamiltonian eigenstates can be further classified as eigenstates of an emergent (in the thermodynamic limit) set of many-body relative translation operators. Using the recently obtained Girvin-MacDonald-Platzman (GMP)\cite{girvin-PhysRevB.33.2481} algebra for Chern insulators \cite{Parameswaran-2011arXiv1106.4025P} for the one-body density operators in the valence band, we build a set of many-body relative translational operators which diagonalize the Hamiltonian eigenstates in the lower manifold in the thermodynamic limit. We then establish an analytic mapping between the counting of zero-mode states in the FQH $2$-D BZ on the torus (quasiholes) and that of the FCI levels in the low-energy manifold. We show that the counting of quasi-degenerate levels below the gap for the FCI should arise from a folding of the zero-mode states in the Fractional Quantum Hall system at identical filling factor. In the process, we show how to count and separate the quasihole excitations of the Read-Rezayi (RR)\cite{read-PhysRevB.59.8084} series in the Fractional Quantum Hall effect into two-dimensional BZ momentum sectors, and then how to map these into the momentum sectors of the FCI. We provide numerical verification for our analytic results by checking the emergent symmetry at work for Laughlin\cite{laughlin-PhysRevLett.50.1395}, Moore-Read (MR)\cite{Moore1991362} and RR\cite{read-PhysRevB.59.8084} states on the checkerboard model of a Chern insulator: we predict, from the FQH quasihole counting plus our FQH to FCI analytic mapping, the number of non-Abelian quasiholes per momentum sector in the FCI, and then check that it matches the numerical data obtained by exact diagonalization for large system sizes.  Our work then also shows, as a proof of principle, that Non-Abelian FCIs exist. The  emergence of the low-manifold of states with folded FQH counting can be regarded as both a consequence as well as an imprint of the existence of the topological phase. Its existence is rather mysterious, as the Girvin-MacDonald-Platzman one-body algebra from which the many-body generators are built is not valid at large momentum    \cite{Parameswaran-2011arXiv1106.4025P}  and is not in one-to-one mapping with the algebra of the  continuum FQH problem.

The paper is organized as follows. In section $1$ we analyze the interacting Landau level problem on the torus in the continuum and obtain the set of many-body relative and center of mass symmetries that can classify the interacting spectrum by two-dimensional BZ of  $N\times N$ momenta where $N= GCD(N_\phi, N_e)$.  $GCD$ stands for greatest common denominator, $N_\phi$ is the number of fluxes that pierce the FQH torus due to the presence of the magnetic field, while $N_e$ is the number of electrons. This section has no new material, most of the end results having been obtained by Haldane in a classic PRL \cite{haldane-PhysRevLett.55.2095}, but it does derive the physical equations in much greater level of detail. We then re-formulate the translational symmetry in terms of density operators, allowing a closer analogy with the FCI. We also explicitly show how to implement the many-body symmetries to build the FQH Hilbert space. Section $2$ describes the counting of the zero modes of several pseudopotential Hamiltonians per two dimensional BZ momentum. We focus on the Read-Rezayi series and show how the generalized Pauli principle that allows the counting of the quasihole states in the one-dimensional orbital space on the torus can be tuned to give the counting of zero modes in the two-dimensional BZ.  As such, resolving the zero-modes (quasiholes) of model Haldane pseudopotential Hamiltonians in the $2D$ BZ becomes a combinatorics problem of counting partitions.  In Section $3$ we first re-derive the commutator algebra of the projected densities in the lowest band of the Chern insulator on a $N_x \times N_y$ lattice first derived in \cite{Parameswaran-2011arXiv1106.4025P} which reduces, in the long wavelength limit to the GMP algebra for the FQH effect for $N_\phi = N_x \cdot N_y$. We show that the non-commutativity of the projected densities which gives rise to the GMP algebra is a requirement in the Chern insulator, while in the trivial insulator the projected densities are adiabatically continuable to commuting variables in the atomic limit.  We then show that the existence of a GMP algebra superimposed on a lattice of $N_x \times N_y$ sites implies the existence of many-body relative translational operators which classify many-body states by a $2$-dimensional momentum in a reduced BZ of $GCD(N_x,N_e) \times GCD(N_y, N_e)$ momenta. If the FCI phase is identical to that of the FQH, there is an emergent, center of mass degeneracy of $N_x/GCD(N_x,N_e)$ in the $x$ direction and $N_y/GCD(N_y,N_e)$ in the $y$ direction. We show that the counting of states in lower energy manifold of the FCI is a folding of the counting of zero-modes in the $N \times N$ BZ of the FQH times the difference between the center of mass degeneracies. We establish the analytic mapping between the FQH and FCI counting, and give several examples of this counting for several Abelian and non-Abelian RR states based on the generalize Pauli principle. We refer the reader versed in FQH translational symmetries to read this section and look at Figures[\ref{entspeconethirdstateN4particles},   \ref{CountingN2Nx3Ny4}, \ref{CountingN4Nx6Ny6entspect}] for a simple understanding of how the FQH-FCI map works.  In section $4$ we engage in extensive numerical calculations of both energy and entanglement spectra that show the existence of Laughlin, MR and $\mathbb{Z}_3$ RR states on the checkerboard model \cite{sun-PhysRevLett.106.236803} of the Chern insulator. We then show that the counting of quasihole states per momentum sector obtained in numerics matched the one derived by our analytic map for all our large set of data and parameters (lattice aspect ratios, electron numbers) tried. Up to the largest sizes available on today's computers, the numerical data supports our analytic result.

\section{Many-Body Symmetries of the Interacting Electrons In a Magnetic Field}

In this section, we aim to analyze the symmetries of the interacting Hamiltonian of electrons on a two-dimensional torus in the presence of a magnetic field and electron-electron interactions:

\beq
H= \frac{1}{2 m} \sum_j^{N_e} \Pi_j^2 + \frac{1}{2} \sum_{i \ne j}^{N_e} V(\vec{r}_i - \vec{r}_j) \label{ContinuumHamiltonianLandauLevel}
\eneq
with $\Pi_j = - i \hbar \nabla_j - e A(r_j)=- i \hbar \nabla_j +|e| A(r_j)$ the canonical momentum in the presence of a magnetic field. $N_e$ is the number of electrons in the system. We choose not to gauge-fix, and have $\vec{\nabla} \times \vec{A} = \vec{B}$. The positions of the particles, $\{\vec{r}_i\}$, reside on a two-dimensional torus of generators $\vec{L}_1$, $\vec{L}_2$. The Hamiltonian is periodic under translations by these vectors, $V(\vec{r}_i - \vec{r}_j) = (\vec{r}_i - \vec{r}_j + \vec{L}_{1,2})$ and can be written as a sum over the allowed reciprocal vectors $\vec{q}$:
\beq
\sum_{i\ne j}^{N_e} V(\vec{r}_i - \vec{r}_j) = \frac{1}{2\cal{A}}\sum_{\vec{q}}  V(\vec{q}) \sum_{i<j}e^{i \vec{q} \cdot(\vec{r}_i - \vec{r}_j)} 
\eneq 
where ${\cal A}= |\vec{L}_1 \times \vec{L}_2|$ is the area of the unit cell, and $\vec{q} =m \vec{q}_1 + n \vec{q}_2$, $m,n \in {\mathbb{Z}}$ and $\vec{q}_1 = \frac{2\pi}{{\cal A}} \vec{L}_2 \times \hat{z}$, $\vec{q}_2 = -\frac{2\pi}{{\cal A}} \hat{z} \times  \vec{L}_1$. We now analyze the single and many-body translational symmetries of the above Hamiltonian.

\subsection{Guiding center coordinates and translational symmetries of the $1$-body problem}
 
We  first consider $N_e=1$ problem and try to find the symmetries of the $1$-body problem. A perfectly good translational operator could be $e^{i \vec{a} \cdot \vec{ \Pi}}$ - which would translate the single-body wavefunction by $\vec{a}$. However, this operator does not commute with the Hamiltonian because
\beq
[\Pi_{\alpha} , \Pi_{\beta}] = - i \hbar F_{\alpha \beta}
\eneq (where $F_{\alpha \beta} = \partial_\alpha A_\beta - \partial_\beta A_\alpha$ is the (magnetic) field strength applied on the sample). Hence the Hamiltonian wavefunctions cannot acquire quantum numbers under this operator and one finds an operator which commutes with Hamiltonian:
\beq
K_\alpha = \Pi_\alpha - \frac{\hbar}{l^2} (\hat{z} \times \vec{r})_\alpha
\eneq where $ l = \sqrt{\hbar/e B}$ is the magnetic length. The commutation relations read:
\beq
[K_\alpha,  K_\beta]= 2i \frac{\hbar^2}{l^2} \epsilon_{\beta z \alpha}  - i \hbar e (\partial_\alpha A_\beta - \partial_\beta A_\alpha) = i \frac{\hbar^2}{l^2} \epsilon_{z\alpha \beta}
\eneq  where we have used  $B_\theta = \frac{1}{2} \epsilon_{\theta \alpha \beta} F_{\alpha \beta}$ as the uniform magnetic field applied on the sample. 
 By direct calculation, we have for the commutators:
\beq
[K_\alpha, \Pi_\beta] =0 \rightarrow [K_\alpha, H] =0
\eneq $\vec{K}$ is called guiding center momentum. Hence the operator that implements the magnetic translation is 
\beq
T(\vec{a} ) = \exp(\frac{i}{\hbar} \vec{a}\cdot \vec{K})
\eneq If translations by different vectors would mutually commute, we could form momentum eigenstates of the system. However, they do not. Using the operator relation $e^{A+ B} =e^{A} e^B e^{-\frac{1}{2} [A,B]}$, valid when $[A, [A,B]]= [B, [A,B]]=0$ (which is the case here as the commutator of two guiding center momenta is a constant for uniform $\vec{B}$), we find:
\beq
T(\vec{a} +\vec{b}) = T(\vec{b}) T(\vec{a}) e^{-\frac{1}{2\hbar^2} a_\alpha b_\beta [K_\alpha, K_\beta]}= T(\vec{b}) T(\vec{a}) e^{-\frac{i}{2 l^2} \hat{z} \cdot(\vec{a} \times \vec{b})}
\eneq 
which leads to
\beq
[T(\vec{a}), T(\vec{b})] = -2 \sin({\frac{1}{2 l^2} \hat{z} \cdot(\vec{a} \times \vec{b})} ) T(\vec{a} +\vec{b}) 
\eneq
This is called the magnetic translation algebra or Girvin-Plazmann-MacDonald (GMP) algebra. This algebra leads to the quantization of flux passing through the lattice: going around the unit cell must give us the identity (same condition as commutators of $\vec{L}_1$, $\vec{L}_2$ translations must commute). This GMP algebra arises in several contexts in the quantum Hall effect and has the interpretation of a quantum deformation of the classical algebra of area-preserving diffeomorphisms on the plane as well as that of magnetic translations in a uniform field as discussed. Integer quantum Hall states are invariant under area preserving diffeomorphisms. In finite size, the Girvin-Plazmann-MacDonald algebra is the Lie algebra of $U(N_\Phi)$. Combinations (powers and products) of the operators of the algebra commute with the interacting Hamiltonian and can hence be simultaneously diagonalized giving the good quantum numbers of the problem. Our scope is to find the maximal set of such good quantum numbers for a translationally invariant interacting electrons  in the presence of a magnetic field.

We now particularize (without loss of generality, since one can always deform the BZ to a rectangle) to $\vec{L}_1 = L_x \hat{x}, \vec{L}_2 = L_y \hat{y}$:
\beq
T(L_x \hat{x}) T(L_y \hat{y})T(-L_x \hat{x}) T(-L_y \hat{y}) = 1= e^{\frac{i}{l^2} L_x L_y}  
\eneq Hence $L_x L_y = {2 \pi}{l^2} N_\phi$ or, for a non-rectangular lattice, $\frac{1}{l^2} \hat{z}\cdot(\vec{L}_1 \times \vec{L}_2) = 2 \pi N_\phi$. For this quantization condition, we have that $[T(\vec{L}_1), T(\vec{L}_2)]= 0$. 
 $N_\phi$ is the number of magnetic flux quanta passing through a unit cell. As a spoiler for the Chern insulator section, notice that this is the same as the number of sites $N_s$ on a square lattice with lattice constant $a_0=\sqrt{2 \pi l^2}$ and number of sites in the $x$ direction $L_x/a_0$ and in the $y$ direction $L_y/a_0$. Hence the magnetic field to Chern insulator lattice analogy is: $L_x/\sqrt{2 \pi l^2} \rightarrow N_x, L_y/\sqrt{2 \pi l^2} \rightarrow N_y, N_\phi \rightarrow N_s$.  Observe that $L_x/\sqrt{2 \pi l^2} \rightarrow N_x, L_y/\sqrt{2 \pi l^2} \rightarrow N_y, N_\phi \rightarrow N_s$ are all integers.

\subsection{Many-Body Translational Symmetries}

We now focus on the interacting problem. The many-body problem is characterized by translation operators for each $i$'th particle, $T_i(\vec{a})$. Translational operators of different particles commute:
\beq
[T_i(\vec{a}), T_j(\vec{b})]= -2  \delta_{ij}  T_i(\vec{a} +\vec{b}) \sin({\frac{1}{2 l^2} \hat{z} \cdot(\vec{a} \times \vec{b})})
\eneq  Physical quantities are invariant under magnetic translations \emph{of any particle $i$} by a linear combination of multiples of $L_x, L_y$:
\beq
\vec{L}_{mn} =   \{ m \vec{L}_1 + n \vec{L}_2, \;\;\;\; m, n \in {\mathbb{Z}} \} 
\eneq Following Haldane \cite{haldane-PhysRevLett.55.2095}, primitive translations $\vec{L}_{mn}$ are those for which  $\lambda \vec{L}_{mn}$, $0<\lambda<1$ are not lattice vectors - hence $(m,n)$ have no common divisors. Choose a potential 
$V(r+ L_a) = V(r)$ such that the whole Hamiltonian commutes with the translation operators $T_i(\vec{L}_1), T_i(\vec{L}_2)$, for each particle $i =1 \ldots N_e$. This means that the wavefunctions of the Hamiltonian are eigenstates of the translation operator up to a phase:
\beq
T_i(\vec{L}_j) \psi_\alpha (\vec{r}_i) = \psi_\alpha (\vec{r}_i + \vec{L}_j) = e^{i \theta^i_j} \psi_{\alpha}(\vec{r}_i)
\eneq where $i$ is the particle index, $j=1,2$ and $\theta^i_j$ are the eigenvalues. In the wavefunction $\psi_\alpha (\vec{r}_i)$, we have suppressed the position of all other particles but the $i$'th, as they do not get translated by $T_i$. Let  $T_i(\vec{L}_{mn})$ be the translation operator that translates the $i$'th particle by the primitive translation $\vec{L}_{mn} = m \vec{L}_1 + n \vec{L}_2$:
\beq
T_i(\vec{L}_{mn}) \ket{\psi_{\alpha}} = e^{i \theta_{mn}^i} \ket{\psi_{\alpha}}
\eneq The eigenvalue $\theta_{mn}^i$, which we will encounter again, can be expressed in terms of the eigenvalues of the primitive translations by $\vec{L}_1, \vec{L}_2$,  $\theta^i_1,\theta^i_2$: 
\beq
T_i(\vec{L}_{mn}) = T_i(m \vec{L}_1 + n \vec{L}_2) = T_i(m \vec{L}_1) T_i(n \vec{L}_2) e^{i \hat{z} \cdot (\vec{L}_1 \times \vec{L}_2) \frac{m n}{2 l^2}} = (T_i(\vec{L}_1))^m (T_i(\vec{L}_2))^n  e^{i \pi N_\phi m n}
\eneq 
Hence the eigenvalue of translation by $\vec{L}_{mn}$ is $\theta_{mn}^i = \pi mn N_\phi + m \theta_1^i + n \theta_2^i$. Since eigenstates are symmetric or antisymmetric under exchange of identical particles $i,j$, we have that 
$e^{i \theta_{mn}^i} = e^{i \theta_{mn}^j} = e^{i \theta_{mn}}$: the eigenvalues of the translation operators for particle $i$ do not depend on the particle chosen. The Hilbert space of the problem is separated in different sectors labeled by $\theta_1, \theta_2$. At this point, the analysis of the symmetries of the problem under translational invariance by each particle has run its course. Haldane  \cite{haldane-PhysRevLett.55.2095} showed how to now introduce the many-body formalism for the translation operators. We expand his description in detail below.

Consider $N_e$ electrons on a torus pierced by $N_\phi$ fluxes, such that the filling factor $\nu = N_e/N_\phi$ is a rational number $p/q$, with $(p,q)=1$ relatively prime. Let $N_e = p N$, $N_\phi = q N$ where $GCD(N_e, N_\phi) = N$. One thing we can hope from a many-body formulation of the translation operators is to separate the many-body problem into its center of mass part and a relative coordinate part. The center of mass operator moves each and every particle by the same amount:
\beq
T_{\rm CM}(\vec{a}) = \prod_{i=1}^{N_e} T_i(\vec{a})
\eneq The center of mass  translation by an arbitrary quantity commutes with the Hamiltonian but does not commute with the single-particle translation operators $T_i(\vec{L}_j)$ and hence does not keep us in a Hilbert space specified by $\theta_1, \theta_2$.  To find the center of mass operators act within the same Hilbert  space,  we look for all $\vec{a}$ for which the center of mass operator $T_{\rm CM}(\vec{a})$  commutes with every $T_i(\vec{L}_j)$. Since operators of different particles commute, this is tantamount to imposing the constraint:
\beq
[T_i(\vec{a}), T_i(\vec{L}_j)] = - 2 T_i(\vec{a}+ \vec{L}_j) \sin({\frac{1}{2 l^2} \hat{z} \cdot(\vec{a} \times \vec{L_j})}) =0  \label{commutatorofcenterofmasswithsingleparticle}
\eneq  
Hence the condition for vanishing commutator is $\hat{z} \cdot(\vec{a} \times \vec{L_j}) = 2 l^2 \pi r$. Since $\vec{L}_1, \vec{L}_2$ span the $2$-D plane, it is clear that any vector in the plane can take the form $\vec{a}= \alpha \vec{L}_1 + \beta \vec{L}_2$ where $\alpha, \beta$ are real numbers. Taking $j=2$ in Eq[\ref{commutatorofcenterofmasswithsingleparticle}] we have: $\alpha \hat{z} \cdot (\vec{L}_1\times \vec{L}_2) = 2 \pi l^2 r = \alpha  2 \pi l^2 N_{\phi}$. Hence $\alpha = r/ N_\phi$, and similarly - by taking $j=1$, for $\beta$.  Hence the most general center of mass motion which leaves the system invariant is $T_{\rm CM} (\vec{L}_{mn}/N_\phi)$. All other forms of center of mass translation change the Hilbert space of the system.

Our purpose now is to separate the motion of the system into a center of mass motion and a relative motion of the particles. Towards this, Haldane  \cite{haldane-PhysRevLett.55.2095}  defined the relative translation operator $\tilde{T}_i$ acting on particle $i$ so that the motion of that particle is compensated by the motion of all other particles in the opposite direction:

\beq
\tilde{T}_i(\vec{a}) = \prod_{j=1}^{N_e} T_i(\vec{a}/N_e) T_j (-\vec{a}/N_e)  =  T_i((N_e-1)\vec{a}/N_e) \prod_{j=1, \; j \ne i }^{N_e} T_j (-\vec{a}/N_e)
\eneq Since $T_i(\vec{a}) T_i(-\vec{a})=1$, the relative translation operator of the system has the property that $\prod_{i=1}^{N_e}  \tilde{T}_i (\vec{a}) = 1$. When $\tilde{T}_i(\vec{a})$ applied to a function of $\vec{r}_i -\vec{ r}_j$ ($i\ne j$), it translates the function by $\vec{a}$; when applied to a function of $\vec{r}_j -\vec{ r}_k$ ($k,j, \ne i$), it does nothing. Due to the periodicity of the interaction term, only $\tilde{T}_i(\vec{L}_{mn})$ commute with the Hamiltonian. We have introduced the center of mass operator and the relative translation operator, so it seems natural that the total translation operator factorizes in a product of the two:
\beq
T_i(\vec{a}) = T_{\rm CM}(\frac{\vec{a}}{N_e}) \tilde{T}_i(\vec{a})   = \prod_{j= 1, \;\; j \ne i}^{N_e} T_j(\frac{\vec{a}}{N_e}) T_j(-\frac{\vec{a}}{N_e}) T_i(\frac{\vec{a}}{N_e}) (T_i(\frac{\vec{a}}{N_e}))^{N_e-1}
\eneq 
The relative translation operator commutes with any center of mass translation for any $\vec{a}, \vec{b}$:$[\tilde{T}_i(\vec{a}),T_{\rm CM}(\vec{b})] =0$, a clear indication that both of them are diagonalizable at the same time. To review the bidding, we have introduced center of mass translations $T_{\rm CM}(\vec{a})$ which commute with the Hamiltonian, found the ones ($T_{\rm CM}(\vec{L}_{mn}/N_\phi)$) which commute with the single particle translation operators $T_i(\vec{L}_j)$ - even though, as we will see, the $T_{\rm CM}(\vec{L}_{mn}/N_\phi)$ do not necessarily commute between themselves and found the relative momentum operators which commute with the Hamiltonian and with the center of mass translations.

We would like to simultaneously diagonalize the relative translation operators and the Hamiltonian, and are hence after the maximal set of $\tilde{T}_i(\vec{a})$  which commute with each other $[\tilde{T}_i(\vec{a}) , \tilde{T}_j (\vec{b})]=0$ and which act in the same Hilbert space, i.e. which commute with $T_j(\vec{L}_{mn})$. What are the $\vec{a}, \vec{b}$ that satisfy these equations?
To solve these constraints, it is simple to expand $\vec{a} = \alpha \vec{L}_1 + \beta \vec{L}_2$, which is always possible, with $\alpha, \beta$ real numbers. From the commutator or relative translations with the single particle translations of primitive lattice vectors, we have, for $j=i$:

\begin{eqnarray}
& 0 =[\tilde{T}_i(\vec{a}) , {T}_i (\vec{L}_{mn})] = \prod_{m=1, \; m \ne i }^{N_e} T_m (-\frac{\vec{a}}{N_e}) [T_i( \frac{(N_e-1)\vec{a}}{N_e}),  {T}_i (\vec{L}_{mn})]=\nonumber \\ &    = - 2 i  \prod_{m=1, \; m \ne i }^{N_e} T_m (-\frac{\vec{a}}{N_e}) T_i (\frac{(N_e-1)\vec{a}}{N_e} + \vec{L}_{mn})\sin(\frac{1}{2 l^2} \frac{N_e-1}{N_e}  \hat{z} \cdot (\vec{a} \times \vec{L}_{mn}))
\end{eqnarray}  
Hence we find: $\frac{(N_e-1)}{N_e} \hat{z} \cdot (\vec{a} \times \vec{L}_{mn}) = 2 \pi l^2 r$ where $r$ is an integer. To find the constraints on $\alpha, \beta$, we can pick $\vec{L}_{mn} = \vec{L}_1, \vec{L}_2$ to obtain: $\frac{(N_e-1)}{N_e} N_\phi \alpha = r$ ,  $\frac{(N_e-1)}{N_e} N_\phi \beta= r$. As we are at filling $N_e/N_\phi = p/q$, we have: $(N_e-1) q \alpha = p r$, $(N_e-1) q \beta = p r$ where $r$ is any integer.  Also from the commutator or relative translations with the single particle translations of primitive lattice vectors, for $j \ne i$, we have:
\begin{eqnarray}
& 0 =[\tilde{T}_j(\vec{a}) , {T}_i (\vec{L}_{mn})]   =\nonumber \\ &=  T_j( \frac{(N_e-1)\vec{a}}{N_e}) \prod_{m=1, \; m \ne i, j }^{N_e} T_m (-\frac{\vec{a}}{N_e}) [T_i (-\frac{\vec{a}}{N_e}),  {T}_i (\vec{L}_{mn})]   = \nonumber \\ &=- 2 i T_j( \frac{(N_e-1)\vec{a}}{N_e}) \prod_{m=1, \; m \ne i, j }^{N_e} T_m (-\frac{\vec{a}}{N_e})T_i (- \frac{\vec{a}}{N_e} + \vec{L}_{mn})\sin(\frac{1}{2 l^2} \frac{-1}{N_e}  \hat{z} \cdot (\vec{a} \times \vec{L}_{mn}))
\end{eqnarray} 
The conditions the above equation gives are: $q \alpha = p r$, $ q \beta = p r$ where $r$ is any integer, consistent with, but more restrictive than the first set of conditions. We now impose the condition that the $0= [\tilde{T}_i(\vec{a}) , \tilde{T}_j (\vec{b})] $. For $i=j$ we have:

\begin{eqnarray}
&\tilde{T}_i(\vec{a})  \tilde{T}_i (\vec{b}) = T_i(\frac{N_e-1}{N_e} \vec{a})  T_i(\frac{N_e-1}{N_e} \vec{b})  \prod_{l \ne i}^{N_e} T_j(-\frac{1}{N_e} \vec{a}) T_j(-\frac{1}{N_e} \vec{b})   = \nonumber \\ 
&= e^{-\frac{i}{2 l^2} \hat{z}\cdot(\vec{a}\times \vec{b}) (\frac{(N_e-1)^2}{N_e^2}  + (N_e-1)\frac{1}{N_e^2})}  T_i(\frac{N_e-1}{N_e} (\vec{a}+\vec{b}))\prod_{l \ne i}^{N_e} T_j(-\frac{1}{N_e} (\vec{a}+\vec{b})) = \nonumber \\
 &= e^{-\frac{i}{2 l^2} \hat{z}\cdot(\vec{a}\times \vec{b}) (\frac{(N_e-1)}{N_e} )}  T_i(\frac{N_e-1}{N_e} (\vec{a}+\vec{b}))\prod_{l \ne i}^{N_e} T_l(-\frac{1}{N_e} (\vec{a}+\vec{b}))
\end{eqnarray} 
For $\tilde{T}_i(\vec{a})  \tilde{T}_i (\vec{b}) $ to equal $\tilde{T}_i(\vec{b})  \tilde{T}_i (\vec{a}) $, we need that $\hat{z}\cdot(\vec{a}\times \vec{b}) (\frac{(N_e-1)}{N_e} ) = 2 \pi l^2 r$. If $\vec{a}= \alpha \vec{L}_1, \vec{b}= \beta \vec{L}_2$ we have: $\frac{(N_e-1)}{N_e} \alpha \beta N_\phi = r$ and hence $(N_e -1 )q \alpha \beta = p r$ for $r$ some integer. For $i\ne j$ we have:

\begin{eqnarray}
&\tilde{T}_i(\vec{a})  \tilde{T}_j (\vec{b}) = T_i (\frac{N_e-1}{N_e} \vec{a} - \frac{1}{N_e} \vec{b})T_j (\frac{N_e-1}{N_e} \vec{b} - \frac{1}{N_e} \vec{a}) \prod_{l \ne i,j} T_l(-\frac{1}{N_e} (\vec{a}+\vec{b})) e^{-\frac{i}{2 l^2} \hat{z}\cdot(\vec{a}\times \vec{b}) )[ - 2(N_e-1) \frac{1}{N_e^2} + (N_e-2) \frac{1}{N_e^2} ]} 
\end{eqnarray} 
and hence (by taking $\tilde{T}_j(\vec{b})  \tilde{T}_i (\vec{a})$ and requiring the vanishing of the commutator),   we must set
$\hat{z}\cdot(\vec{a}\times \vec{b})[ - 2(N_e-1) \frac{1}{N_e^2} + (N_e-2) \frac{1}{N_e^2} ] = 2 \pi l^2 r$ 
If $\vec{a}= \alpha \vec{L}_1, \vec{b}= \beta \vec{L}_2$ we have $ q \alpha \beta = p r$.

We now bring the four sets of constraints on $\alpha, \beta$ together. The conditions for the set ${a}$ for which the relative translation operators commute between themselves and commute with the $T_i(\vec{L}_{mn})$ are: $q\alpha =  p r_1, \;\; q \beta = p r_2, \;\; q \alpha \beta = p r_3, \;\;\; r_1, r_2, r_3 \in {\mathbb{Z}}$ If $q$ is a prime number, then we can show that $\alpha, \beta = p$: we substitute $\alpha, \beta$ from the first two equations in the third to obtain $p r_1 r_2 = q r_3$. Since $p, q$ are relatively prime then $r_3 = p r$ and $r_1 r_2 = q r$ where $r\in {\mathbb{Z}}$. But since $q$ is prime, then $r_1 r_2= q r$ implies $r_1 = q r'$ where $r' \in {\mathbb{Z}}$ (or vice-versal for $r_2$).  We then plug this in $q\alpha = p r_1$ to obtain $\alpha = p r'$ and so $\alpha$ is an integer proportional to $p$. The smallest value (which gives the largest set of $\vec{a} =\alpha \vec{L}_1$) of this is $\alpha =p$. Then $q\beta = r_3 = p r$ and hence $\beta = p$. The set of vectors $\vec{a}$ for which relative translation operators commute with each other \emph{and} with the single particle  momenta is $\vec{a}= {p \vec{L}_{mn}}$ if $q$ is a prime number.  In passing, we note that if $q$ is not a prime, other possibilities arise for the set of maximally commuting relative translations. The simplest case is to assume $q= q_1^2$, i.e. $q$ is a perfect square, but same situation occurs whenever $q$ is not prime. Then if we choose $\alpha =\beta = p/q_1$ we have $r_1= r_2= q_1, r_3= p$ - all integers and hence the relative translation operators commute with themselves and with the one-body translation operators by $\vec{L}_{mn}$. This then gives $\vec{a} = \frac{p}{q_1} \vec{L}_{mn}$. This, however, corresponds to a different choice of resolving the groundstate  center of mass degeneracy, as will be now shown. The center of mass operators $T_{\rm CM}(\vec{L}_{mn}/N_\phi)$ commute with the the $T_i(\vec{L}_{mn})$ but not between themselves (so they cannot all be simultaneously diagonalized):
\begin{eqnarray}
& T_{\rm CM}(\vec{L}_{mn}/N_s) T_{\rm CM}(\vec{L}_{m'n'}/N_s)= \prod_{i=1}^{N_e} [ T_i( (\vec{L}_{mn} + \vec{L}_{m'n'})/N_s)  e^{-\frac{i}{2 l^2} \frac{1}{N_s^2} (m n' - n m')  \hat{z} \cdot (\vec{L}_1 \times \vec{L}_2) }]= \nonumber \\ &=  e^{-i \pi \frac{N_e}{N_s} (m n' - n m') }  \prod_{i=1}^{N_e}  T_i( (\vec{L}_{mn} + \vec{L}_{m'n'})/N_s)=  e^{-i \pi \frac{p}{q} (m n' - n m') }  \prod_{i=1}^{N_e}  T_i( (\vec{L}_{mn} + \vec{L}_{m'n'})/N_s) 
\end{eqnarray} 
As a hint of the $q$-fold degeneracy of the spectrum, for $q$ a prime number, we see that $T_{\rm CM}^q(\vec{L}_{mn}/N_s)$ commutes with $T_{\rm CM}(\vec{L}_{mn})$.  Moreover, if $q$ is not a prime number, and it is, for example, a perfect square $q=q_1^2$, we see that  a set of mutually commuting operators  is $T_{\rm CM}^{q_1}(\vec{L}_{mn})$. 

We now try to diagonalize the maximum commuting set of relative momentum operators $\tilde{T}_i(p \vec{L}_{mn})$ which commute with themselves and also with the $T_i(\vec{L}_{mn})$ operators. We would like to find its eigenvalues, labeled by a $2$-momentum $\vec{k}$. Following Haldane, we make the assumption that the many-body state (which is an eigenstate of $\tilde{T}_i(p \vec{L}_{mn})$) experiences, when acted on by $\sum_{i=1}^{N_e} \exp(i \vec{Q}\cdot \vec{r}_i)$, an increase in its momentum by $\vec{Q}$, as long as $\vec{Q}$ is a reciprocal lattice momentum $\exp(i \vec{Q}\cdot \vec{L}_{mn}) =1$. In other words, let the eigenstates of $\tilde{T}_i(p \vec{L}_{mn})$ be $\ket{\psi(\vec{k})}$, with eigenvalue $\lambda_{\vec{k}}$  
\beq
\tilde{T}_i(p \vec{L}_{mn}) \ket{\psi(\vec{k})} = \lambda_{\vec{k}} \ket{\psi(\vec{k})},\;\;\;\; \sum_i e^{i \vec{Q}\cdot \vec{r}_i} \ket{\psi(\vec{k})} = \ket{\psi(\vec{k}+\vec{Q})}
\eneq By applying $\tilde{T}_i(p \vec{L}_{mn})$ on the $\sum_i e^{i \vec{Q}\cdot \vec{r}_i} \ket{\psi(\vec{k})} $ we obtain:
\begin{eqnarray}
& \lambda_{\vec{k}+ \vec{Q}} \ket{\psi(\vec{k}+\vec{Q})} = \tilde{T}_i(p \vec{L}_{mn})\sum_j^{N_e} e^{i \vec{Q}\cdot \vec{r}_j} \ket{\psi(\vec{k})} =\nonumber \\
 &=  T_i( \frac{N_e-1}{N_e} p \vec{L}_{mn}) \prod_{l \ne i} T_l (-\frac{p \vec{L}_{mn}}{N_e})  (e^{i \vec{Q} \cdot \vec{r}_i} + \sum_{j \ne i}^{N_e} e^{i \vec{Q}\cdot \vec{r}_j}) \ket{\psi(\vec{k})} =\nonumber \\
  &=   ( e^{- i \vec{Q} \frac{p \vec{L}_{mn}}{N_e}} \sum_{l \ne i}^{N_e} e^{i \vec{Q} \cdot \vec{r}_l}  + e^{i \vec{Q}\frac{N_e -1}{N_e} p \vec{L}_{mn}} e^{i \vec{Q} \cdot \vec{r}_i }) \tilde{T}_i(p \vec{L}_{mn}) \ket{\psi(\vec{k} )}= \nonumber \\
 &= e^{- i \frac{p}{N_e} \vec{Q} \cdot \vec{L}_{mn}} \sum_j^{N_e} e^{i \vec{Q}\cdot \vec{r}_j} \lambda_{\vec{k}}  \ket{\psi(\vec{k}) } =  e^{- i \frac{p}{N_e} \vec{Q} \cdot \vec{L}_{mn}} \lambda_{\vec{k}}  \ket{\psi(\vec{k}+\vec{Q}) } 
\end{eqnarray} 
We have then $\lambda_{\vec{k}+\vec{Q}} = \lambda_{\vec{k}} e^{-i \frac{p}{N_e} \vec{Q}\cdot \vec{L}_{mn}}$ and hence $\lambda_{\vec{k}} = D e^{- i \frac{p}{N_e} \vec{k} \cdot \vec{L}_{mn}}$.  $D$ can only be a phase independent on $\vec{k}$.  The constant $D$ can be found by requiring that the $\vec{k}=0$ state remains invariant under all translations. 
\begin{eqnarray}
D  \ket{\psi(\vec{k}=0)}= \tilde{T}_i( p \vec{L}_1) \ket{\psi(\vec{k}=0)} = \tilde{T}_i( -p \vec{L}_2) \ket{\psi(\vec{k}=0)} = \tilde{T}_i( p \vec{L}_2 - p \vec{L}_1) \ket{\psi(\vec{k}=0)}   
\end{eqnarray} 
However, the last term $\tilde{T}_i( p \vec{L}_2 - p \vec{L}_1)$ can be re-expressed in terms of the product of the translation operators of $p \vec{L}_2$ and $p \vec{L}_1$. We have
\begin{eqnarray}
& \tilde{T}_i(p \vec{L}_2) \tilde{T}_i(-p \vec{L}_1) = T_i(\frac{N_e-1}{N_e} p \vec{L}_2) T_i(-\frac{N_e-1}{N_e} p \vec{L}_1)  \prod_{l \ne i} T_l(-\frac{1}{N_e} p \vec{L}_2) T_l(\frac{1}{N_e} p \vec{L}_1)   = \nonumber \\
 & = T_i(\frac{N_e-1}{N_e} p( \vec{L}_2- \vec{L}_1)) \prod_{l \ne i} T_l(-\frac{1}{N_e} p( \vec{L}_2- \vec{L}_1)) e^{\frac{i}{2 l^2} \frac{(N_e-1)^2}{N_e^2} p^2 \hat{z}\cdot(\vec{L}_2 \times \vec{L}_1)} e^{\frac{i}{2 l^2} \frac{N_e -1}{N_e^2} p^2 \hat{z} \cdot (\vec{L}_2 \times \vec{L}_1)  } = \nonumber \\
 &=  \tilde{T}_i( p \vec{L}_2- p \vec{L}_1) e^{\frac{i}{2 l^2} \frac{N_e-1}{N_e} p^2 (- 2 \pi l^2 N_s)}=  \tilde{T}_i( p \vec{L}_2- p \vec{L}_1) e^{\frac{i}{2 l^2} \frac{N_e-1}{N_e} p^2 (- 2 \pi l^2 N_s)} = \tilde{T}_i( p \vec{L}_2- p \vec{L}_1) e^{-i (N_e-1) p q \pi }
\end{eqnarray} 
This yields the equation $D  \ket{\psi(\vec{k}=0)} = D^2 e^{i (N_e-1) p q \pi }\ket{\psi(\vec{k}=0)}$ which gives $D = (-1)^{(N_e-1) p q}$. After establishing the constant $D$, we would like to now find the possible values that $\vec{k}$ can take. We remark that $N_e = p N$ and $N_\phi = q N$ and hence:
\beq
 \tilde{T}_i(p \vec{L}_{mn})=  T_i( \frac{N_e-1}{N}  \vec{L}_{mn}) \prod_{l \ne i} T_l (-\frac{\vec{L}_{mn}}{N})  
\eneq 
We know that $T_i ({\vec{L}_{mn}})$ has fixed eigenvalue $\theta_{mn}$ (which for identical particles does not depend on $i$). This means that the $N$'th power of the relative $\tilde{T}_i(p \vec{L}_{mn})$ operator has a fixed eigenvalue:
\beq
 \tilde{T}_i(p \vec{L}_{mn})^N = T_i((N_e-1) \vec{L}_{mn}) \prod_{l \ne i}^{N_e} T_l(-\vec{L}_{mn}) = e^{i \theta_{mn}(N_e-1) } \prod_{l \ne i}^{N_e} e^{-i \theta_{mn}} =1 = D^{N} e^{- i \vec{k} \cdot \vec{L}_{mn}}
\eneq 
Having determined the constant $D$, we have $D^N = (-1)^{qN_e (N_e -1)}=1$. Hence $e^{-i \vec{k} \cdot \vec{L}_{mn}}=1$, and we have as solutions $\vec{k} \cdot \vec{L}_1 = 2\pi i$, $\vec{k} \cdot \vec{L}_2  = 2 \pi j$, $i,j \in Z$. We now must ask how many of these solutions represent unique eigenvalues of $\tilde{T}_i(p \vec{L}_{mn})$.  Its eigenvalues are $(-1)^{pq (N_e-1)} \exp({-i \vec{k}\cdot \vec{L}_{mn}/N})$, and are different for  $\vec{k} \cdot \vec{L}_1 = 2\pi i$, $\vec{k} \cdot \vec{L}_2  = 2 \pi j$, $i,j \in [1 \ldots N]$. Hence the BZ of the relative translation operators is made up of $N \times N$ values of the momentum.  An alternative way of presenting the resulting momenta is to say that the eigenvalues of the relative momentum $\tilde{T}_i (p \vec{L}_{mn})$ are $e^{-i  \vec{k} \cdot \vec{L}_{mn}/N}$, with $\vec{k} l = \sqrt{ \frac{2 \pi}{ N_s \lambda}} (s- s_0, \lambda(t -t_0))$ where $s,t =1 \ldots N$ and $\lambda= L_x/L_y$ is the aspect ratio, while the $s_0, t_0$ are the quantum numbers belonging to zero momentum: $\exp(2\pi i s_0/N) = \exp(2 \pi t_0/N) = (-1)^{p q(N_e-1)}$.  This exhausts our discussion of the relative translation operators.

Having fixed $\theta_1, \theta_2, k$, we now ask if there are any other degeneracies? The answer is yes. Physically, this is because $\theta_1, \theta_2$ define the single-particle Hilbert space while $k$ is the principal symmetry quantum number of the relative wavefunction $\ket{\psi_{rel}}$. Left untouched so far is the center of mass translational symmetry of the problem, or the center of mass wavefunction $\ket{\psi_{CM}}$. In principle, we have diagonalized only the $\tilde{T}_i(p \vec{L}_{mn})$ but there are missing center of mass operators, $T_{\rm CM}(\vec{L}_{mn}/N_s)$ which also commute with the $T_{i}(\vec{L}_{mn})$ operators and keep us in the same Hilbert space; also, any center of mass translation operator commutes with the relative translation operator, and hence we are well on our way towards finding other commuting operators.  \emph{However, crucially, two center of mass translations $T_{\rm CM} (\vec{L}_{mn}/N_s)$ and $T_{\rm CM}(\vec{L}_{m'n'}/N_s)$ do not commute with each other} and hence they cannot be simultaneously diagonalized. We hence must find the maximum set of $T_{\rm CM}(\vec{L}_{mn}/N_s)$ that can be diagonalized, which, per the above is equal to the maximally commuting set of $T_{\rm CM}(\vec{L}_{mn}/N_s)$.  We can find out how many $T_{\rm CM}(\vec{L}_{mn}/N_s)$ are self-commuting by either brute-force calculation or by a smart argument. We start with the brute-force calculation of the commutator:
\beq
[T_{\rm CM}(\frac{\vec{L}_{mn}}{N_s}), T_{\rm CM}(\frac{\vec{L}_{m'n'}}{N_s})] = e^{-\frac{i}{2 l^2} \frac{\hat{z} \cdot (\vec{L}_{mn} \times \vec{L}_{m'n'})}{N_s^2} N_e } \prod_{i =1}^{N_e} T_i(\frac{\vec{L}_{mn}+ \vec{L}_{m'n'}}{N_s}) = e^{-i \pi\frac{p}{q} (m n' - n m')} \prod_{i =1}^{N_e} T_i(\frac{\vec{L}_{mn}+ \vec{L}_{m'n'}}{N_s}) 
\eneq which vanishes iff $(mn'-n m')/q \in {\mathbb{Z}}$. Hence there are $q$ values possible. If we require $\ket{\psi_{CM}}$ to be an eigenstate of $T_{\rm CM}(L^0/N_s)$ where $L^0$ is some particular primitive translation, the set of all center of mass translations that commute with $T_{\rm CM}(\vec{L}^0/N_s)$ is given by $\{ T_{\rm CM}((q \vec{L}_{mn} + r \vec{L}^0)/N_s)\}$ with $r=0,1, \ldots q-1$. Indeed, the commutator
\beq
[ T_{\rm CM}((q \vec{L}_{mn} + r \vec{L}^0)/N_s),  T_{\rm CM}((q \vec{L}_{m'n'} + r \vec{L}^0)/N_s)] =0
\eneq As Haldane mentions  \cite{haldane-PhysRevLett.55.2095}, this is only one of the few resolutions of the center of mass degeneracy.  The smart and quick  argument which reveals the $q$-fold degeneracy is the following: we know that
\beq
T_i(p \vec{L}_{mn}) = T_{\rm CM}(p \vec{L}_{mn}/N_e) \tilde{T}_i(p \vec{L}_{mn})
\eneq and hence the eigenvalue of
\beq
(T_{\rm CM}(\vec{L}_{mn}/N_s))^q= T_{\rm CM}(p \vec{L}_{mn}/N_e) =T_i(p \vec{L}_{mn})  \tilde{T}_i( -p \vec{L}_{mn})
\eneq is fixed once we have diagonalized $\theta_1, \theta_2, k$ to be $(-1)^{pq (N_e-1)} \exp({i p \theta_{mn} } )\exp( i q k\cdot \vec{L}_{mn}/N_s)$. Since only the $q$'th power of the eigenvalue is fixed, we see that there must be $q$ center of mass operators (of different eigenvalues). If the eigenvalue of $T_{\rm CM}(\vec{L}^0/N_s)$ is $\lambda$, then the eigenvalues of the set of maximally commuting center of mass operators $\{ T_{\rm CM}((q \vec{L}_{mn} + r \vec{L}^0)/N_s)\}$ with $r=0,1, \ldots q-1$ have eigenvalues $\lambda^r (-1)^{pqr (N_e-1)} \exp({i p \theta_{mn} } )\exp( i q k\cdot \vec{L}_{mn}/N_s)$. This completes the many-body theoretical symmetry analysis of the spectrum of Fractional Quantum Hall states. The section so far did not contain new material, although we believe and hope that the detailed and expanded description of the calculations present in  \cite{haldane-PhysRevLett.55.2095}  is useful to the reader. 

\subsection{Building the Hilbert space}

On the torus, each Landau level has an identical number of states that it can accommodate. The operator for an electron at position $(x,y)$ is $\psi(x,y) = \sum_{m, j} \phi_{m,j}(x,y) c_{m,j}$ where $c_{m,j}$ is the annihilation operator of an electron of momentum $j$ in the $m$'th Landau level, and the $\phi_{m,j}$ are the single-particle orbitals:
\beq
\phi_{m,j}(x,y) = \sum_{k \in {\mathbb{Z}}} e^{\frac{2 \pi }{L_y}(j+ k N_\phi)(x+ i y)} e^{-\frac{x^2}{2 l^2}}e^{-\frac{1}{2} \left(\frac{2 \pi l}{L_y} \right)^2(j + k N_\phi)^2} H_m(\frac{2 \pi l}{L_y} (j+ k N_\phi) - \frac{x}{l}))
\eneq $H_m$ is the Hermite polynomial. In the above, we have picked the Landau gauge $\vec{A} = B x \hat{y}$. While the single-particle orbitals given above are not normalized, it is crucial to notice that the normalization factor depends only on the Landau level index $m$ and not on the orbital momentum $j$. This is a feature of the torus geometry, on the sphere the single particle orbitals depend on the angular momentum quantum number. 
 For the lowest Landau level, to which we particularize, $H_0=1$. The translational properties of the single-particle orbitals (for a rectangular lattice $\vec{L}_1 = L_x \hat{x}, \vec{L}_2= L_y \hat{y}$) are trivially obtained:
\beq
\phi_{m,j}(x,y + L_y) = \phi_{m,j}(x,y), \;\;\;\; \phi_{m,j} (x+ L_x, y) = e^{i \frac{2 \pi}{L_y} N_\phi y} \phi_{m,j}(x, y)
\eneq 
In order to obtain the action of the relative translation operators on the many-body states, we note the translational properties under translations by $(N_e-1) p \vec{L}_{mn}/N_e$ and by $- p \vec{L}_{mn}/N_e$:
\beq
\phi_{m,j} (x, y+ \frac{N_e- 1}{N_e} p  L_y )  = \phi_{m,j} (x, y- \frac {1}{N_e} p n L_y ) = e^{-i \frac{2\pi j }{N}} \phi_{m,j}(x,y) \label{translationalsymmetriesonsingleparticleorbitals1}
\eneq
\beq
\phi_{m,j} (x+\frac{N_e- 1}{N_e} p  L_x, y) = e^{i \frac{2 \pi}{L_y} q(N_e-1) y } \phi_{m, j+q}(x,y), \;\; \phi_{m,j} (x-\frac{1}{N_e} p  L_x, y) = e^{ -i \frac{2 \pi}{L_y} q y } \phi_{m, j+q}(x,y) \label{translationalsymmetriesonsingleparticleorbitals2}
\eneq 
In the Landau gauge used, the guiding center translation operator  $T(\vec{L}) $ is related to the usual translation operator $t(\vec{L}) = \exp{\vec{L} \cdot \vec{\nabla}}$ by  $T(\vec{L}) = \exp(\frac{i}{l^2} (L_x y+ \frac{1}{2} L_x L_y)) t(\vec{L})$. On a one-body state $T(\vec{L}) \ket{m,j} = \int\int dx dy T(\vec{L}) \ket{x,y} \langle x,y\ket{m,j} = \int\int dx dy \exp({\frac{i}{l^2} (L_x (y+L_y) + \frac{1}{2} L_x L_y)}   \ket{x+ L_x,y+ L_y}) \langle x,y\ket{m,j}  $ where $\langle x,y\ket{m,j}   = \phi_{m,j}(x,y)$. By switching variables in the integral, we have: $T(\vec{L}) \ket{m,j} = \int\int dx dy \exp({\frac{i}{l^2} (L_x y + \frac{1}{2} L_x L_y)}   \ket{x,y}) \langle x - L_x,y- L_y\ket{m,j} $. Using the properties of Eq[\ref{translationalsymmetriesonsingleparticleorbitals1},  \ref{translationalsymmetriesonsingleparticleorbitals2}], we can prove the following:
\begin{eqnarray}
&\tilde{T}_i(pL_y \hat{y}) \ket{j_1, \ldots, j_{N_e}} =e^{i\frac{2 \pi}{N} \sum_{i=1}^{N_e} j_i} \ket{j_1, \ldots, j_{N_e}}
\end{eqnarray} This was possible because there is no $L_x$ in the $\vec{L}= p L_y \hat{y}$ translation, and hence the factor $ \exp({\frac{i}{l^2} (L_x y + \frac{1}{2} L_x L_y)}  =1$.  The relative translation operator by $p L_x \hat{x}$, $\tilde{T}_i(pL_x \hat{x})$  has the factor $ \exp({\frac{i}{l^2}\frac{N_e-1}{N_e} p L_x y}) =  \exp(i{\frac{2\pi}{L_y} q({N_e-1}) y_i}) $ if it acts on the particle $i$ in the single-particle decomposition, which cancels the factor $\exp(-i{\frac{2\pi}{L_y} q({N_e-1}) y_i})  $ present in $ \langle x_i - \frac{N_e-1}{N_e} p L_x ,y_i \ket{m,j} $. Similarly, $\tilde{T}_i(pL_x \hat{x})$ there contains operators $T_j(-p L_x \hat{x}/N_e)$. These give a factor  $ \exp(-{\frac{i}{l^2}\frac{1}{N_e} p L_x y_j}) =  \exp(-i{\frac{2\pi}{L_y} q y_j})$ which cancel the factor $ \exp(i{\frac{2\pi}{L_y} q y_j})$  arising from $ \langle x_j - \frac{-1}{N_e} p L_x ,y_j \ket{m,j} $. We hence obtain
\begin{eqnarray}
&\tilde{T}_i(pL_x \hat{x}) \ket{j_1, \ldots, j_{N_e}} = \ket{j_1 + q, \ldots, j_{N_e}+q}
\end{eqnarray}
We hence found that (in the Landau Gauge), the many-body Hilbert space vectors $\ket{j_1, \ldots, j_{N_e}}$ is an eigenstate of $\tilde{T}_i (p L_y \hat{y})$ with eigenvalue $e^{i\frac{2 \pi}{N} \sum_{i=1}^{N_e} j_i} $ dependent on the total momentum $\sum_{i=1}^{N_e} j_1 (\mod N) $. Due to the denominator $N$, the momentum is defined only $\mod N$, which is an explicit way of seeing that the relative momentum BZ is made out of $N$ $k_y$ momenta.  Note however, and this is \emph{essential}, that this does \emph{not} imply that all states with identical  $\sum_{i=1}^{N_e} j_i  (\mod N) $ belong to the same Hilbert space. In fact, only states with all  $\sum_{i=1}^{N_e} j_i  (\mod N_\phi) $  belong to the same Hilbert space. The construction of relative translational symmetry-sorted Hilbert space proceeds as follows: First, write down all possible states $\ket{j_1,\ldots,, j_{N_e}}$ with no constraint other than no double occupancy of orbitals in the case of fermions.  Now sort the Hilbert space into different sectors given by the constraint that each sector contains terms with identical $\sum_{i=1}^{N_e} j_i  (\mod N_\phi)$. Not two states having different  $\sum_{i=1}^{N_e} j_i  (\mod N_\phi)$ can be coupled by a momentum-conserving Hamiltonian. Since $N_\Phi = q N$, there will, in general, be more several sectors, with different  $\sum_{i=1}^{N_e} j_i  (\mod N_\phi)$, which have the same  $\sum_{i=1}^{N_e} j_i  (\mod N)$. It is hence a mistake to first sort the Hilbert space first by sectors with identical $\sum_{i=1}^{N_e} j_i  (\mod N)$: this will result in a much larger number of free-many-body states per sector, because the states have yet to be sorted by $\sum_{i=1}^{N_e} j_i  (\mod N_\Phi)$. Once we have sorted the states by $\sum_{i=1}^{N_e} j_i  (\mod N_\phi)$, for the elements in one sector (which has a momentum $\exp(-i  k_y L_y/N)= \exp( i 2 \pi \sum_{i=1}^{N_e} j_i/N)\exp(i \pi p q(N_e-1))$) we must implement the translational symmetry in the other direction. As the translation operator $\tilde{T}_i (p L_x \hat{x})$ takes $\ket{j_1, \ldots, j_{N_e}}$ into $\ket{j_1 + q, \ldots, j_{N_e}+q}$ (which crucially, has a the identical $\sum_{i=1}^{N_e} (j_i +q) (\mod N_\phi) =(q N_e +  \sum_{i=1}^{N_e} j_i) (\mod N_\phi)= \sum_{i=1}^{N_e} j_i (\mod N_\phi)$, and same thing with $N_\phi$ replaced by $N$, so $k_y$ remains unchanged), we must form the orbits of this operator. This operation goes as follows:  for every element of the set $\ket{j_1, \ldots, j_{N_e}}$ form all the elements $\ket{j_1 +  q k, \ldots, j_{N_e} + q k}$ with $k\in {\mathbb{Z}}$. These elements form the orbit of $\tilde{T}_i (p L_x \hat{x})$. Let the number of elements in an orbit be $Z$. We can form eigenstates of the $\tilde{T}_i (p L_x \hat{x})$ by taking combinations of the  states in an orbit:
\beq
\ket{ k_x, k_y} = \sum_{k=0}^{Z} e^{i \frac{2 \pi n }{N}  k}  \ket{j_1 +  q k, \ldots, j_{N_e} + q k}
\eneq where $n$ is an integer. The $k_y$ momentum of these states has been already established by the equation $\exp(- i k_y L_y/N)= \exp( i 2 \pi \sum_{i=1}^{N_e} j_i/N)\exp(i \pi p q(N_e-1))$. To establish the $k_x$ momentum, we note that 

\beq
\tilde{T}_i (p L_x \hat{x}) \ket{ k_x, k_y} =  e^{ i \frac{2 \pi n }{N}  }  \ket{ k_x, k_y}  = (-1)^{pq (N_e-1)} \exp(- i k_x L_x/N)\ket{ k_x, k_y} 
\eneq and hence the $\exp(-i  k_x L_x/N)= \exp( i 2 \pi n/N)\exp(i \pi p q(N_e-1))$, where $n \in [0,\ldots,N-1]$ is an integer. These are the basis states explicitly translationally invariant with the relative translation operator. They allow a large simplification of numerical computation and explain the degeneracies observed in the spectrum of any translationally invariant Hamiltonian. We will come back in more detail to the explicit construction of a translationally invariant Hilbert space in the section \ref{countingzero}.

\subsection{The Density Algebra in the Lowest Landau Level}

The results above can be put in an equivalent form when written in terms of the projected density algebra. In a similar way to how one defined the guiding center momentum $K_\alpha = \Pi_\alpha - \frac{\hbar}{l^2} (\hat{z} \times \vec{r})_\alpha$, a guiding center coordinate $R_\alpha$ (one-body operator) can be defined:
\beq
R_\alpha = r_\alpha - \frac{l^2}{\hbar} (\vec{\Pi} \times \hat{z})_\alpha, \;\;\;\;\;\; \vec{R} = \frac{l^2}{\hbar} \hat{z} \times \vec{K}
\eneq where $r$ is the position of the particle. The guiding center coordinate is then related to the guiding center position and one can obtain the similar commutations: $[\Pi_\alpha, R_\beta]=0$, $[R_\alpha, R_\beta] = i l^2 \epsilon_{\alpha \beta z}$. We can not build a guiding center density operator:
\beq
\rho(\vec{q})  =e^{i \vec{q} \cdot \vec{R}}
\eneq It is known that this operator has several interesting properties. First, it satisfies the GMP algebra:
\beq
\rho(\vec{q}) \rho(\vec{q}') = \rho(\vec{q} + \vec{q}') e^{-\frac{i}{2} l^2 (\vec{q} \times \vec{q}') \cdot \hat{z}},
\eneq 
which is not surprising as it is just a re-statement of the guiding center translation operator:
\beq
T(\vec{L}) = \rho(\frac{\hat{z} \times \vec{L}}{l^2})
\eneq The relation between the translation operators and densities means that \emph{all} the many-body translation operators obtained in the previous section can be expressed in terms of the density operators. If we define the two fundamental lattice momenta $\vec{q}_1 = 2 \pi \vec{L}_2/{\cal A}, \vec{q}_2 = -2 \pi \vec{L}_1/{\cal A}$,  where ${\cal A}$ is the area of the BZ ${\cal A}= 2 \pi l^2 N_\phi$, we obtain $T(\vec{L}_{mn}) = (\rho(\vec{q}_{nm}))^{N_\phi}$ with $q_{mn} = m\vec{q}_1 + n\vec{q}_2$, and $[(\rho(\vec{q}_{nm}))^{N_\phi}, H]=0$.  From here on, one can just copy the many-body formulation of the translation operators to obtain it in terms of the guiding center densities, and define a center of mass density $\bar{\rho}(\vec{q}_{nm})= \prod_{i=1}^{N_e} \rho_i (\vec{q}_{nm}) $ and a set of self-commuting relative density operators $\tilde{\rho}_i(p N_\phi \vec{q}_{nm}) =\rho_i((N_e-1) p N_\phi \vec{q}_{nm}/N_e) \prod_{j \ne i}^{N_e} \rho_j(- p N_\phi \vec{q}_{nm}/N_e)$ which can be simultaneously diagonalized to obtain the relative momentum quantum numbers, an identical copy of the situation encountered in the translation operator example. 

It is crucial to understand how the guiding center density relates to the usual density operator $\rho_0(\vec{q})= \exp(i \vec{q} \cdot \vec{r})$. It is easy to prove that, up to a factor, the guiding center density equals the projection of the usual density operator into the lowest landau level:
\beq
P e^{i \vec{q} \cdot \vec{r}}P = e^{- \frac{q^2 l^2}{4}} \rho(\vec{q})
\eneq In other words, the guiding center density is the projection of the usual density into the lowest band of the Landau level insulator. Let us re-capitulate the bidding. The many-body translational symmetries can be expressed in terms of a set of many-body density operators that obey a GMP algebra and who are the projection of the regular density operators in the lowest band of the insulator.

\section{Counting of Zero Modes for the Read-Rezayi and Other Model States (and Quasiholes) per Momentum Sector on the Torus} 
\label{countingzero}

In this section, we present a procedure which allows for the computation of the number of zero modes (groundstates and quasiholes) of many model FQH states resolved per each $k_x, k_y$ momentum sector of the relative translational operators $\tilde{T}_i (p L_x \hat{x}) , \tilde{T}_i (p L_y \hat{y})$.  We aim to obtain this counting  \emph{without} diagonalizing the Hamiltonians for which these model states are exact zero modes. We aim to give a combinatoric procedure for counting the states that is valid and can be performed for any number of particles.  All states whose quasihole counting satisfied a generalized Pauli principle in orbital space are amenable to our construction. In a future section, we will use this counting and the one numerically obtained from the FCI to check the emergent symmetry proposed in the FCI.

\begin{figure}[tbp]
\includegraphics[width=5.5in]{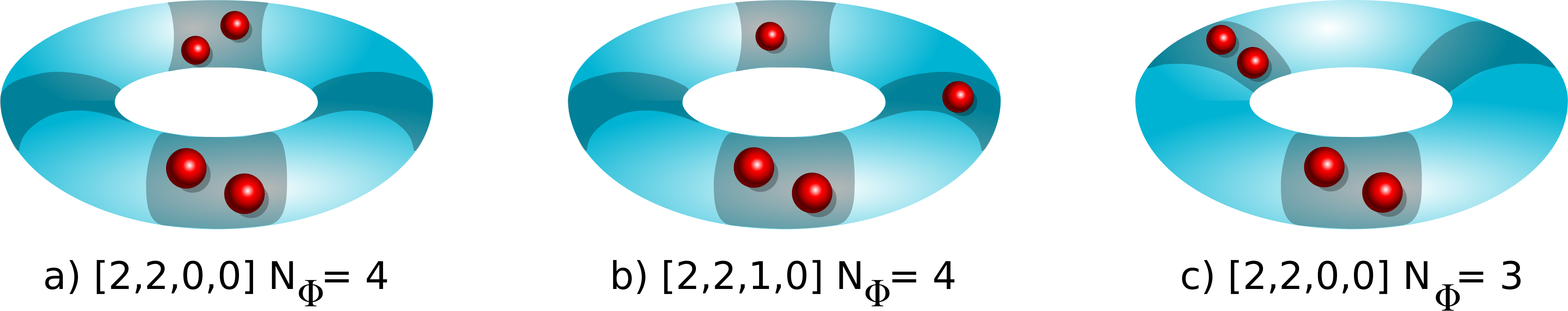}
\caption{A schematic view of the partitions on the torus geometry. The shaded regions are the orbitals (4 for figures $a$ and $b$, 3 for figure $c$). Among the three examples, only the situation $a$ corresponds to a $(2,2)$-admissible partition on the torus. On the sphere geometry both $a$ and $c$ would be $(2,2)$-admissible partitions.}\label{torusoccupation}
\end{figure}

We work with a system of $N_{\phi}$ orbitals. If that system was the sphere manifold,  the problem of counting the zero modes of certain pseudopotential Hamiltonians can be related to the problem of counting partitions satisfying certain conditions. The same conditions, subject to another periodicity constraint, are valid on the torus. We now state and justify these conditions, which have been obtained in previous studies \cite{bernevig-08prl246802,haldaneUCSB2006,bergholtz-06prb081308,Bergholtz-2006-04-L04001,ardonne-2008-04-P04016,seidel-PhysRevLett.95.266405,seidel-PhysRevLett.97.056804,hermanns-PhysRevB.83.241302,seidel-PhysRevB.84.085122}.  Fermionic/bosonic many-body wavefunctions of $N_e$ particles  on any manifold can be expressed as linear combinations of Fock states in the occupation number basis of the single-particle orbitals. Each Fock state $|\lambda\rangle$ can be labeled either  by the list of occupied orbitals, $\lambda$,  or by the occupation number configuration, $n(\lambda)$. 
$\lambda =[\lambda_1, \lambda_2, \ldots, \lambda_{N_e}]$ is an ordered partition of (the $z$ angular momentum  $L_z^{tot}$ on the sphere and) the one-dimensional momentum $K = \sum_{i=1}^{N_e} \lambda_i  (\mod N_\phi)$ on the torus  into $N_e$ parts and each orbital with index $\lambda_j$ is occupied in the Fock state.  We number the orbitals on the torus from $0$ to $N_\phi-1$, and hence each $\lambda_i$ belongs to this interval. By definition, $\lambda_i\geq\lambda_j \textrm{ if } i<j$, and re-ordering of the basis acquires the appropriate number of negative signs depending on whether the state is bosonic or fermionic. 
$n(\lambda)$ is the occupation number configuration. It is defined as $n(\lambda)=\{n_j(\lambda), j=0,\ldots N_\phi\}$, where $n_j(\lambda)$ is the occupation number of the single-particle  orbital with (angular)  $1-D$ momentum $j$. 
In the unnormalized  basis in the lowest Landau level on the torus,
\begin{align}\label{eq:mon}
\langle z_1,\ldots, z_N | \lambda\rangle&= S[\phi_{0,\lambda_1}(z_1)\cdot \ldots \cdot \phi_{0,\lambda_N} (z_N)]
\end{align} 
where $S$ is the process of symmetrization/anti-symmetrization over all indices $i,j$ such that $\lambda_i\neq \lambda_j$.  In this section, we will repeatedly use special kinds of partition --- the so-called $(k,r)$-admissible partition.  For bosons, a $(k,r)$-admissible partition labels a configuration whose occupation configuration has no more than $k$ particles in $r$ consecutive orbitals. For fermions, a $(k,r)$-admissible partition labels a configuration whose occupation number has no more than $k$ particles in $k+r$ consecutive orbitals \emph{and} not more than two particles in two consecutive orbitals. For example, $[2,2,0,0]$ is a $(2,2)$-admissible bosonic partition on the torus of $N_\phi=4$, while $2,1,0,0$ is not. On the torus, due to its periodicity (the orbital of $1D$ momentum $0$ is identified with the one of momentum $N_\phi$) $[2,2,0,0]$ is not a $(2,2)$-admissible partition on the torus of $N_\phi=3$, as depicted in Fig. \ref{torusoccupation}.For obvious reasons, we call these admissibility rules generalized Pauli principles.  Partitions satisfying Pauli generalized principles as the $(k,r)$ one above, or other more complex ones play a prominent role in the physics of the FQH as it was proved that they count the Hilbert space of the quasiholes of many model FQH liquids. The number of $(k,2)$-admissible partitions count the number of quasiholes of the  $\mathbb{Z}_k$ RR state, while the generic $(k,r)$ case counts the number of Jack polynomials states. On the sphere, other concepts such as squeezing, etc are important, but they are not on the torus and we will not discuss them further, referring the reader to the existing literature \cite{bernevig-PhysRevLett.100.246802}
On the sphere and on the torus, the bosonic RR sequence at filling $\nu=k/2$ here (see Sec.~\ref{countingzero} for other model states), are the unique, highest density zero-mode wavefunctions of $(k+1)$-body pseudopotential Hamiltonians\cite{simon-PhysRevB.75.195306}. Recent work\cite{bernevig-08prl246802} has shown the RR bosonic wavefunctions $\psi$ to be Jack polynomials $J^\alpha_{\lambda_0}$ indexed by a parameter $\alpha=-(k+1)$ and the densest-possible $(k,2)$-admissible `root' configuration\cite{stanley1989} $n(\lambda_0)= \{k0k0k0\ldots k0k \}$. On the torus, no extension of the Jack polynomials exist, but the correspondence between zero mode and partition counting is still valid. One can imagine it does, as the admissible partitions implement a generalized Pauli exclusion statistics, which should be independent of the manifold on which the state resides. On the torus, at filling $k/2$, $N_\phi= 2 N_e/k$ for the groundstate and  bosonic $(k,2)$-admissible partitions, we obtain the correct degeneracy ($k+1$ for the $\mathbb{Z}_k$ RR groundstates on the torus) by just counting the occupation number configurations with no more than $k$-particles in $2$ consecutive orbitals $[k\;\; 0\;\; k\;\;0....\;\;k\;\;0]$, $[k-1\;\; 1\;\; k-1\;\;1....\;\;k-1\;\;1]$, $[k-2\;\; 2\;\; k-2\;\;2....\;\;k-2\;\;2]$, and so on till $[0\;\; k\;\; 0\;\;k....\;\;0\;\;k]$. Quasihole counting proceeds identically. Counting states through partitions in this way is usually referred to as the "thin"-torus limit, although we stress that this counting also applies for the sphere (where it can be analytically proved through the Jack-polynomial mapping), as well as for the finite size torus.   

\begin{table}[htbp]
\centering
\begin{tabular}{| l |c|c|c |c|c|c|ccccccccc }
\toprule
¥ $\# \backslash \lambda$\;\;\;\;¥  ¥  0 ¥ 1 ¥ 2 ¥ 3 ¥ 4 ¥ 5 ¥ 6 ¥ 7 ¥ 8 ¥ 9 ¥10 ¥11 ¥\;\;\;\;\;\;$[\lambda_1,\lambda_2]$ ¥ \;\;\; $\sum \lambda_i$  ¥\;\;\;  $\sum \lambda_i$ mod $N_\phi$ ¥ \;\;\;$\sum \lambda_i$ mod $N$ ¥ \\     \hline
 ¥ 1 \;\;\;\;\;\;\;\;\;¥ 1 ¥ 0 ¥ 0 ¥ 1 ¥ 0 ¥ 0 ¥ 0 ¥ 0 ¥ 0 ¥ 0 ¥ 0 ¥ 0 ¥\;\;\;\;\;\;\;\;\;$[3, 0]$ \;¥\;\;\;\;\;\;\;\;  3 ¥ \;\;\;\;\;\;\;\;\;\;\;\;\; 3 ¥ \;\;\;\;\;\;\;\;\;\;\;\;\;\;\;\;\;\;\;\;\;\; 1 ¥ \\    \hline
 ¥ 2 \;\;\;\;\;\;\;\; ¥ 1 ¥ 0 ¥ 0 ¥ 0 ¥ 1 ¥ 0 ¥ 0 ¥ 0 ¥ 0 ¥ 0 ¥ 0 ¥ 0 ¥\;\;\;\;\;\;\;\;\;$[4, 0]$ \;¥\;\;\;\;\;\;\;\;  4 ¥ \;\;\;\;\;\;\;\;\;\;\;\;\; 4 ¥ \;\;\;\;\;\;\;\;\;\;\;\;\;\;\;\;\;\;\;\;\;\; 0 ¥ \\     \hline
    ¥ 3 \;\;\;\;\;\;\;\; ¥ 1 ¥ 0 ¥ 0 ¥ 0 ¥ 0 ¥ 1 ¥ 0 ¥ 0 ¥ 0 ¥ 0 ¥ 0 ¥ 0 ¥\;\;\;\;\;\;\;\;\;$[5, 0]$ \;¥\;\;\;\;\;\;\;\;  5 ¥ \;\;\;\;\;\;\;\;\;\;\;\;\; 5 ¥ \;\;\;\;\;\;\;\;\;\;\;\;\;\;\;\;\;\;\;\;\;\; 1 ¥ \\    \hline
   ¥ 4 \;\;\;\;\;\;\;\; ¥ 1 ¥ 0 ¥ 0 ¥ 0 ¥ 0 ¥ 0 ¥ 1 ¥ 0 ¥ 0 ¥ 0 ¥ 0 ¥ 0 ¥\;\;\;\;\;\;\;\;\;$[6, 0]$ \;¥\;\;\;\;\;\;\;\;  6 ¥ \;\;\;\;\;\;\;\;\;\;\;\;\; 6 ¥ \;\;\;\;\;\;\;\;\;\;\;\;\;\;\;\;\;\;\;\;\;\; 0 ¥ \\    \hline
   ¥ 5 \;\;\;\;\;\;\;\; ¥ 1 ¥ 0 ¥ 0 ¥ 0 ¥ 0 ¥ 0 ¥ 0 ¥ 1 ¥ 0 ¥ 0 ¥ 0 ¥ 0 ¥\;\;\;\;\;\;\;\;\;$[7, 0]$ \;¥\;\;\;\;\;\;\;\;  7 ¥ \;\;\;\;\;\;\;\;\;\;\;\;\; 7 ¥ \;\;\;\;\;\;\;\;\;\;\;\;\;\;\;\;\;\;\;\;\;\; 1 ¥ \\    \hline
     ¥ 6 \;\;\;\;\;\;\;\; ¥ 1 ¥ 0 ¥ 0 ¥ 0 ¥ 0 ¥ 0 ¥ 0 ¥ 0 ¥ 1 ¥ 0 ¥ 0 ¥ 0 ¥\;\;\;\;\;\;\;\;\;$[8, 0]$ \;¥\;\;\;\;\;\;\;\;  8 ¥ \;\;\;\;\;\;\;\;\;\;\;\;\; 8 ¥ \;\;\;\;\;\;\;\;\;\;\;\;\;\;\;\;\;\;\;\;\;\; 0 ¥ \\    \hline
       ¥ 7 \;\;\;\;\;\;\;\; ¥ 1 ¥ 0 ¥ 0 ¥ 0 ¥ 0 ¥ 0 ¥ 0 ¥ 0 ¥ 0 ¥ 1 ¥ 0 ¥ 0 ¥\;\;\;\;\;\;\;\;\;$[9, 0]$ \;¥\;\;\;\;\;\;\;\;  9 ¥ \;\;\;\;\;\;\;\;\;\;\;\;\; 9 ¥ \;\;\;\;\;\;\;\;\;\;\;\;\;\;\;\;\;\;\;\;\;\; 1 ¥ \\    \hline
         ¥ 8 \;\;\;\;\;\;\;\; ¥ 0 ¥ 1 ¥ 0 ¥ 0 ¥ 1 ¥ 0 ¥ 0 ¥ 0 ¥ 0 ¥ 0 ¥ 0 ¥ 0 ¥\;\;\;\;\;\;\;\;\;$[4, 1]$ \;¥\;\;\;\;\;\;\;\;  5 ¥ \;\;\;\;\;\;\;\;\;\;\;\;\; 5 ¥ \;\;\;\;\;\;\;\;\;\;\;\;\;\;\;\;\;\;\;\;\;\; 1 ¥ \\    \hline
           ¥ 9 \;\;\;\;\;\;\;\; ¥ 0 ¥ 0 ¥ 1 ¥ 0 ¥ 0 ¥ 1 ¥ 0 ¥ 0 ¥ 0 ¥ 0 ¥ 0 ¥ 0 ¥\;\;\;\;\;\;\;\;\;$[5, 2]$ \;¥\;\;\;\;\;\;\;\;  7 ¥ \;\;\;\;\;\;\;\;\;\;\;\;\; 7 ¥ \;\;\;\;\;\;\;\;\;\;\;\;\;\;\;\;\;\;\;\;\;\; 1 ¥ \\     \hline
             ¥ 10 \;\;\;\;\;\; ¥ 0 ¥ 0 ¥ 0 ¥ 1 ¥ 0 ¥ 0 ¥ 1 ¥ 0 ¥ 0 ¥ 0 ¥ 0 ¥ 0 ¥\;\;\;\;\;\;\;\;\;$[6, 3]$ \;¥\;\;\;\;\;\;\;\;  9 ¥ \;\;\;\;\;\;\;\;\;\;\;\;\; 9 ¥ \;\;\;\;\;\;\;\;\;\;\;\;\;\;\;\;\;\;\;\;\;\; 1 ¥ \\    \hline
               ¥ 11 \;\;\;\;\;\; ¥ 0 ¥ 0 ¥ 0 ¥ 0 ¥ 1 ¥ 0 ¥ 0 ¥ 1 ¥ 0 ¥ 0 ¥ 0 ¥ 0 ¥\;\;\;\;\;\;\;\;\;$[7, 4]$ \;¥\;\;\;\;\;\;\;\;  11 ¥ \;\;\;\;\;\;\;\;\;\;\; 11 ¥ \;\;\;\;\;\;\;\;\;\;\;\;\;\;\;\;\;\;\;\; 1 ¥ \\    \hline
                 ¥ 12 \;\;\;\;\;\; ¥ 0 ¥ 0 ¥ 0 ¥ 0 ¥ 0 ¥ 1 ¥ 0 ¥ 0 ¥ 1 ¥ 0 ¥ 0 ¥ 0 ¥\;\;\;\;\;\;\;\;\;$[8, 5]$ \;¥\;\;\;\;\;\;\;\;  13 ¥ \;\;\;\;\;\;\;\;\;\;\; 1 ¥ \;\;\;\;\;\;\;\;\;\;\;\;\;\;\;\;\;\;\;\;\;\; 1 ¥ \\    \hline
                   ¥ 13 \;\;\;\;\;\; ¥ 0 ¥ 0 ¥ 0 ¥ 0 ¥ 0 ¥ 0 ¥ 1 ¥ 0 ¥ 0 ¥ 1 ¥ 0 ¥ 0 ¥\;\;\;\;\;\;\;\;\;$[9, 6]$ \;¥\;\;\;\;\;\;\;\;  15 ¥ \;\;\;\;\;\;\;\;\;\;\; 3 ¥ \;\;\;\;\;\;\;\;\;\;\;\;\;\;\;\;\;\;\;\;\;\; 1 ¥ \\    \hline
                     ¥ 14 \;\;\;\;\;\; ¥ 0 ¥ 0 ¥ 0 ¥ 0 ¥ 0 ¥ 0 ¥ 0 ¥ 1 ¥ 0 ¥ 0 ¥ 1 ¥ 0 ¥\;\;\;\;\;\;\;\;\;$[10, 7]$ \;¥\;\;\;\;\;\;  17 ¥ \;\;\;\;\;\;\;\;\;\;\; 5 ¥ \;\;\;\;\;\;\;\;\;\;\;\;\;\;\;\;\;\;\;\;\;\; 1 ¥ \\   \hline
                       ¥ 15 \;\;\;\;\;\; ¥ 0 ¥ 0 ¥ 0 ¥ 0 ¥ 0 ¥ 0 ¥ 0 ¥ 0 ¥ 1 ¥ 0 ¥ 0 ¥ 1 ¥\;\;\;\;\;\;\;\;\;$[11, 8]$ \;¥\;\;\;\;\;\;  19 ¥ \;\;\;\;\;\;\;\;\;\;\; 7 ¥ \;\;\;\;\;\;\;\;\;\;\;\;\;\;\;\;\;\;\;\;\;\; 1 ¥ \\    \hline
                         ¥ 16 \;\;\;\;\;\; ¥ 0 ¥ 1 ¥ 0 ¥ 0 ¥ 0 ¥ 1 ¥ 0 ¥ 0 ¥ 0 ¥ 0 ¥ 0 ¥ 0 ¥\;\;\;\;\;\;\;\;\;$[5, 1]$ \;¥\;\;\;\;\;\;\;\;  6 ¥ \;\;\;\;\;\;\;\;\;\;\;\;\; 6 ¥ \;\;\;\;\;\;\;\;\;\;\;\;\;\;\;\;\;\;\;\;\;\; 0 ¥ \\    \hline
                           ¥ 17 \;\;\;\;\;\; ¥ 0 ¥ 0 ¥ 1 ¥ 0 ¥ 0 ¥ 0 ¥ 1 ¥ 0 ¥ 0 ¥ 0 ¥ 0 ¥ 0 ¥\;\;\;\;\;\;\;\;\;$[6, 2]$ \;¥\;\;\;\;\;\;\;\;  8 ¥ \;\;\;\;\;\;\;\;\;\;\;\;\; 8 ¥ \;\;\;\;\;\;\;\;\;\;\;\;\;\;\;\;\;\;\;\;\;\; 0 ¥ \\    \hline
                             ¥ 18 \;\;\;\;\;\; ¥ 0 ¥ 0 ¥ 0 ¥ 1 ¥ 0 ¥ 0 ¥ 0 ¥ 1 ¥ 0 ¥ 0 ¥ 0 ¥ 0 ¥\;\;\;\;\;\;\;\;\;$[7, 3]$ \;¥\;\;\;\;\;\;\;\;  10 ¥ \;\;\;\;\;\;\;\;\;\;\; 10 ¥ \;\;\;\;\;\;\;\;\;\;\;\;\;\;\;\;\;\;\;\; 0 ¥ \\    \hline
             ¥ 19 \;\;\;\;\;\; ¥ 0 ¥ 0 ¥ 0 ¥ 0 ¥ 1 ¥ 0 ¥ 0 ¥ 0 ¥ 1 ¥ 0 ¥ 0 ¥ 0 ¥\;\;\;\;\;\;\;\;\;$[8, 4]$ \;¥\;\;\;\;\;\;\;\;  12 ¥ \;\;\;\;\;\;\;\;\;\;\; 0 ¥ \;\;\;\;\;\;\;\;\;\;\;\;\;\;\;\;\;\;\;\;\;\; 0 ¥ \\     \hline
              ¥ 20 \;\;\;\;\;\; ¥ 0 ¥ 0 ¥ 0 ¥ 0 ¥ 0 ¥ 1 ¥ 0 ¥ 0 ¥ 0 ¥ 1 ¥ 0 ¥ 0 ¥\;\;\;\;\;\;\;\;\;$[9, 5]$ \;¥\;\;\;\;\;\;\;\;  14 ¥ \;\;\;\;\;\;\;\;\;\;\; 2 ¥ \;\;\;\;\;\;\;\;\;\;\;\;\;\;\;\;\;\;\;\;\;\; 0 ¥ \\    \hline
                ¥ 21 \;\;\;\;\;\; ¥ 0 ¥ 0 ¥ 0 ¥ 0 ¥ 0 ¥ 0 ¥ 1 ¥ 0 ¥ 0 ¥ 0 ¥ 1 ¥ 0 ¥\;\;\;\;\;\;\;\;\;$[10, 6]$ \;¥\;\;\;\;\;\;  16 ¥ \;\;\;\;\;\;\;\;\;\;\; 4 ¥ \;\;\;\;\;\;\;\;\;\;\;\;\;\;\;\;\;\;\;\;\;\; 0 ¥ \\    \hline
                  ¥ 22 \;\;\;\;\;\; ¥ 0 ¥ 0 ¥ 0 ¥ 0 ¥ 0 ¥ 0 ¥ 0 ¥ 1 ¥ 0 ¥ 0 ¥ 0 ¥ 1 ¥\;\;\;\;\;\;\;\;\;$[11, 7]$ \;¥\;\;\;\;\;\;  18 ¥ \;\;\;\;\;\;\;\;\;\;\; 6 ¥ \;\;\;\;\;\;\;\;\;\;\;\;\;\;\;\;\;\;\;\;\;\; 0 ¥ \\    \hline
                  ¥ 23 \;\;\;\;\;\; ¥ 0 ¥ 1 ¥ 0 ¥ 0 ¥ 0 ¥ 0 ¥ 1 ¥ 0 ¥ 0 ¥ 0 ¥ 0 ¥ 0 ¥\;\;\;\;\;\;\;\;\;$[6, 1]$ \;¥\;\;\;\;\;\;\;\;  7 ¥ \;\;\;\;\;\;\;\;\;\;\;\;\; 7 ¥ \;\;\;\;\;\;\;\;\;\;\;\;\;\;\;\;\;\;\;\;\;\; 1 ¥ \\    \hline
                   ¥ 24 \;\;\;\;\;\; ¥ 0 ¥ 0 ¥ 1 ¥ 0 ¥ 0 ¥ 0 ¥ 0 ¥ 1 ¥ 0 ¥ 0 ¥ 0 ¥ 0 ¥\;\;\;\;\;\;\;\;\;$[7, 2]$ \;¥\;\;\;\;\;\;\;\;  9 ¥ \;\;\;\;\;\;\;\;\;\;\;\;\; 9 ¥ \;\;\;\;\;\;\;\;\;\;\;\;\;\;\;\;\;\;\;\;\;\; 1 ¥ \\    \hline
                    ¥ 25 \;\;\;\;\;\; ¥ 0 ¥ 0 ¥ 0 ¥ 1 ¥ 0 ¥ 0 ¥ 0 ¥ 0 ¥ 1 ¥ 0 ¥ 0 ¥ 0 ¥\;\;\;\;\;\;\;\;\;$[8, 3]$ \;¥\;\;\;\;\;\;\;\;  11 ¥ \;\;\;\;\;\;\;\;\;\;\; 11 ¥ \;\;\;\;\;\;\;\;\;\;\;\;\;\;\;\;\;\;\;\; 1 ¥ \\    \hline
                     ¥ 26 \;\;\;\;\;\; ¥ 0 ¥ 0 ¥ 0 ¥ 0 ¥ 1 ¥ 0 ¥ 0 ¥ 0 ¥ 0 ¥ 1 ¥ 0 ¥ 0 ¥\;\;\;\;\;\;\;\;\;$[9, 4]$ \;¥\;\;\;\;\;\;\;\;  13 ¥ \;\;\;\;\;\;\;\;\;\;\; 1 ¥ \;\;\;\;\;\;\;\;\;\;\;\;\;\;\;\;\;\;\;\;\;\; 1 ¥ \\    \hline
                      ¥ 27 \;\;\;\;\;\; ¥ 0 ¥ 0 ¥ 0 ¥ 0 ¥ 0 ¥ 1 ¥ 0 ¥ 0 ¥ 0 ¥ 0 ¥ 1 ¥ 0 ¥\;\;\;\;\;\;\;\;\;$[10, 5]$ \;¥\;\;\;\;\;\;  15 ¥ \;\;\;\;\;\;\;\;\;\;\; 3 ¥ \;\;\;\;\;\;\;\;\;\;\;\;\;\;\;\;\;\;\;\;\;\; 1 ¥ \\     \hline
                  ¥ 28 \;\;\;\;\;\; ¥ 0 ¥ 0 ¥ 0 ¥ 0 ¥ 0 ¥ 0 ¥ 1 ¥ 0 ¥ 0 ¥ 0 ¥ 0 ¥ 1 ¥\;\;\;\;\;\;\;\;\;$[11, 6]$ \;¥\;\;\;\;\;\;  17 ¥ \;\;\;\;\;\;\;\;\;\;\; 5 ¥ \;\;\;\;\;\;\;\;\;\;\;\;\;\;\;\;\;\;\;\;\;\; 1 ¥ \\    \hline
                  ¥ 29 \;\;\;\;\;\; ¥ 0 ¥ 1 ¥ 0 ¥ 0 ¥ 0 ¥ 0 ¥ 0 ¥ 1 ¥ 0 ¥ 0 ¥ 0 ¥ 0 ¥\;\;\;\;\;\;\;\;\;$[7, 1]$ \;¥\;\;\;\;\;\;\;\;  8 ¥ \;\;\;\;\;\;\;\;\;\;\;\;\; 8 ¥ \;\;\;\;\;\;\;\;\;\;\;\;\;\;\;\;\;\;\;\;\;\; 0 ¥ \\    \hline
                   ¥ 30 \;\;\;\;\;\; ¥ 0 ¥ 0 ¥ 1 ¥ 0 ¥ 0 ¥ 0 ¥ 0 ¥ 0 ¥ 1 ¥ 0 ¥ 0 ¥ 0 ¥\;\;\;\;\;\;\;\;\;$[8, 2]$ \;¥\;\;\;\;\;\;\;\;  10 ¥ \;\;\;\;\;\;\;\;\;\;\; 10 ¥ \;\;\;\;\;\;\;\;\;\;\;\;\;\;\;\;\;\;\;\; 0 ¥ \\    \hline
                    ¥ 31 \;\;\;\;\;\; ¥ 0 ¥ 0 ¥ 0 ¥ 1 ¥ 0 ¥ 0 ¥ 0 ¥ 0 ¥ 0 ¥ 1 ¥ 0 ¥ 0 ¥\;\;\;\;\;\;\;\;\;$[9, 3]$ \;¥\;\;\;\;\;\;\;\;  12 ¥ \;\;\;\;\;\;\;\;\;\;\; 0 ¥ \;\;\;\;\;\;\;\;\;\;\;\;\;\;\;\;\;\;\;\;\;\; 0 ¥ \\    \hline
                     ¥ 32 \;\;\;\;\;\; ¥ 0 ¥ 0 ¥ 0 ¥ 0 ¥ 1 ¥ 0 ¥ 0 ¥ 0 ¥ 0 ¥ 0 ¥ 1 ¥ 0 ¥\;\;\;\;\;\;\;\;\;$[10, 4]$ \;¥\;\;\;\;\;\;  14 ¥ \;\;\;\;\;\;\;\;\;\;\; 2 ¥ \;\;\;\;\;\;\;\;\;\;\;\;\;\;\;\;\;\;\;\;\;\; 0 ¥ \\    \hline
                      ¥ 33 \;\;\;\;\;\; ¥ 0 ¥ 0 ¥ 0 ¥ 0 ¥ 0 ¥ 1 ¥ 0 ¥ 0 ¥ 0 ¥ 0 ¥ 0 ¥ 1 ¥\;\;\;\;\;\;\;\;\;$[11, 5]$ \;¥\;\;\;\;\;\;  16 ¥ \;\;\;\;\;\;\;\;\;\;\; 4 ¥ \;\;\;\;\;\;\;\;\;\;\;\;\;\;\;\;\;\;\;\;\;\; 0 ¥ \\     \hline
                  ¥ 34 \;\;\;\;\;\; ¥ 0 ¥ 1 ¥ 0 ¥ 0 ¥ 0 ¥ 0 ¥ 0 ¥ 0 ¥ 1 ¥ 0 ¥ 0 ¥ 0 ¥\;\;\;\;\;\;\;\;\;$[8, 1]$ \;¥\;\;\;\;\;\;\;\;  9 ¥ \;\;\;\;\;\;\;\;\;\;\;\;\; 9 ¥ \;\;\;\;\;\;\;\;\;\;\;\;\;\;\;\;\;\;\;\;\;\; 1 ¥ \\    \hline
                  ¥ 35 \;\;\;\;\;\; ¥ 0 ¥ 0 ¥ 1 ¥ 0 ¥ 0 ¥ 0 ¥ 0 ¥ 0 ¥ 0 ¥ 1 ¥ 0 ¥ 0 ¥\;\;\;\;\;\;\;\;\;$[9, 2]$ \;¥\;\;\;\;\;\;\;\;  11 ¥ \;\;\;\;\;\;\;\;\;\;\; 11 ¥ \;\;\;\;\;\;\;\;\;\;\;\;\;\;\;\;\;\;\;\; 1 ¥ \\    \hline
                   ¥ 36 \;\;\;\;\;\; ¥ 0 ¥ 0 ¥ 0 ¥ 1 ¥ 0 ¥ 0 ¥ 0 ¥ 0 ¥ 0 ¥ 0 ¥ 1 ¥ 0 ¥\;\;\;\;\;\;\;\;\;$[10, 3]$ \;¥\;\;\;\;\;\;  13 ¥ \;\;\;\;\;\;\;\;\;\;\; 1 ¥ \;\;\;\;\;\;\;\;\;\;\;\;\;\;\;\;\;\;\;\;\;\; 1 ¥ \\     \hline
                    ¥ 37 \;\;\;\;\;\; ¥ 0 ¥ 0 ¥ 0 ¥ 0 ¥ 1 ¥ 0 ¥ 0 ¥ 0 ¥ 0 ¥ 0 ¥ 0 ¥ 1 ¥\;\;\;\;\;\;\;\;\;$[11, 4]$ \;¥\;\;\;\;\;\;  15 ¥ \;\;\;\;\;\;\;\;\;\;\; 3 ¥ \;\;\;\;\;\;\;\;\;\;\;\;\;\;\;\;\;\;\;\;\;\; 1 ¥ \\    \hline
                     ¥ 38 \;\;\;\;\;\; ¥ 0 ¥ 1 ¥ 0 ¥ 0 ¥ 0 ¥ 0 ¥ 0 ¥ 0 ¥ 0 ¥ 1 ¥ 0 ¥ 0 ¥\;\;\;\;\;\;\;\;\;$[9, 1]$ \;¥\;\;\;\;\;\;\;\;  10 ¥ \;\;\;\;\;\;\;\;\;\;\; 10 ¥ \;\;\;\;\;\;\;\;\;\;\;\;\;\;\;\;\;\;\;\; 0 ¥ \\    \hline
                      ¥ 39 \;\;\;\;\;\; ¥ 0 ¥ 0 ¥ 1 ¥ 0 ¥ 0 ¥ 0 ¥ 0 ¥ 0 ¥ 0 ¥ 0 ¥ 1 ¥ 0 ¥\;\;\;\;\;\;\;\;\;$[10, 2]$ \;¥\;\;\;\;\;\;  12 ¥ \;\;\;\;\;\;\;\;\;\;\; 0 ¥ \;\;\;\;\;\;\;\;\;\;\;\;\;\;\;\;\;\;\;\;\;\; 0 ¥ \\     \hline
                  ¥ 40 \;\;\;\;\;\; ¥ 0 ¥ 0 ¥ 0 ¥ 1 ¥ 0 ¥ 0 ¥ 0 ¥ 0 ¥ 0 ¥ 0 ¥ 0 ¥ 1 ¥\;\;\;\;\;\;\;\;\;$[11, 3]$ \;¥\;\;\;\;\;\;  14 ¥ \;\;\;\;\;\;\;\;\;\;\; 2 ¥ \;\;\;\;\;\;\;\;\;\;\;\;\;\;\;\;\;\;\;\;\;\; 0 ¥ \\    \hline
                  ¥ 41 \;\;\;\;\;\; ¥ 0 ¥ 1 ¥ 0 ¥ 0 ¥ 0 ¥ 0 ¥ 0 ¥ 0 ¥ 0 ¥ 0 ¥ 1 ¥ 0 ¥\;\;\;\;\;\;\;\;\;$[10, 1]$ \;¥\;\;\;\;\;\;  11 ¥ \;\;\;\;\;\;\;\;\;\;\; 11 ¥ \;\;\;\;\;\;\;\;\;\;\;\;\;\;\;\;\;\;\;\; 1 ¥ \\    \hline
                   ¥ 42 \;\;\;\;\;\; ¥ 0 ¥ 0 ¥ 1 ¥ 0 ¥ 0 ¥ 0 ¥ 0 ¥ 0 ¥ 0 ¥ 0 ¥ 0 ¥ 1 ¥\;\;\;\;\;\;\;\;\;$[11, 2]$ \;¥\;\;\;\;\;\;  13 ¥ \;\;\;\;\;\;\;\;\;\;\; 1 ¥ \;\;\;\;\;\;\;\;\;\;\;\;\;\;\;\;\;\;\;\;\;\; 1 ¥ \\    \hline \toprule
\end{tabular}
\caption{Admissible $(1,3)$ configurations (not more than $1$ particle in $3$ consecutive orbitals) for $N_e=2$ particles in $N_\phi=12$ orbitals. The total number of such configurations  ($42$) equals the number of zero modes of the Haldane pseudopotential Hamiltonian whose highest density groundstate is the Laughlin state. We have $N=2$ and hence two $k_y$ possible momenta, $0, \pi/L_y$}\label{Laughlinconfigsoftwoparticlesintwelveorbitals}
\end{table}

The counting of quasiholes for the Jack polynomial states proceeds as follows. First, in the $1D$ momentum on the torus, write down all the $(k,r)$ partitions ($r=2$ gives the RR series) $\lambda = [\lambda_1,\ldots, \lambda_{N_e}]$ possible on the torus (remembering that the torus is periodic). Their number is the \emph{total} number of the quasihole states for that flux. We now aim to resolve the partitions by assuming they represent states and check their properties under the relative translation operators. It is trivial to first resolve $k_y$. To obtain all the states  (unresolved in $k_x$) at a certain $k_y$ we sort the partitions based on the values of $\sum_{i=1}^{N_e} \lambda_i  (\mod N_\phi)$. Partitions with the same  $\sum_{i=1}^{N_e} \lambda_i  (\mod N_\phi)$ belong to the same $k_y$, whose value given by $\exp(- i k_y L_y/N)= \exp( i 2 \pi \sum_{i=1}^{N_e} \lambda_i/N)\exp(i \pi p q(N_e-1))$. Call such a set of partitions $\{\lambda^{k_y}\}$. We now must implement the translational symmetry in the $x$ direction with momentum $k_x$, which we do as in the previous section. From all the partitions $\{\lambda^{k_y}\}$ we again form the orbits  by taking all partitions $[\lambda_1 +q s, \ldots, \lambda_{N_e} + qs]$, forming the orbit of number of elements $Z$ and form the states $\sum_{s=0}^{Z} \exp(i 2 \pi n s/N)  [\lambda_1 +qs, \ldots, \lambda_{N_e} + qs]$ with momentum  $\exp(-i  k_x L_x/N)= \exp( i 2 \pi n/N)\exp(i \pi p q(N_e-1))$. Several important things must be taken into account: in the elements of one orbit, $[\lambda_1 +qs, \ldots, \lambda_{N_e} + qs]$, one needs to bring them to the same form of \emph{decreasing} elements. Note that this is not trivial as $\lambda_i + qs $ is defined $\mod N_\phi$, and that it could  involve sign changes for fermions.  Also, for some $n$'s the linear combination of the orbit's partitions might not exist, as some states make up single orbits. We will see this in the example below. For other orbits, the number of elements might be smaller than $N$, in which case again not all $k_x$ momenta can be present. By building the states $\ket{k_x, k_y}$ one can clearly see which momenta are missing. After momentum resolving, we can easily count the number of states in each $k_x, k_y$ sector. Note that while this procedure is constructive (we do not have a theoretical expression  for the number of quasiholes per sector, but only a way to construct them), it is an analytic map from a series of admissible partitions in $1D$ to a two-dimensional relative momentum BZ.

A complete understanding of how the counting of quasiholes per sector works comes from working out an example. We will try to resolve the counting of quasihole states of a $\nu=1/3$ Laughlin state of two electrons $N_e=2$ on a torus of $N_\phi=12$. The quasihole states are the zero-modes of the Laughlin quasiholes.  Per our analysis above, the spectrum of any translationally invariant Hamiltonian has a (note $N_e/N_\phi= 1/6$ is the filling of the system; the degeneracy comes from it, but we are looking at the quasiholes of the $\nu=1/3$ Laughlin state, and the counting will reflect that)   six-fold degeneracy. We will aim to show that the counting of zero-modes satisfies this center-of-mass degeneracy. Note that the value of $N_e/N_\phi$ fixes the degeneracy while the specified FQH state whose quasiholes we try to find $\nu=1/3$ fixes the Pauli principle responsible for the counting. To start, we write down \emph{all} the $(1,3)$-admissible partitions in $N_\phi=12$ orbitals of $N_e=2$ particles. There are $42$ such configurations and they are given in Table[\ref{Laughlinconfigsoftwoparticlesintwelveorbitals}]. Since $N_e/N_\phi=1/6$, by our theory, the spectrum should be $6$-fold center of mass degenerate. The non-degenerate states should be resolved by $k_y$ and $k_x$ relative translation momenta, which each take $GCD(N_e, N_\phi)=2$ values $0, 2 \pi/L_y$ and $0, 2\pi/L_x$. Per our procedure, we first sort the state into ones with values $\sum_{i=1}^{N_e} \lambda_i (\mod N_{\phi})$. Table[\ref{Laughlinconfigsoftwoparticlesintwelveorbitalssortedbyorbitals}] presents the partitions sorted in such way (they can obviously also be read from the $5$'th column of  Table[\ref{Laughlinconfigsoftwoparticlesintwelveorbitals}]

\begin{table}[htbp]
\centering
\begin{tabular}{| l |c|c|c |c|c|c|ccccccccc }
\toprule
¥  $\sum \lambda_i$ mod $N_\phi$ ¥ \;\;\;\;\;\;\;\;\;\;\;\;\;\;\;\;\;\;\;\;\;\;\;\;$[\lambda_1,\lambda_2]$ ¥\;\;\;\;\;\;\;\;\;\;\;\;\;\;\;\;\;\; \;\;\;\;\;\;$\sum \lambda_i$ mod $N$ ¥ \\     \hline
¥  \;\;\;\;\;\;\;\;\;\;0 ¥\;\;\;\;\;\;\;\;\;\;\;\;\;\;\;\;\;\;\;\;\;\;\;\;\;\;\;\;$[8,4]; [9,3]; [10,2]$ ¥ \;\;\;\;\;\;\;\;\;\;\;\;\;\;\;\;\;\;\;\;\;\;\;\;0  ¥ \\     \hline
¥  \;\;\;\;\;\;\;\;\;\;1 ¥\;\;\;\;\;\;\;\;\;\;\;\;\;\;\;\;\;\;\;\;\;\;\;$[8,5]; [9,4]; [10,3]; [11,2]$ ¥ \;\;\;\;\;\;\;\;\;\;\;\;\;\;\;\;\;\;1   ¥ \\     \hline
¥  \;\;\;\;\;\;\;\;\;\;2 ¥\;\;\;\;\;\;\;\;\;\;\;\;\;\;\;\;\;\;\;\;\;\;\;\;\;\;\;\;$[9 ,5 ]; [10 ,4 ]; [11 ,3 ]$ ¥ \;\;\;\;\;\;\;\;\;\;\;\;\;\;\;\;\;\;\;\;\;\;0  ¥ \\     \hline
¥  \;\;\;\;\;\;\;\;\;\;3 ¥\;\;\;\;\;\;\;\;\;\;\;\;\;\;\;\;\;\;\;\;\;\;\;$[3,0]; [9,6]; [10 ,5 ]; [11,4]$ ¥ \;\;\;\;\;\;\;\;\;\;\;\;\;\;\;\;\;\;1   ¥ \\     \hline
¥  \;\;\;\;\;\;\;\;\;\;4 ¥\;\;\;\;\;\;\;\;\;\;\;\;\;\;\;\;\;\;\;\;\;\;\;\;\;\;\;\;$[4 ,0 ]; [10 ,6 ]; [11 ,5 ]$ ¥ \;\;\;\;\;\;\;\;\;\;\;\;\;\;\;\;\;\;\;\;\;\;0  ¥ \\     \hline
¥  \;\;\;\;\;\;\;\;\;\;5 ¥\;\;\;\;\;\;\;\;\;\;\;\;\;\;\;\;\;\;\;\;\;\;\;$[5,0]; [4,1];[10,7]; [ 11,6 ]$ ¥ \;\;\;\;\;\;\;\;\;\;\;\;\;\;\;\;\;\;1   ¥ \\     \hline
¥  \;\;\;\;\;\;\;\;\;\;6 ¥\;\;\;\;\;\;\;\;\;\;\;\;\;\;\;\;\;\;\;\;\;\;\;\;\;\;\;\;$[6,0 ]; [5 ,1 ]; [11 ,7 ]$ ¥ \;\;\;\;\;\;\;\;\;\;\;\;\;\;\;\;\;\;\;\;\;\;\;\;0  ¥ \\     \hline
¥  \;\;\;\;\;\;\;\;\;\;7 ¥\;\;\;\;\;\;\;\;\;\;\;\;\;\;\;\;\;\;\;\;\;\;\;$[7,0]; [5,2]; [11 ,8 ]; [6,1]$ ¥ \;\;\;\;\;\;\;\;\;\;\;\;\;\;\;\;\;\;\;\;1   ¥ \\     \hline
¥  \;\;\;\;\;\;\;\;\;\;8 ¥\;\;\;\;\;\;\;\;\;\;\;\;\;\;\;\;\;\;\;\;\;\;\;\;\;\;\;\;$[8 ,0 ]; [6 ,2 ]; [7 ,1 ]$ ¥ \;\;\;\;\;\;\;\;\;\;\;\;\;\;\;\;\;\;\;\;\;\;\;\;\;\;0  ¥ \\     \hline
¥  \;\;\;\;\;\;\;\;\;\;9 ¥\;\;\;\;\;\;\;\;\;\;\;\;\;\;\;\;\;\;\;\;\;\;\;$[6,3]; [7,2]; [8 , 1]; [9,0]$ ¥ \;\;\;\;\;\;\;\;\;\;\;\;\;\;\;\;\;\;\;\;\;\;1   ¥ \\     \hline
¥  \;\;\;\;\;\;\;\;\;\;10 ¥\;\;\;\;\;\;\;\;\;\;\;\;\;\;\;\;\;\;\;\;\;\;\;\;\;\;\;\;$[ 7,3 ]; [8 ,2 ]; [9 ,1 ]$ ¥ \;\;\;\;\;\;\;\;\;\;\;\;\;\;\;\;\;\;\;\;\;\;\;\;0  ¥ \\     \hline
¥  \;\;\;\;\;\;\;\;\;\;11 ¥\;\;\;\;\;\;\;\;\;\;\;\;\;\;\;\;\;\;\;\;\;\;\;$[7,4]; [8,3]; [9 ,2 ]; [10,1]$ ¥ \;\;\;\;\;\;\;\;\;\;\;\;\;\;\;\;\;\;1   ¥ \\     \hline
\toprule
\end{tabular}
\caption{Partitions $[\lambda_1, \lambda_2]$ sorted by identical $\sum_i \lambda_i \mod N_\phi$.}\label{Laughlinconfigsoftwoparticlesintwelveorbitalssortedbyorbitals}
\end{table}

Notice that the $\sum_{\lambda_i} \mod  N =0 $ sector (which is the equivalent of the $k_y=0$ sector, as $(-1)^{pq (N_e-1)}=1$ has $3$ elements while the $\sum_{\lambda_i} \mod  N =1 $ sector (the $k_y= 2\pi/L_y$ sector) has $4$ elements.  Notice how the counting of each of the $k_y=0, \pi/L_y$ sectors is $q=6$-fold degenerate, as the theoretical symmetry analysis mentions. We now would like to implement the $x$ translational symmetry to find the counting of Laughlin quasiholes per momentum sector. The same result will be obtained no matter which of the $6$-fold degenerate sectors we pick. 

We first look at $k_y=0$ sector with the states mapped by the partitions $[8,4], [9,3],[10,2]$. To form the orbits we must look at all elements $[\lambda_1+ qs, \lambda_2+ qs] =[\lambda_1+ 6s, \lambda_2+ 6s]$. The orbit of $[8,4]$ contains $[14, 10]\equiv [2,10]$ (and obviously the orbit of $[10,2]$ contains $[16,8]\equiv [4,8]$). Hence the two states $[8,4],[10,2]$ can be used to build momentum eigenstates:
\beq
\ket{n, k_y=0} =[8,4] + e^{i \frac{2 \pi  n}{N} } [14,10]=[8,4] + e^{i\frac{2 \pi  n}{N} } [2,10] = [8,4] - e^{i \pi  n } [10,2] 
\eneq where we have used $N=2$ and $[2,10] = -[10,2]$ because the particles are fermions. The action of the relative translation operator gives: 
\beq
\tilde{T}_i(p L_x \hat{x}) \ket{0, k_y} = [14,10]- [16,8] = [2,10]-[4,8]= \ket{0,k_y}
\eneq and hence this state has momentum $k_x=0$.  Similarly, the state $\ket{1,k_y}$ has momentum $k_x = 2\pi/L_x$. Out of the original $3$ states at momentum $k_y=0$, we now have to analyze $[9,3]$. As it is a single state, it forms an orbit by itself. This is indeed true, as $[9+6, 3+6] = [15,9] \equiv [3,9]= -[9,3]$. Due to the fermionic sign, this is state at momentum $k_x= 2 \pi/L_x$. We have now determined that we have $1$ zero mode state at momentum $k_x,k_y =0,0$ and $2$ at momentum $k_x,k_y = 2\pi/L_x, 2\pi/L_y$.  Note that of course, our choice of which states to pick for the $k_y=0$ momentum is irrelevant, giving the same counting. If instead we had picked the set of $k_y=0$ states $[9,5], [10,4],[11,3]$, then $[9,5] - [11,3]$ would be the state at $k_x=0$, while $[9,5]+[11,3]$ and $[10,4]$ would be the state at momentum $k_x=2 \pi/L_x$. This completes the symmetry analysis of the zero modes of $2$ particles in $N_\phi=12$ of the Haldane pseudopotential Hamiltonian for the Laughlin state at $k_y=0$.

We now look at $k_y=2 \pi/L$ sector with the states denoted by the partitions $[8,5],[9,4],[10,3],[11,2]$. The orbit of $[8,5]$ is $[2,11]\equiv -[11,2]$, while the orbit of $[9,4]$ is $[3,10]\equiv -[10,3]$. Hence the state $[8,5]-[11,2]$ has momentum $k_x=0$ while $[8,5]+[11,2]$ has $k_x=2 \pi/L_x$. Identically, $[9,4] - [10,3]$ and$[9,4]+ [10,3]$ have momenta $k_x=0, \pi/L_x$. Hence there are $2$ states for $k_x=0$ and $2$ states for $k_x=2\pi/L_x$. The same counting would be obtained if we picked any of the other states at $k_y=2\pi/L_y$. We have now given a combinatoric prescription for counting the states per momentum sector of the relative translation operators. For $2$-particles in $12$ orbitals the quasiholes of the Laughlin state have the following quantum numbers: $2$ states in each of the $k_x,k_y$ equal to $0, 2\pi/L_y$, $2\pi/L_x, 0$, and $2\pi/L_x, 2\pi/L_y$ momenta, and $1$ state in the $0,0$ momentum. There is $6$-fold center of mass degeneracy which multiplies these $7$ states to obtain the total of $42$ quasihole states.

As a last simple example, let us look at resolving the quantum numbers of the bosonic groundstates of the Haldane pseudopotential for the MR interaction for $N_e=4$ particles in $N_\phi=4$ fluxes. As $N_e/N_\phi=1$ there is no center of mass degeneracy and the BZ of the relative translation operators is a $4\times 4$ momenta. We first write down all the $(k,r)= (2,2)$ partitions (whose counting corresponds to the bosonic MR state) of $4$ particles in $4$ orbitals. They are, in occupation number $2020$, $0202$, $1111$, totaling $3$, the correct non-Abelian degeneracy of the MR state, and they correspond to the partitions $\lambda=[2,2,0,0], [3,3,1,1], [3,2,1,0]$ respectively. We first sort the states by $\sum_{i=1}^4 \lambda_i (\mod 4)$,  which gives $0, 0, 2$. The  $\sum_{i=1}^4 \lambda_i (\mod 4) = 2$ state is $[3,2,1,0]$ has $e^{- i k_y L_y/4} = e^{i 2\pi  \cdot 2/4}\cdot(-1)$, the $-$ sign coming from $pq(N_e-1)$ being odd. Hence $k_y=0$. We now find its $k_x$: $\tilde{T}_i(p L_x \hat x) [3,2,1,0] = [4,3,2,1]\equiv [0,3,2,1]  \equiv [3,2,1,0]$ by virtue of the bosonic nature of the state. Hence $e^{-i k_x L_x/N} =-1$ or $k_x=4\pi /L_x$. There is $1$ state at $k_x,k_y = 4 \pi/L_x, 0$. The two other states $[2,2,0,0],[3,3,1,1]$ belong to the $k_y= 4 \pi/L_y$ sector and to the same orbit. We hence form the states: 
\begin{eqnarray}
&\ket{0,k_y} = [2,2,0,0] + [3,3,1,1],\;\;\;\;  \ket{1,k_y} = [2,2,0,0] +e^{i \frac{\pi}{2}}  [3,3,1,1] = [2,2,0,0] + i [3,3,1,1] \nonumber \\ & \ket{2,k_y} = [2,2,0,0] -  [3,3,1,1] ,\;\;\;\;  \ket{3,k_y} = [2,2,0,0] +e^{i \frac{3 \pi}{2}}  [3,3,1,1] = [2,2,0,0] - i [3,3,1,1] 
\end{eqnarray}
It is clear that two states $[2,2,0,0]$, $[3,3,1,1]$ cannot be made into the $4$ independent momentum states above. Indeed, this is because $\ket{1, k_y}, \ket{3,k_y}$ are not momentum eigenstates; for example $\tilde{T}_i (p L_x\hat{x}) \ket{1, k_y} = i \ket{3,k_y}$. The only two momentum eigenstates are $\ket{0, k_y}$ at momentum $k_x=4 \pi/L_x$ and $\ket{2, k_y}$, at momentum $k_x =  2\pi (4 \mod N)/L_x  = 0$. We now found that our combinatoric counting principle predicts that the zero modes of the Haldane pseudopotential for which the MR wavefunction is the groundstate  are resolved by the relative translation operators into momenta $k_x,k_y$ equal to $(4 \pi/L_x, 0)$, $(0, 4 \pi/L_y)$ and $4 \pi/l_x, 4 \pi/L_y)$. An exact diagonalization of the Hamiltonian confirms this.

 It is immediately obvious that the current example can be easily generalized to all the $(k,r)$ states. Our procedure allows for a combinatoric approach to resolving the number of quasiholes per momentum sector just by counting partitions. It goes beyond the usual counting on the $1D$-torus momentum which can only give the total number of quasiholes per $1$ D momentum.
More examples and data for Abelian and non-Abelian states will be presented in the section which uncovers the mapping of FQH to FCI.

\section{Emergent Many-Body Symmetries in the Fractional Chern Insulator}

In this section we show the appearance of many-body translational symmetry operators in the FCI. We first re-derive the result, obtained in \cite{Parameswaran-2011arXiv1106.4025P} and rederived in \cite{goerbig-2012epjb}, that the long-wavelength algebra of the projected density operators in a Chern insulator has a GMP form when the Berry curvature variation is not large. We show that the non-commutativity of the projected density matrices is required by the nonzero Chern number in the nontrivial Chern insulator.  We then show that the problem of a Hamiltonian with such a projected density algebra can be mapped into a FQH problem of a translationally invariant Hamiltonian  \emph{superimposed} on a background lattice of $N_x\times N_y(= N_\phi)$ sites. We then show that this problem admits relative translational operators and momenta which are a folding of the ones obtained in the FQH case in the continuum.

\subsection{One-body projected density algebra of an insulator}

In this subsection, we re-derive the result obtained in \cite{Parameswaran-2011arXiv1106.4025P}. We work with translationally invariant one-body Hamiltonians:
\beq
H = \sum_{i,j, \alpha, \beta} c_{i\alpha}^\dagger h_{i-j; \alpha, \beta} c_{j \beta}
\eneq where $\alpha, \beta$ contain orbital and spin indices. We call $\alpha$ an orbital index.  We now quickly introduce some conventions. Our  Fourier transform sign convention is $c_{k\alpha} = \frac{1}{\sqrt{N_s}} \sum_j e^{- i k j} c_{j\alpha}$  where by $k j$ we mean $\vec{k} \cdot \vec{j}$ and $N_s$ is the number of sites on a lattice with aspect ratio $N_x \times N_y$. The Bloch Hamiltonian matrix is: $h_{\alpha \beta}(k) = \frac{1}{\sqrt{N_s}} \sum_r  e^{- i k r}  h_{r; \alpha, \beta} $ The Bloch Hamiltonian can be separated into normal mode operators $\gamma_k^n$ at momentum $k$ of the band $n$ $H = \sum_{k, n} E_n(k) \gamma_{k}^{n \dagger} \gamma_{k}^{n}$ where the normal modes can be written as a matrix rotation of the original electron operators $\gamma_k^n = \sum_\beta u^{n \star}_{k, \beta} c_{k, \beta}$. The elements of the matrix  are $u^{n \star}_{k, \beta}$ which form the  eigenstates of the Bloch Hamiltonian  $\sum_\beta h_{\alpha \beta}(k) u^{n }_{k, \beta}= E_n(k) u^{n}_{k, \alpha}$.
We then define the projector into the band $n$ at momentum $k$ by
\beq
P_{nk} = \gamma_k^{n \dagger} \ket{0}{\bra{0}} \gamma_k^n
\eneq
Fractional topological insulators are usually constructed and observed in models with fractionally filled bands whose bandwidth is very small, such that interactions and not the kinetic energy dominate the physics. The ideal example of such an insulator is the flat band one-body deformation  of any Hamiltonian insulator $h(k)$:
\beq
H_{FB} =- \sum_{n_1, E_{n_1,k}<\mu}  P_{n_1, k} +  \sum_{n_2, E_{n_2,k}>\mu}  P_{n_2, k} 
\eneq From now on, we will consider the physics of the (fractionally) occupied bands and look only at projectors into the occupied bands. We define the density operator per orbital  $\alpha$ at site $j$, $\rho_j^\alpha = c_{j \alpha}^\dagger c_{j \alpha}$ whose Fourrier transform reads: $\rho_q^\alpha =  \frac{1}{\sqrt{N_s}} \sum_j e^{- i q j} \rho_j^\alpha = \frac{1}{\sqrt{N_s}} \sum_k c^\dagger_{\alpha,k} c_{\alpha, k+q}$. Notice that  we do not shift the sum to put $k+q/2, k-q/2$ because, while perfectly suitable in the continuum, that would change the boundary conditions on a lattice at which $q_i$ is quantized in units of $2\pi/N_i$ $(i=x,y)$

As we are interested in the low-energy physics in the fractionally filled bands, we define the projected density operator $\tilde{\rho}^{n_1, \alpha, n_2}_{k_1, q, k_2} = P_{n_1, k_1} \rho^{\alpha}_{q} P_{n_2, k_2}$ which has the following form: $\tilde{\rho}^{n_1, \alpha, n_2}_{k_1, q, k_2}  = \frac{1}{\sqrt{N_s}} u_{\alpha, k_1}^{n_1 \star} u_{\alpha, k_1+ q}^{n_2}\delta_{k_2, k_1+q}  \gamma_{k_1}^{n_1 \dagger} \ket{0}{\bra{0}} \gamma_{k_1+q}^{n_2}$ The full projected density involves a summation over the orbital $\alpha$,  and over the projection operators $P = \sum_{k,n} P_{n,k}$:
\beq
\tilde{\rho}_q =\frac{1}{\sqrt{N_s}}  \sum_{n_1, n_2, \alpha, k }  u_{\alpha, k}^{n_1 \star} u_{\alpha, k+ q}^{n_2}  \gamma_{k}^{n_1 \dagger} \ket{0}{\bra{0}} \gamma_{k+q}^{n_2}
\eneq  Taking of two densities at momenta $q$ and $w$, we obtain:
\begin{widetext}
\beq
[\rho_q, \rho_w]=\frac{1}{N_s} \sum_{k, n_1,n_2,m_2, \alpha, \beta}  ( u_{\alpha, k_1}^{n_1 \star} u_{\alpha, k+q}^{n_2} u_{\beta, k+q}^{n_2 \star} u_{\beta, k+ w +q}^{m_2} - u_{\beta, k}^{n_1 \star} u_{\beta, k+ w}^{n_2}  u_{\alpha, k+w}^{n_2 \star} u_{\alpha, k+w+q}^{m_2} )  \gamma_{k}^{n_1 \dagger} \ket{0}{\bra{0}} \gamma_{k+q+w}^{m_2} \label{commutatorofdensitiesinChernInsulator}
\eneq
\end{widetext}
We particularize to the low-wavelength continuum limit $q<< 1$ (in lattice constant units). On a discrete lattice, low $q$ represents a small integer combination of $2\pi/N_i$. 
In this limit, a Taylor expansion gives:
\beq
u_{\alpha, k+q}^n = u_{\alpha, k}^n + q_i \partial_i  u_{\alpha, k}^n  + \frac{1}{2} q_i q_j \partial_i \partial_j u_{\alpha, k}^n 
\eneq where $\partial$ denote discrete momentum derivatives. We expand the difference in Eq[\ref{commutatorofdensitiesinChernInsulator}] in the wavelength limit, and use the identities $\sum_\alpha u_{\alpha, k_1}^{n_1 \star} u_{\alpha, k_1}^{n_2} = \delta_{n_1, n_2}$, $\sum_\alpha (\partial_i u_{\alpha, k_1}^{n_1 \star}) u_{\alpha, k_1}^{n_2} +  u_{\alpha, k_1}^{n_1 \star} \partial_i  u_{\alpha, k_1}^{n_2}  =0$ and
\beq
A_{ij} (2 (\partial_i u_{\alpha, k_1}^{n_1 \star}) (\partial_j u_{\alpha, k_1}^{n_2}) + (\partial_i \partial_j  u_{\alpha, k_1}^{n_1 \star})u_{\alpha, k_1}^{n_2}+  u_{\alpha, k_1}^{n_1 \star} (\partial_j \partial_j u_{\alpha, k_1}^{n_2}) =0
\eneq (valid for $A_{ij}$ a symmetric tensor), to find, after tedious algebra (presented in Appendix 1), that in the long wavelength limit:
\begin{widetext}
\beq
\sum_{n_2} u_{\alpha, k}^{n_1 \star} u_{\alpha, k+q}^{n_2} u_{\beta, k+q}^{n_2 \star} u_{\beta, k+ w +q}^{m_2} - u_{\beta, k}^{n_1 \star} u_{\beta, k+ w}^{n_2}  u_{\alpha, k+w}^{n_2 \star} u_{\alpha, k+w+q}^{m_2} \approx  \frac{i}{2} (q_i w_j - w_i q_j) F_{ij}^{n_1 m_2}
\eneq
\end{widetext} where $F_{ij}^{n_1, m_2}$ is the non-Abelian field strength: $F_{ij}^{n_1, m_2} = \partial_i A_j^{n_1, m_2}- \partial_j A_i^{n_1, m_2} - i [A_i, A_j]^{n_1, m_2}$ with the Berry potential $A_j^{n_1n_2} = -i \sum_\beta u_\beta^{n_1 \star} \partial_j u_\beta^{m_2}$ the non-Abelian field strength. The reference \cite{Parameswaran-2011arXiv1106.4025P} hence finds:

\beq
[\rho_q, \rho_w]=-\frac{i}{2} (q_i w_j - w_i q_j) \frac{1}{N_s} \sum_{k, n_1,m_2} F_{ij}^{n_1, m_2}  \gamma_{k}^{n_1 \dagger} \ket{0}{\bra{0}} \gamma_{k+q+w}^{m_2} \label{nonabelianmatrix}
\eneq
 The difference with the result of \cite{Parameswaran-2011arXiv1106.4025P} is only apparent because we are working in the low wavelength limit $q,w$ small and we kept terms to second order in $q,w$, to the order in perturbation theory that we are working at, we have that $\sum_b u_b^{\alpha \star}(k_+) u_b^\alpha(k_-)\approx \sum_b u_b^{\alpha \star}(k_+) u_b^\alpha(k)= 1$ which proves the equivalence between the formula in \cite{Parameswaran-2011arXiv1106.4025P} and the above.  The smallest $q_x, w_y$ are $q_x= 2 \pi/N_x$ and $q_y = 2\pi/w_y$.

 Before we investigate the full implications of this result, we point out three important results. First, the commutator of two densities has to be non-zero in a nontrivial Chern insulator. That is so because the Chern number of the insulator can be expressed as a trace over the density commutator, as shown in Appendix $2$. Second, the commutator of densities at the smallest available lattice momentum $2\pi/(N_x a), 2\pi/(N_y a)$ with $a$ the lattice constant is identical to the commutator of the projected $X$ and $Y$ lattice position operators, as shown in Appendix $3$. The trace of this commutator is known to give the Chern number. Third, as a bonus, the expression of the Chern number in terms of the commutator of the projected $X,Y$ coordinates allows for a direct derivation of the Chern number of the continuum Landau level, which was not possible with the usual TKNN formula for the Chern number which involve derivatives over the two momenta. We show this in Appendix $4$

\subsection{Girvin-MacDonald-Platzman Algebra for Insulators with Smooth Berry Curvature}

We now particularize to two-band models (insulators with one band below and above the gap) or to many-band insulators where the non-Abelian components of the field strength can be neglected. We observe that the commutator in Eq[\ref{nonabelianmatrix}] does not form an algebra as the right-hand side is not easily expressible in terms of a density operator. However, the commutator algebra has an eerie similarity to a GMP algebra at long wavelength. As pointed out in \cite{Parameswaran-2011arXiv1106.4025P},  if the local Berry curvature can be replaced by its average:
\beq
F_{ij} (k) =\epsilon_{ijm} B_m(k) =\epsilon_{ij} \frac{\int_{\text{BZ}} d^2k B_m(k)}{\int_{\text{BZ}} d^2k} = \frac{2 \pi C}{{(2\pi/a)^2}} \hat{z}
\eneq 
where $C$ is the band Chern number. We then observe that the field strength can be taken out of the sum  in Eq[\ref{nonabelianmatrix}], and, since to zeroth order in $q+w$, $\rho_{q+w} = \sum_k \gamma_{k}^{n \dagger} \ket{0}{\bra{0}} \gamma_{k+q+w}^{m} + O(q+w)$. The commutator algebra then becomes:
\beq 
[\rho_q, \rho_w]= -i( \vec{q} \times \vec{w} )\cdot \hat{z} \frac{2 \pi C}{(2\pi/a)^2} \rho_{q+w}
\eneq which is identical to the  GMP algebra at small $q$. Several comments are now necessary: first, in a two-band insulator, it is impossible to have a constant Berry curvature due to the no-hair theorem\cite{podolskiandavronprivatecommunication}, although this seems possible in insulators with four or more bands \cite{podolskiandavronprivatecommunication}. However, we will from now on assume that this is indeed possible, and assume the validity, at low-energies of the GMP algebra for Chern insulators. We then observe that the density algebra for the Chern insulator is identical for low momenta to the one for the Landau level QH problem if:
\beq
( \vec{q} \times \vec{w} )\cdot \hat{z} \frac{2 \pi }{(2\pi/a)^2}  =( \vec{q}_1 \times \vec{q}_2)\cdot \hat{z} l^2
\eneq where $q,w$ are the smallest values of the momentum on the lattice: $\vec{q}= 2\pi/(N_xa) \hat{x}, \vec{w} = 2\pi/(N_y a) \hat {y}$ while $\vec{q}_1 = 2 \pi \vec{L}_2/{\cal A}, \vec{q}_2 = -2\pi \vec{L}_1/{\cal A}$ are the smallest momenta of the QH torus. ${\cal A}=(\vec{L}_1 \times \vec{L}_2 )\cdot \hat{z} = 2\pi l^2 N_\phi$ is the area of the continuum torus BZ, ${(2\pi/a)^2}$ is the area of the BZ on the lattice, and we specialized, without loss of generality to a square lattice. Since the LLL has Chern number $1$ (see Appendix $4$), we also took $C=1$. We hence have $( \vec{q}_1 \times \vec{q}_2)\cdot \hat{z} l^2 = (2\pi)^2 l^2/{\cal A} = 2\pi/N_\phi = ( \vec{q} \times \vec{w} )\cdot \hat{z} \frac{2 \pi C}{(2\pi/a)^2}  = 2\pi /(N_x N_y)$ which leads us to the equivalence of the GMP algebras for a Chern insulator with Chern number $1$ on a lattice with $N_x, N_y$ sites in the $x,y$ directions and the algebra of the QH effect on a torus pierced by $N_\phi$ flux quanta \emph{if}:
\beq 
N_\phi = N_x \cdot N_y
\eneq An important remark should be made here. As pointed out in  \cite{Parameswaran-2011arXiv1106.4025P}, it is important to realize that there is no one-to-one mapping between the Chern insulator density algebra and the FQH one. First, the FCI density GMP algebra is valid only at long-wavelengths, and it will be our \emph{assumption} that as system sizes reach thermodynamic limit, its application transcends the long-wavelength limit and it becomes an approximate symmetry of the system at \emph{all} wavelengths. This is indeed justified posteriori by our extensive numerical checks, which prove the existence of center of mass degeneracies in the Chern insulator which would be possible only if the many-body translational operators obeyed GMP algebra for \emph{all} wavelengths. Second, the number of density generators in the FCI and FQH differ ($N_s$ in the FCI, $N_s^2$ in the FQH) \cite{Parameswaran-2011arXiv1106.4025P}, but, as was pointed out in \cite{Parameswaran-2011arXiv1106.4025P}, there are many instances when the significant physics is governed only by the $N_s$ operators in the FCI.

\subsection{Fractional Quantum Hall to Fractional Chern Insulator Mapping}

We now found that the constant Berry curvature Chern insulator density algebra is identical to that of electrons in a uniform magnetic field with number of fluxes $N_\phi = N_x \cdot N_y$. Hence the translational symmetries of the two problems should share similarities. To complete the analogy, since the Chern insulator is defined on a lattice, and hence we need to introduce a lattice to the FQH problem. We are then led to consider the translational symmetries of the Hamiltonian in magnetic field:

\beq
H= \frac{1}{2 m} \sum_j^{N_e} \Pi_j^2 +\frac{1}{2{\cal A}}\sum_{\vec{q}}  V(\vec{q}) \sum_{i<j}e^{i \vec{q} \cdot(\vec{r}_i - \vec{r}_j)}  + V_0 \sum_{i=1}^{N_e} \sum_{m = 1}^{N_x} \sum_{n=1}^{N_y} \delta(\vec{r}_i - m l_x \hat{x} - n l_y \hat{y}) \label{latticeHamiltonianLandauLevel}
\eneq
with $\Pi_j = - i \hbar \nabla_j - e A(r_j)=- i \hbar \nabla_j +|e| A(r_j)$ the canonical momentum in the presence of a magnetic field. $N_e$ is the number of electrons in the system. We choose not to gauge-fix, and have $\vec{\nabla} \times \vec{A} = \vec{B}$. The positions of the particles, $\{\vec{r}_i\}$, reside on a two-dimensional torus of generators $\vec{L}_1$, $\vec{L}_2$. The Hamiltonian is periodic under translations by these vectors, $V(\vec{r}_i - \vec{r}_j) = (\vec{r}_i - \vec{r}_j + \vec{L}_{1,2})$ and can be written as a sum over the allowed reciprocal vectors $\vec{q}$. ${\cal A}= |\vec{L}_1 \times \vec{L}_2|$ is the area of the sample, and $\vec{q} =m \vec{q}_1 + n \vec{q}_2$, $m,n \in {\mathbb{Z}}$ and $\vec{q}_1 = \frac{2\pi}{{\cal A}} \vec{L}_2 \times \hat{z}$, $\vec{q}_2 = -\frac{2\pi}{{\cal A}} \hat{z} \times  \vec{L}_1$. We choose to work on a square lattice, without any loss of generality. If the Hamiltonian of Eq[\ref{latticeHamiltonianLandauLevel}] was made out of only the first two terms, the symmetry analysis would have been identical to the one presented in Section $2$ and the number of states of any model FQH  state per momentum sector would have been obtainable by our constructive procedure of implementing the relative translational symmetries in both $x,y$ directions on the $(k,r)$-generalized Pauli principle states. 

The last term of Eq[\ref{latticeHamiltonianLandauLevel}]  implements the lattice with $N_x$ sites in the $x$ direction and $N_y$ sites in the $y$. The lattice constants in the $x,y$ directions are $l_x= L_1/N_x , \; l_y = L_2/N_y$. It makes electrons energetically favorable (for large negative $V_0$) to sit on lattice sites, and it makes the momenta in the $x,y$ direction take the discretized set of values $ (\vec{q} \cdot \hat{x}, \vec{q}\cdot \hat{y}) = (2\pi i/(N_xl_x), 2 \pi j/(N_y l_y))$ with $i =0, \ldots, N_x, \; j =0, \ldots, N_y$, which also serve as the momenta of the many-body states. The problem defined by Eq[\ref{latticeHamiltonianLandauLevel}]  is similar to that of the Chern insulator: it has the same density algebra in the long wavelength limit, the same lattice and the same translationally invariant Hamiltonian. The two-body Hamiltonian can be replaced by an $n$-body Hamiltonian, as long as it is translationally invariant. A physical assumption is  then the fact that the low-energy spectrum of the FQH and FCI problems (when the FCI problem has a topologically ordered groundstate) will be similar. More to the point, we take it as a proof of the topological nature of the FCI state that the low-energy manifold of states has the identical counting, per momentum sector to that of the FQH states at the same filling \emph{in the presence of the lattice }in Eq[\ref{latticeHamiltonianLandauLevel}]. We now find what that counting is.

The presence of the lattice influences the symmetries of the problem. The set of continuum translation operators $T_i(\vec{a})$ are not valid unless $\vec{a}$ is a multiple of the lattice constants $l_x\hat{x}, l_y\hat{y}$. Hence translational symmetry operators are $T_i(m_x l_x \hat{x}), T_i(m_y l_y \hat{y})$. Although the one-body terms of the Eq[\ref{latticeHamiltonianLandauLevel}]  commute with  $T_i(l_x \hat{x}), T_i(l_y \hat{y})$,  due to the many-body interacting term (which does not commute with those operators), the Hamiltonian still commutes only with $T_i(N_x l_x \hat{x}), T_i(N_y l_y \hat{y})$. The Hilbert space is described by the two twist parameters:
\beq
T_i(N_x l_x \hat{x})\ket{\psi} =e^{i \theta_x} \ket{\psi},\;\;\;\; T_i(N_y l_y \hat{y})\ket{\psi} =e^{i \theta_y} \ket{\psi}
\eneq As the number of  fluxes per unit cell plaquette is $ N_\phi/(N_x N_y) = 1$,  the translational operators of unit cell length commute: $[T_i(l_x \hat{x}), T_j(l_y \hat{y})] = - 2\delta_{ij} T_i (l_x \hat{x} + l_y \hat{y}) \sin(l_x l_y /2 l^2)=0$ since $l_x l_y /2l^2= L_xL_y/2 l^2 N_xN_y = 2 \pi l^2 N_\phi/2l^2 N_x N_y= \pi$. The center of mass translation operator $T(\vec{a}) = \prod_{i=1}^{N_e} T_i(\vec{a})$ commutes with the translation operator $T_i(l_x \hat{x})$ or $T_i(l_y \hat{y})$ if $\vec{a}=l_x \hat{x}$ or $l_y \hat{y}$. Notice that the operator $T(\vec{L}_1/N_\phi)$ (or $T(\vec{L}_2/N_\phi)$), which commuted with $T_i(\vec{L}_2)$, (or $T_i(\vec{L}_1)$ respectively) in the continuum limit, does not appear here because $L_1/N_\phi= L_1/(N_x N_y)= l_x/N_y$ is smaller than the permissible lattice constant. Hence the center of mass translation operators commute between themselves and the\emph{exact} $q$-fold degeneracy apparent in the continuum FQH is reduced to a one-fold degeneracy of any sector (in other words, we will not have exact degeneracies anymore). However, \emph{if} by adding the lattice we still remain in the FQH state, then remnants of the original degeneracies and spectrum in the continuum should appear even after adding the lattice. In particular, we know the counting of the degenerate zero-mode quasihole states (for FQH series such as RR) in the continuum, where they are zero-modes of pseudopotential Hamiltonians. We now try to find what the counting on the lattice momentum.

As we know the counting of the FQH per continuum  momentum of the relative translational operators, we now try to implement the relative translational symmetries of the problem. The relative translation operators in the continuum, $\tilde{T}_i (\vec{a}) = T_i((N_e-1)\vec{a}/N_e)\prod_{j\ne i}^{N_e} T_i(- \vec{a}/N_e)$, make sense only if $a/N_e$ is a multiple of $l_x$ or $l_y$, and hence the relative translation operators by the smallest amount are:
\beq
\tilde{T}_i(N_e l_x \hat{x}) = T_i((N_e-1)l_x \hat{x}) \prod_{j \ne i} T_j (- l_x \hat{x})
\eneq and similarly for $l_y$. We now ask two questions : do these operators commute with $T_i (\vec{L}_1), T_i(\vec{L}_2)$  \emph{and} do they commute with the Hamiltonian of the problem? The answer to the first question is an unequivocal yes, as $L_1 l_y/ 2 l^2 = L_x L_y/2 l^2 N_y= \pi N_\phi/N_y= \pi N_x$). The answer to the last question is \emph{no}, because $\tilde{T}_i(N_e l_x \hat{x})$ commutes with $\exp(i \vec{q}\cdot(\vec{r}_j - \vec{r}_l))$ when $j,l \ne i$ as it translates both coordinates by the same amount but does not commute with $\exp(i \vec{q}\cdot(\vec{r}_i - \vec{r}_l))$
\beq 
\tilde{T}_i(N_e l_x \hat{x}) e^{i \vec{q} (\vec{r}_i - \vec{r}_l)} = e^{i \vec{q} (\vec{r}_i + (N_e-1) l_x \hat{x})} e^{-i \vec{q}\cdot (\vec{r}_l- l_x \hat{x})}e^{i \vec{q} (\vec{r}_i - \vec{r}_l)} \tilde{T}_i(N_e l_x \hat{x}) = e^{i \vec{q}\cdot \hat{x} N_e l_x}e^{i \vec{q} (\vec{r}_i - \vec{r}_l)} \tilde{T}_i(N_e l_x \hat{x}) 
\eneq 
since $\vec{q} \cdot \hat{x}= 2\pi j/N_x l_x $, $j=0,1,\ldots N_x$,  and hence $N_el_x =2\pi N_e/N_x$; an identical situation obviously occurs for $\tilde{T}_i(N_e l_y \hat{y})$. As $N_e, N_x, N_y$ are integers,  we now define several important ratios and common denominators:
\beq
GCD(N_e, N_x)= N_{0x},\;\;\;GCD(N_e, N_y)= N_{0y},\;\;\; \frac{N_e}{N_x} = \frac{p_x}{q_x}, \;\;\; \frac{N_e}{N_y} = \frac{p_y}{q_y}
\eneq The $q_x, q_y$ are integers and should not be mistaken for the components of a $q$ momentum vector. Fortunately, this mistake is hard to do from the context. There are obvious relations between the above and $N=GCD(N_e, N_\phi), \;\; N_e/N_\phi= p/q$, but we will come to this later. It is then clear to see which relative translation operators commute with the Hamiltonian:
\beq
\tilde{T}_i(q_x N_e l_x \hat{x}) = T_i(q_x (N_e-1)l_x \hat{x}) \prod_{j \ne i} T_j (-q_x  l_x \hat{x}), \;\;\;\; \tilde{T}_i(q_y N_e l_y \hat{y}) = T_i(q_y (N_e-1)l_y \hat{y}) \prod_{j \ne i} T_j (-q_y  l_y \hat{y})
\eneq These operators commute with both translations and the Hamiltonian. They also commute between themselves and hence can be diagonalized (remember that in the Chern insulator, the commutations are only approximate, low-energy properties, as the GMP algebra is valid only at long wavelength).  We now ask how many eigenvalues of these operators are independent. We use the same reasoning as in the continuum limit and assume that the many-body state experiences, when acted upon by $\sum_{i=1}^{N_e}\exp(i \vec{Q}\cdot \vec{r}_i)$, an increase in its momentum by $\vec{Q}$ as long as $\vec{Q}$ is an allowed lattice momentum $\exp({i Q_x L_1}) = \exp({ Q_y L_2})=1$. The eigenvalues $\lambda_k$
\beq
\tilde{T}_i (q_x N_e l_x \hat{x}+ q_y N_e l_y \hat{y}  ) \ket{\psi(k)} = \lambda(k) \ket{\psi(k)}
\eneq  
can be easily  obtained by following the continuum proof closely and satisfy $\lambda(k) = D \exp({- i (k_y q_y l_y - k_x q_x l_x)})$. The coefficient $D$ can be determined by requiring that the $k=0$ state has maximum lattice symmetry, and hence $D\ket{\psi(k=0)} = \tilde{T}_i( q_y N_e l_y \hat{y}) \ket{\psi(k=0)} = \tilde{T}_i( - q_x N_e l_x \hat{x}) \ket{\psi(k=0)} = \tilde{T}_i(  q_y N_e l_y \hat{y} - q_x N_e l_x \hat{x}) \ket{\psi(k=0)} $. From here we obtain that $D= (-1)^{q_x q_y N_e (N_e -1)}=1$ (notice the difference from the continuum case, in which $D$ could be both $1,-1$). To obtain the values of different relative momenta, we notice that:
\beq
(\tilde{T}_i (q_x N_e l_x \hat{x}))^{N_{0x}} = (\tilde{T}_i (q_y N_e l_y \hat{y}))^{N_{0y}} = 1
\eneq which implies $\exp(- i k_x q_x N_{0x} l_x) =\exp(- i k_y q_y N_{0y} l_y)=1$. Expectedly, $k_x = 2\pi j/ N_x l_x$ and $k_y = 2\pi j/ N_y l_y$ with $j=1,2....$. Now that we obtained the values of the different momenta, we can ask how many of these eigenvalues are distinct. The eigenvalues appear as $\exp(i k_x q_x l_x) =\exp(i 2 \pi j q_x /N_x )=\exp(i 2 \pi j /N_{0x} )$  and similarly for the $y$ component, so the number of distinct momenta resolving the relative translation symmetry is:
\beq
k_x = \frac{2\pi j }{N_x l_x}, \;\; j=0, \ldots, N_{0x}-1;\;\;\;\;\;\;\; k_y = \frac{2\pi j }{N_y l_y}, \;\; l=0, \ldots, N_{0y}-1
\eneq 
This defines a $N_{0x}\times N_{0y}$ BZ. The momenta of the relative translational operators of the FCI, $k_x = \frac{2\pi j }{N_x l_x}, \;\; j=0, \ldots, N_{0x}-1;\;\;\;\;\;\;\; k_y = \frac{2\pi j }{N_y l_y}, \;\; l=0, \ldots, N_{0y}-1$ are in fact foldings of the $k_x = \frac{2\pi j }{L_x}, \;\; j=0, \ldots, N;\;\;\;\;\;\;\; k_y = \frac{2\pi j }{L_y}, \;\; l=0, \ldots, N$ momenta of the FQH. To show this, we first note that both $N_{0x}, N_{0y}$ divide $N$. Indeed, since $GCD(N_x, N_e) = N_{0x}$, $GCD(N_y, N_e) = N_{0y}$, and $(N_s, N_e) = (N_x N_y, N_e) = N$, we have that $GCD(N_x N_y/N_{0x}, N_e/N_{0x}) = N/N_{0x}= GCD(q_x N_y, p_x N_{0x})$, which proves that $N/N_{0x}$ is an integer and similarly for $N/N_{0y}$. With $N_e/N_s = p/q$, we have $N_e = p N= p_x N_{0x}$ and hence $N= p_x N_{0x}/p = p_y N_{0y}/p$ we have that the FCI momentum quantum numbers $k_x, k_y$  of the relative translation operator are a $p_x/p, p_y/p$ folding of the FQH momentum quantum numbers in the $x,y$ directions respectively. We have hence taken care of the relative momentum quantum numbers.

In the FQH in the continuum there is a center of mass degeneracy of $q$. This center of mass degeneracy comes from the non-commutativity of the continuum center of mass translation operators $T_{\rm CM}(\vec{L}_{mn}/N_s)$.  On the lattice, we also know that since:
\beq 
\tilde{T}_i (q_x N_e l_x \hat{x}) \cdot T_{\rm CM}(q_x l_x \hat{x}) = T_i (q_x N_e l_x \hat{x})
\eneq
the eigenvalue of $T_{\rm CM}(q_x l_x \hat{x})$ is fixed once $k_x$ and $\theta_x$ are known (similarly for $T_{\rm CM}({q_y l_y  \hat{y})}$). Since we have already resolved the motion in the $N_{0x} \times N_{0y}$ Brillouin zone, we have a remainder of $q_x \times q_y$ momenta to reach the full lattice BZ of $N_x \times N_y$. $T_{\rm CM}(l_x\hat{x}), T_{\rm CM}(l_y\hat{y})$ form a maximally commuting set of center of mass operators which can be simultaneously diagonalized (note that in the FQH $T_{\rm CM}(\vec{L}_{mn}/N_s)$ could not be simultaneously diagonalized) and which commute with the Hamiltonian, so no exact degeneracies are generically present. However, if the system is in the same universality class as the FQH, the low-energy modes will then exhibit a center of mass $q_x$ quasi-degeneracy in the $x$ direction and $q_y$ quasi-degeneracy in the $k_y$ direction. This degeneracy, as well as the relative momentum, make up the full BZ of the $N_x \times  N_y$ lattice. Note that in the lattice example, all translation operators commute with each other (courtesy of unit flux per unit cell), so in effect we could just diagonalize those. However, we choose to build the relative translational operators because in the continuum case we already know the counting per the momentum quantum number attached with these operators in the FQH. To obtain the counting of the eigenstates per FCI relative momentum, one needs to fold the FQH relative momentum BZ. This folding now needs to be supplemented by the miss-match in the center of mass degeneracy. In the FQH, the center of mass degeneracy is $q$; in the FCI, it is $q_x q_y$. In Appendix $5$ we show that $q/(q_x q_y)$ is an integer number. The counting of states in the low manifold, sorted, per FCI relative momentum sector is a folding of the FQH  relative momentum BZ times the mismatch in center of mass degeneracy $q/(q_x q_y)$. 

In conclusion, the diagonalization of a Chern insulator at fractional rational filling should exhibit, in the topologically ordered phase, a low-energy manifold of states with counting which encodes the topological character of the state. The low-energy manifold of states is a result of the emergent translational symmetry of the system that the long wavelength GMP algebra implies.  For a $N_x \times N_y$ lattice with $N_e$ electrons, the states in this low manifold will experience an emergent center of mass degeneracy, with the spectrum being $q_x \times q_y$ quasi-degenerate (in the $k_x$, $k_y$ direction). If the state developing in the system is at filling $p/q$, the low-energy manifold will contain either the groundstates or the quasihole excitations of this state. Once the center of mass degeneracy is resolved, the spectrum of the FCI per relative momentum sector is a folding of the spectrum of the FQH relative momentum BZ.  We present clear examples of this procedure in the next subsection for the FCI at  $1/3$- Abelian topological order before presenting numerical data on the Abelian and non-Abelian FCI states.

\subsection{Two Simple Examples of the FQH to FCI Mapping}

\begin{figure}[tbp]
\includegraphics[width=5.5in]{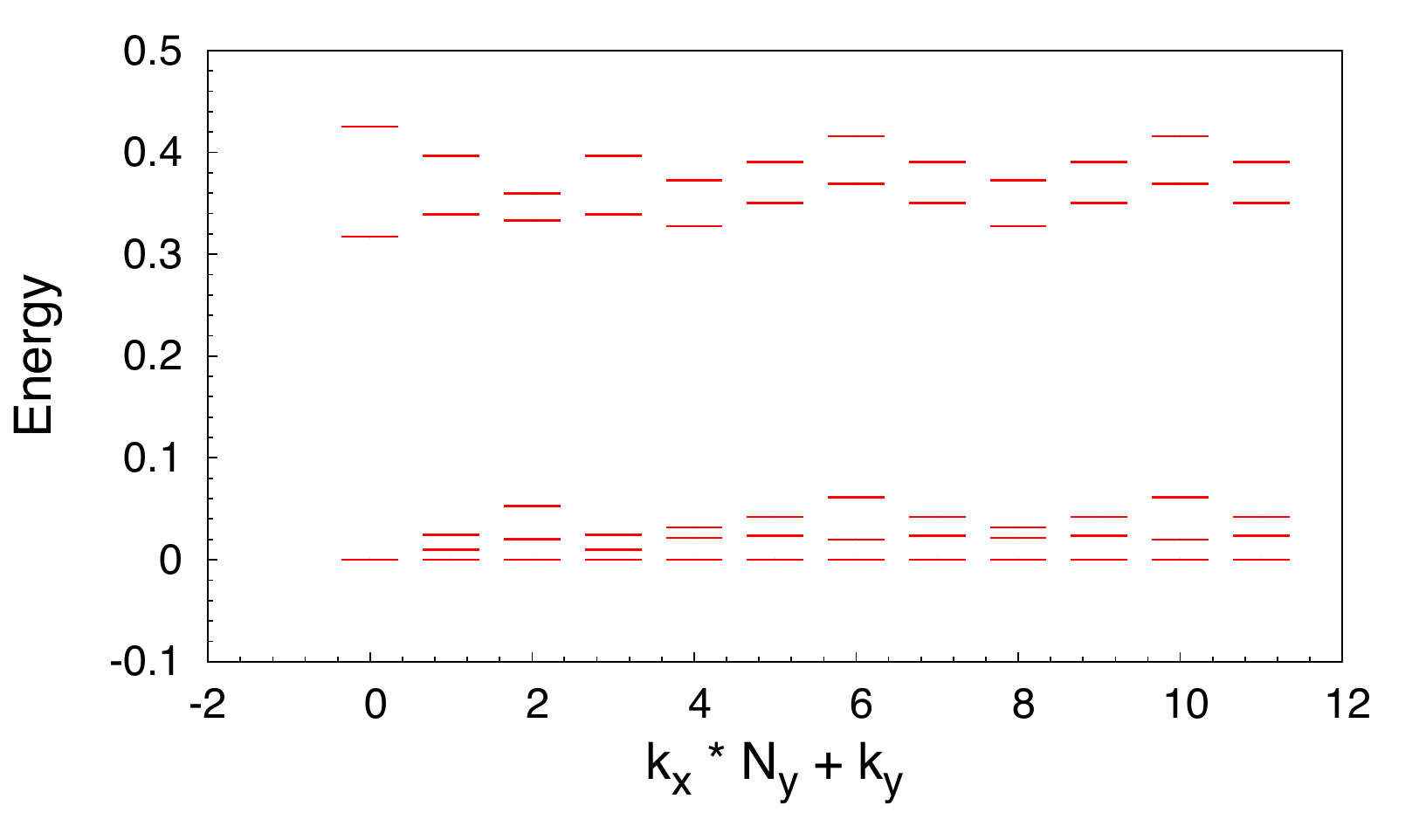}
\caption{Energy spectrum for $N=2, N_x=3, N_y=4$. The total number of states of the low-energy manifold the dotted line is 42. Per momentum sector, the counting is $3$ states if $k_y$ is divisible by $2$ and $4$ states otherwise, irrespective of $k_x$ (there are degeneracies in the spectrum, for example, there are $5$ states at momentum $(0,0)$, the low-energy manifold being $3$-fold degenerate with zero  energy }\label{entspeconethirdstateN4particles}
\end{figure}

\begin{figure}[tbp]
\includegraphics[width=7.3in]{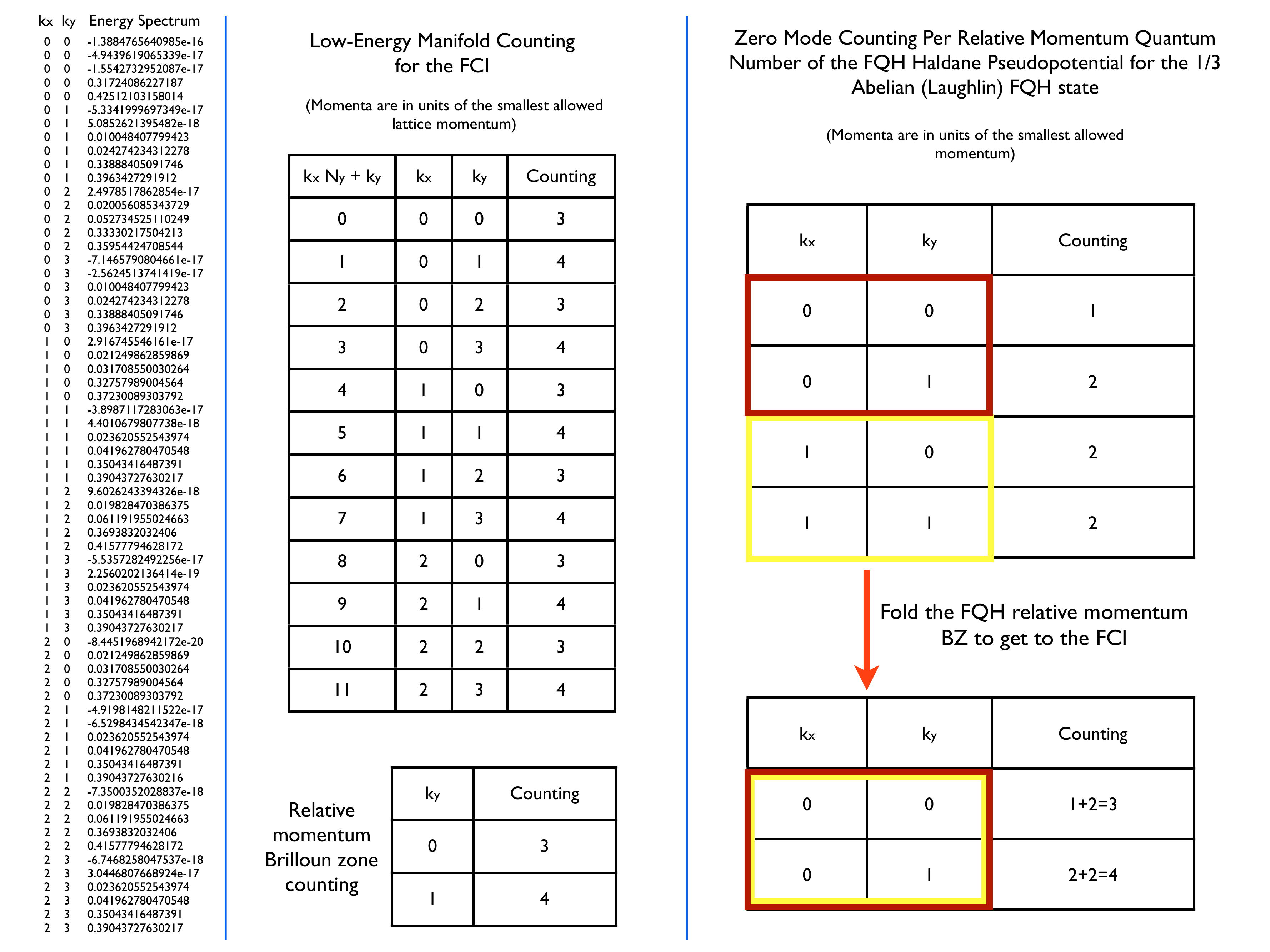}
\caption{Example of the FQH to FCI mapping for $N_e=2$, $N_x=3$,$N_y=4$ state. The spectrum of the one-body flat-band Chern Hamiltonian \cite{sun-PhysRevLett.106.236803} with nearest neighbor interaction is given on the left. The counting of the low-energy manifold of states exhibits clear center  of mass degeneracies and the relative momentum BZ of the FCI is a $1\times 2$ BZ. The counting per momentum sector in this BZ is a folding of the counting of zero modes of the Haldane FQH pseudopotential whose highest density state is the Laughlin state. This implies that the universality class of the FCI state is the same as that of the $\nu=1/3$ state.  }
\label{CountingN2Nx3Ny4}
\end{figure}

We now look at two simple examples of how the analytical construction in the previous sections can be applied to the numerical data in both the energy and the entanglement spectrum. The counting per momentum sector can be obtained from our FQH to FCI mapping. Lets first look at the specific case of such counting for $N_x \times N_y =3 \times 4$ lattice Chern insulator model of \cite{sun-PhysRevLett.106.236803} diagonalized in the flat-band limit \cite{regnault-PhysRevX.1.021014} with $N_e=2$ electrons and nearest neighbor (on-site being forbidden) Hubbard interaction. This interaction favors a $1/3$ Abelian state, which we know occurs in the FQH state for short-range repulsive pseudopotentials. As the state contains only $2$-particles, we expect that the physics is that of $2$-electron $\nu=1/3$ Abelian FQH state plus a number of $6$ quasiholes. The physics also reveals the structure of the pseudopotential two-body interaction in the FCI. The spectrum, shown in Fig[\ref{entspeconethirdstateN4particles}], confirms this. We observe a clear gap between a low-energy manifold of states and a high-energy manifold of spurious states. We focus on the low-energy manifold.  By looking at the numerical data (Fig[\ref{CountingN2Nx3Ny4}], which includes degeneracies), we see that the low-energy manifold exhibits a counting of states of either $3$ or $4$ states per momentum sector, depending on whether $k_y$ is divisible or not by $2$, as can be seen in Fig[\ref{CountingN2Nx3Ny4}]. This clearly shows that the spectrum is $6$-fold degenerate, and more-over, that this degeneracy splits in a $k_x$ - $3$-fold degenerate and a $k_y$- $2$-fold degeneracy. This is consistent to our analysis, for $N_e/N_x= 2/3$, $N_e/N_y = 1/2$ which fix $q_x=3, q_y=2$. Since the $GCD(N_x, N_e)=1$, while the $GCD(N_y, N_e)=1$, the relative momentum BZ is made of $k_x = 0$ and $k_y = 0, 2\pi/(N_y a)$ where $a$ is the lattice constant in the $y$ direction. In this BZ, the counting of states is $3,4$ respectively, while the whole counting in the full BZ  of the FCI lattice is a $3\times 2$ center of mass translation of this in the $x,y$ directions respectively (see Fig[\ref{CountingN2Nx3Ny4}]). In the FQH  effect, $N_e/N_s = 1/6$ and hence the center of mass degeneracy of $q=6$ ($=q_x q_y$) and there is no center of mass degeneracy mismatch between the FQH and the FCI. The FQH relative momentum BZ that remains after the center of mass degeneracy is removed is a  $2\times 2$ BZ ($GCD(N_s, N_e) = 2$). The Laughlin $1/3$ Abelian FQH counting of quasihole states analysis was already presented in Table[\ref{Laughlinconfigsoftwoparticlesintwelveorbitals}]. In that section, we found that  for $2$-particles in $12$ orbitals the quasiholes of the Laughlin state have the following quantum numbers: $2$ states in each of the $k_x,k_y$ equal to $0, 2\pi/L_y$, $2\pi/L_x, 0$, and $2\pi/L_x, 2\pi/L_y$ momenta, and $1$ state in the $0,0$ momentum. There is $6$-fold center of mass degeneracy which multiplies these $7$ states to obtain the total of $42$ quasihole states. Our theory then implies that the counting of the FCI per relative momentum sector is a $p_x/p=2$-folding of the FQH relative momentum BZ. This folding gives a $2+1, 2+2$ counting for the  $k_x = 0$ and $k_y = 0, 2\pi/(N_y a)$ momenta respectively, which perfectly agrees with the numerical data (see Fig[\ref{CountingN2Nx3Ny4}]).

For larger number of particles and quasiholes, the cleanest example of the low-energy manifold of states whose imprint characterizes the topological order is seen in the (particle) entanglement spectrum of a FCI groundstate. The particle entanglement spectrum of the groundstate reveals the physics of the excitations of the system in the particular topologically ordered state of the groundstate. It is more robust than the actual energy spectrum because the latter can exhibit aliasing effects due to other states at nearby fillings. Since the Hamiltonian used to obtain the spectrum is not a "model" Haldane pseudopotential Hamiltonian but a generic Hubbard-like one, adding too many quasiholes to the system can, in the energy spectrum, take us to a state at a different filling (with different quasihole counting). However, the entanglement spectrum will keep the topological order (filling) of the original groundstate, and reveal directly the quasiholes of that state.   For example, if we are looking at the energy spectrum of $4$ particles in $36$ orbitals on the lattice, it is possible that instead of seeing the spectrum of $4$ electrons in a $1/3$ Abelian state with $24$ extra-quasiholes, we could, since the interaction is generic Hubbard, obtain the spectrum of $4$ electrons in a filling-$1/9$ Abelian state or $4$ electrons in a  filling $1/7$ state with $8$ extra-quasiholes. However, the particle  entanglement spectrum of the groundstate with $N=12$ particles in $36$ orbitals (filling $1/3$) would reveal the physics of the excitations of an Abelian $1/3$ state if we look  at the particle entanglement spectrum (described below) of $N_A= 4$ particles versus the rest.

\begin{figure}[tbp]
\includegraphics[width=6.3in]{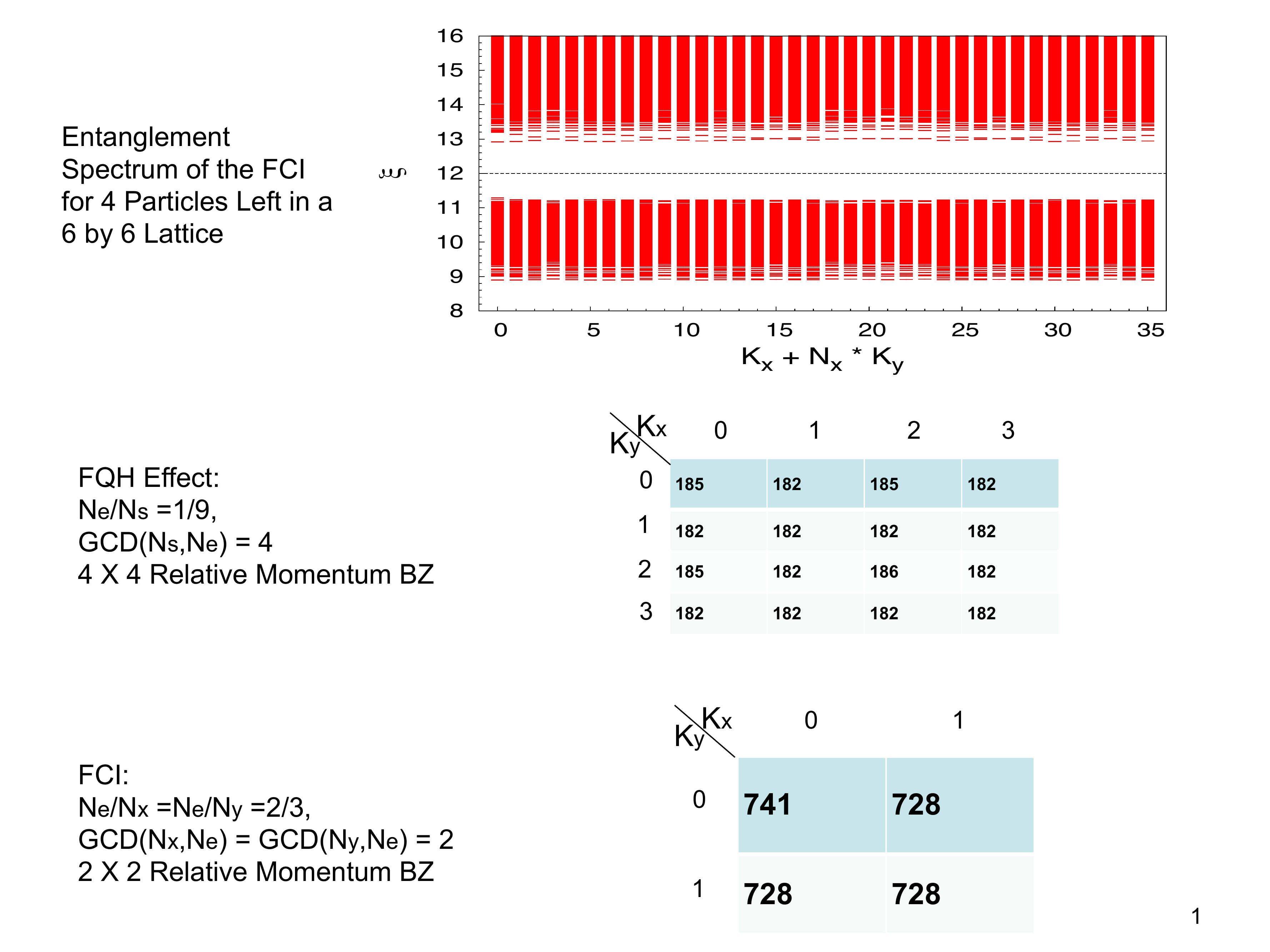}
\caption{Example of the FQH to FCI mapping for $N_e=4$, $N_x=6$,$N_y=6$. The one-body flat-band Chern Hamiltonian \cite{sun-PhysRevLett.106.236803} with nearest neighbor interaction for $12$ particles in a $6 \times 6$ lattice has a $3$-fold quasidegenerate groundstate from which we build the entanglement spectrum shown here. The low-energy manifold (below the dotted line) has $741$ states in momentum sectors where $N_x (\mod 2) = N_y (\mod 2) = 0$ and $728$ elsewhere.  The counting of the low-energy manifold of states exhibits clear $3\times 3$ fold (in $k_x,k_y$) center  of mass degeneracies and the relative momentum BZ of the FCI is a $2 \times 2$ BZ. The counting per momentum sector in this BZ is a folding of the counting of zero modes of the Haldane FQH pseudopotential whose highest density state is the Laughlin state. For example, the $(0,0)$ relative momentum sector of the FCI contains $741$ states which exactly equals the $p_x/p=p_y/p=2$ folding $185+185+185+186$ of the FQH Laughlin quasihole counting. Likewise, the $(0,1)$ FCI relative momentum sector contains $728$ states, which exactly equals the $p_x/p=p_y/p=2$ folding $182+182+182+182$ of the FQH Laughlin quasihole counting.  This implies that the universality class of the FCI state is the same as that of the $\nu=1/3$ state.  }\label{CountingN4Nx6Ny6entspect}
\end{figure}

The spectrum of the Chern insulator at $1/3$ filling has a quasi-degenerate ($3$-fold, as our analysis would imply) groundstate, at momenta consistent with the folding rule (we wont expand on this rule for the groundstate, as we will give the much more complicated quasihole example). The $3$-fold degenerate groundstate of the system contains information about the Abelian Fractional $1/3$ character of the excitation spectrum.  This comes out of the recently developed entanglement spectrum \cite{li-08prl010504,sterdyniak-PhysRevLett.106.100405} which for a single non-degenerated ground state $\ket{\Psi}$ can be defined through the Schmidt decomposition of $\ket{\Psi}$ in two regions $A$, $B$ (not necessarily spatial):

\begin{equation}
\ket{\Psi}=\sum_i e^{-\xi_i / 2} \ket{\Psi^A_i} \otimes \ket{\Psi^B_i}
\label{schmidt}
\end{equation}
\noindent where $\braket{\Psi^A_i}{\Psi^A_j}=\braket{\Psi^B_i}{\Psi^B_j}=\delta_{i,j}$. The $\exp(-\xi_i)$ and $ \ket{\Psi^A_i}$ are the eigenvalues and eigenstates of the reduced density matrix,  $\rho_A={\rm Tr}_B \rho$, where $\rho=\ket{\Psi}\bra{\Psi}$ is the total density matrix. There is no generalization of the Schmidt decomposition (\ref{schmidt}) to degenerate groundstates but definition of the entanglement spectrum (ES) through the reduced density matrix still holds. We want to build a density matrix which commutes with the total translation operators -a desirable feature if we want to sort the $\xi_i$ with respect to the momentum quantum numbers. Such a condition is satisfied \cite{sterdyniak-PhysRevLett.106.100405} by the incoherent summation over the degenerate groundstates $\rho=\frac{1}{3}\sum_{i} \ket{\Psi_i}\bra{\Psi_i}$. Depending on whether the $A$ and $B$ regions are real, momentum or particle space, the ES reveals different aspects of the system's excitations. If the regions $A$, $B$ are regions of particles, it can be proved \cite{sterdyniak-PhysRevLett.106.100405} that the particle entanglement spectrum obtained by tracing over the positions of a set of $B$ particles gives information about the number of quasiholes of the system of $N_A$ particles and number of orbitals identical to that of the untraced system. In the case of the usual FQH, the particle entanglement spectrum of a model state (such as Laughlin, MR, etc) contains an identical number of levels as those of the quasiholes, thereby providing a numerical check of our analytic proof \cite{sterdyniak-PhysRevLett.106.100405}. Away from the model states (the Coulomb ground state, for example) the ES may exhibits an entanglement gap \cite{li-08prl010504,thomale-10pr180502}. It separates a low-energy structure with perfect quasihole counting and and a high entanglement energy nonuniversal part. But a clear and significant gap is not always observed, even for the $\nu=1/3$ Coulomb state.

Surprisingly, for the Fractional Chern insulator, the situation is much better: in a previous work \cite{regnault-PhysRevX.1.021014}, we observed a clear, large entanglement gap between a manifold of  low entanglement energy levels  and a manifold of  high entanglement energy levels. Moreover, the counting of the levels below the gap is identical to the counting of quasiholes of $N_A$ particles in $N_x N_y$ orbitals. We show this for a large-size example. Out of the $3$-fold quasi-degenerate groundstate of $12$ particles on the Chern insulator on the checkerboard lattice with $N_x \times N_y = 6\times 6$, we build the density matrix and construct the entanglement spectrum for $N_A=4$ particles. The one-body Chern insulator model of \cite{sun-PhysRevLett.106.236803} presented in Eq[\ref{onebodymodel}], with a two-body nearest neighbor Hubbard $U=1$ interaction in the flat-band limit is used to generate the $3$-fold quasi-degenerate groundstates. The entanglement spectrum has a gap (see Fig[\ref{CountingN4Nx6Ny6entspect}]); the counting of the low-energy manifold of states  below the dot- ted line is 741 in momentum sectors where $N_x (\mod 2) = N_y (\mod 2) = 0$ and 728 elsewhere. The total number below this line (26325) exactly matches the number of $(1,3)$-admissible configurations of $4$ particles in $36$ orbitals. The counting per relative momentum sector is a folding of that of the zero modes in the FQH state, as the complete analysis of Fig[\ref{CountingN4Nx6Ny6entspect}] shows.

As said in our previous paper \cite{regnault-PhysRevX.1.021014}, we find it very revealing that the FCI groundstates obtained here contain much clearer information (large, clear entanglement gap)  than the groundstates of the Coulomb interaction in the FQH. The entanglement spectrum shows that the groundstates by themselves know of the fractional nature of the excitations in the Fractional Chern insulator. The current clean application of the entanglement spectrum also shows that this quantity is fundamentally useful towards revealing the physics of strongly-correlated states besides the usual FQH model wavefunctions. This is even crucial in this case since, despite several attempts\cite{vaezi-2011arXiv1105.0406V,qi-PhysRevLett.107.126803}, there is no analytical expression of the Laughlin state for the FCI to compare with.

\section{Numerical Proof of Emergent Symmetries in Abelian and Non-Abelian Fractional Chern Insulators}

In the current section, we expand our numerical analysis of FCI states to the non-Abelian case and show that the counting rule obtained in the current analysis is robust. We perform exhaustive numerical calculations of both the energy and the entanglement spectra for several interactions at the specific filling where non-Abelian states are expected. The calculations serve as both a numerical check of our emergent translational symmetry principle, as well as  the first proof of principle that non-Abelian states can be stabilized in a fractionally filled Chern insulator.

\begin{figure}[tbp]
\includegraphics[width=6.3in]{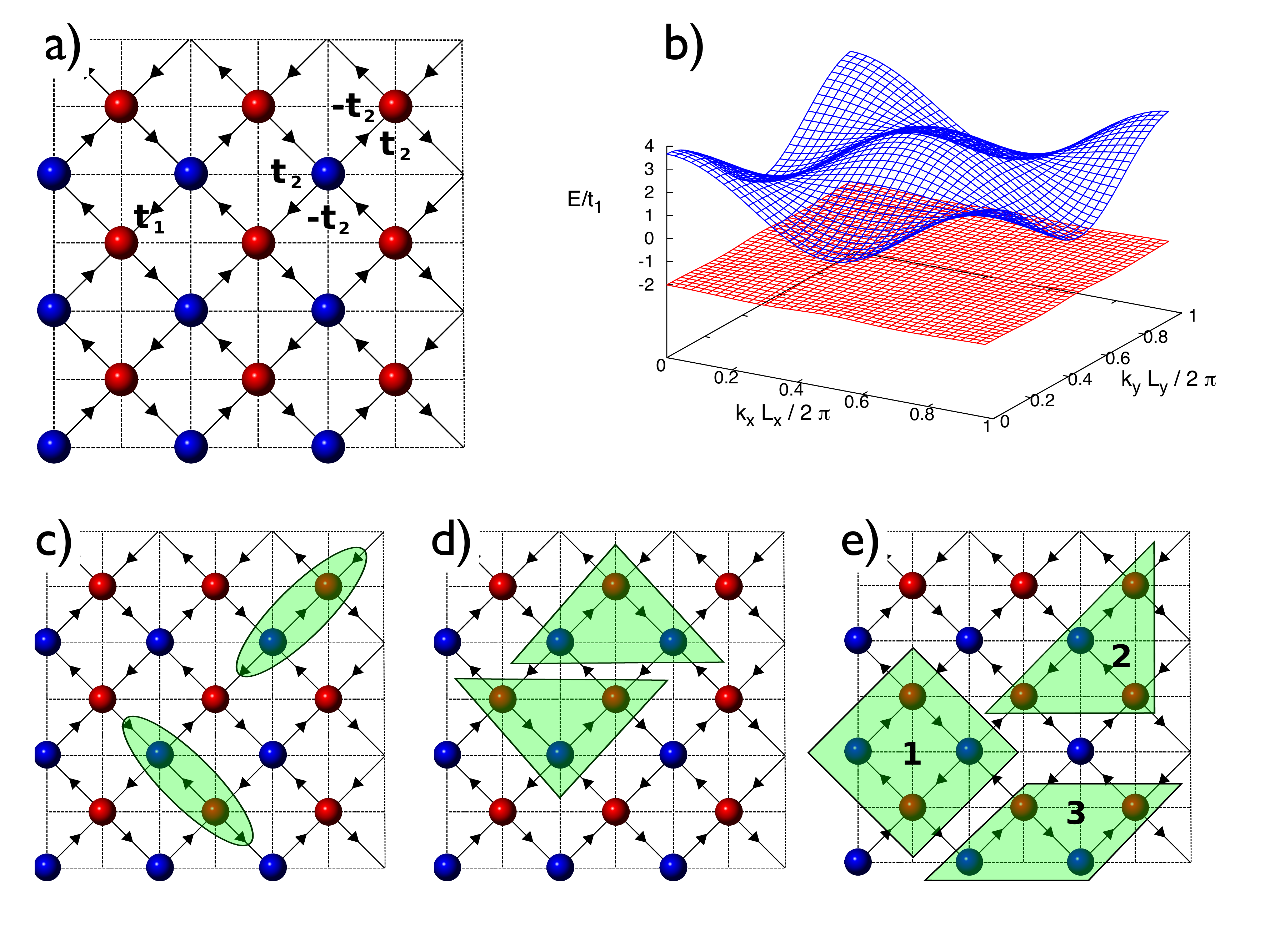}
\caption{a) One-body model hopping amplitudes in the checkerboard Chern insulator model used in our numerical calculations. Direction of the arrow means positive hopping amplitude. b) Dispersion of the model shows that it is an insulator with a nearly flat valence band. Our diagonalizations are in the flat-band limit where the valence band is made completely flat. c) Simplest Hubbard two-body interaction couples nearest neighbor sites used to obtain the Abelian $1/3$ state. d) Simplest three-body Hubbard interaction used to obtain the non-Abelian MR Pfaffian state. Equal weight is given to the three configurations e) Simplest four-body Hubbard interaction configurations used to obtain the non-Abelian $\mathbb{Z}_3$ RR states. Either of the interaction $1,2,3$ can be used, all giving  a RR groundstate but of different gaps (the largest gap being obtained with the interaction 1).  }\label{checkerboardlatticeanddifferentinteractions}
\end{figure}

For our one-body model, we pick the Chern insulator on a checkerboard lattice, first introduced in \cite{sun-PhysRevLett.106.236803,sheng-natcommun.2.389,neupert-PhysRevLett.106.236804}. This model already exhibits weak dispersion of the bands (see Fig[\ref{checkerboardlatticeanddifferentinteractions}]) like several other models \cite{tang-PhysRevLett.106.236802,hu-PhysRevB.84.155116,wang-PhysRevB.84.241103}. But because we work in the flat-band limit, this is not essential to our calculation.  With $A,B$ being the two sites in the unit cell, the one-body, two-band Hamiltonian reads \cite{sun-PhysRevLett.106.236803}:

 \begin{figure}[tbp]
\includegraphics[width=6.3in]{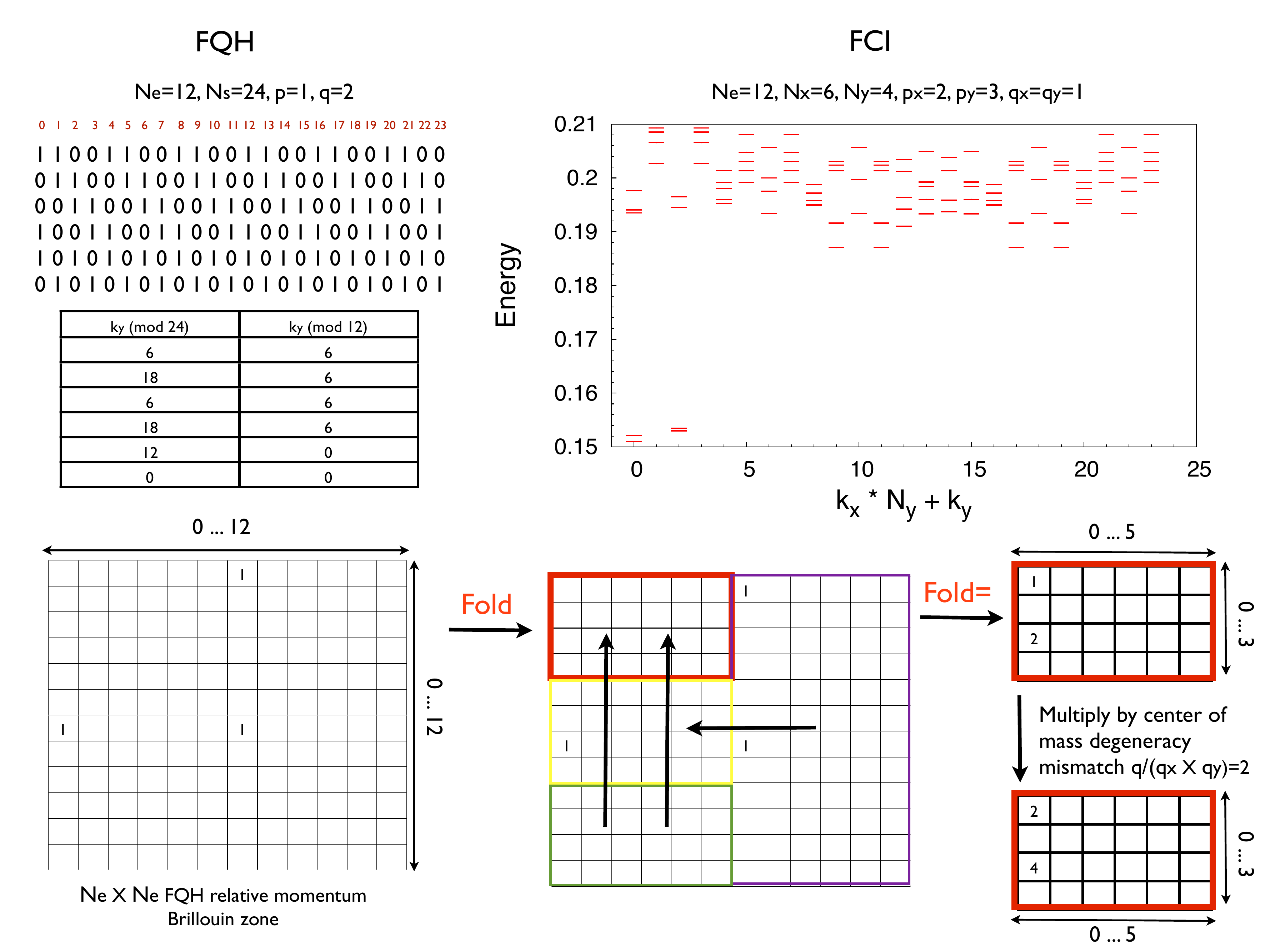}
\caption{The spectrum of the three-body Hubbard interaction at filling $\nu=1/2$ shows two quasi-degenerate states at $(0,0)$ momentum and $4$ states (separated into two quasi-degenerate doublets) at $(0,2)$ (in units of the smallest lattice momentum). The procedure and folding from FQH to FCI is also presented. }\label{Moore-ReadFCIGroundstate}
\end{figure}

\beq H_1 = \sum_{k} (c_{kA}^\dagger , c_{kB}^\dagger) h_1(k)  (c_{kA} , c_{kB})^T, \;\;\;\;\;\;\; h_1(k) = \sum_i d_i (k) \sigma_i 
\eneq
where the $d_i(k)$'s are:
 \beq d_x(k) = 4 t_1 \cos(\phi) \cos(k_x/2) \cos(k_y/2), \;\; d_y(k) = 4 t_1 \sin(\phi) \sin(k_x/2) \sin(k_y/2),\;\; d_z = 2 t_2(\cos(k_x) - \cos(k_y) ) + M\eneq 
In the original model \cite{sun-PhysRevLett.106.236803} there is an additional diagonal term -$4 t_3 \cos (k_x) cos (k_y) $ - which shrinks the dispersion of the bands, thereby making them flatter, but does not matter for the eigenstates; since we are diagonalizing in the flat-band limit, we neglect such a term.  $\phi$ is the phase factor added to the nearest neighbor hopping, while the parameter $M$ is a mass term added in order to drive the transition from a topological Chern insulator (for $M=0$) to a trivial atomic limit insulator when $M\rightarrow  \pm \infty$. The model is always gapped (for $t_1, t_2, \phi$ not vanishing) with the exception of the points $k_x=0, k_y=\pi, M=-4 t_2$ and $k_x=\pi, k_y =0, M=4 t_2$ which are gapless and where the phase transitions between the atomic limits $M\rightarrow  \pm \infty$ and the Chern insulator phase occur. For $|M|< 4 |t_2|$ the filled valence band has a Hall conductance of $1$. The single  particle Hamiltonian matrix has the following symmetries: inversion with identity inversion matrix $h_1(-k_x, -k_y)= h_1(k_x, k_y)$, as well as (at $M=0$) a certain type-of particle hole symmetry coupled with a $C_4$ rotation and a mirror operation:  $\sigma_z h_1(k_x, k_y) \sigma_z = - h_1(k_y, k_x)$. Due to the presence of fractions $k/2$, the model in \cite{sun-PhysRevLett.106.236803} is not in Bloch form. To render it in Bloch form, we perform the gauge transformation $c_{kB} \rightarrow c_{kB} \exp(-i (k_x- k_y)/2)$ to obtain:
\begin{eqnarray}
& h_2(k)= \left( {\begin{array}{cc}
 h_{11}(k) & h_{12}(k)  \\
 h_{12}^\star(k) &  - h_{11}(k) \\
 \end{array} } \right) \nonumber \\ & h_{12}(k) =  t_1e^{i \phi}(1+ e^{i(k_y-k_x)}) + t_1 e^{- i \phi}(e^{i k_y}+ e^{-i k_x}) \nonumber \\
 & h_{11}(k) = 2 t_2 (\cos(k_x) - \cos(k_y)) + M\label{onebodymodel}
 \end{eqnarray} 
The inversion symmetry of $h_1(k)$ translates into a symmetry of $h_2(k)$ given by $U^\dagger(k) h_2(\vec{k}) U(k) = h_2(-\vec{k})$ with $U(k)$ a diagonal $2 \times 2$ unitary matrix with $1, e^{-i (k_x- k_y)/2}$ on the diagonal. To eliminate the effect of the band curvature, and to allow the filled particles to democratically sample the whole BZ, we always work in the flat-band limit of a topological insulator. This corresponds to keeping the single particle eigenstates of $h_2(k)$ but putting the energy of the occupied bands to be an arbitrary energy $\pm E_0$ where $E_0>0$. At the Hamiltonian level, we transform from $h_2(k) = E_-(k) P_-(k) + E_+(k) P_+(k)$ to $h_2^{FB}(k) = -E_0 P_-(k) + E_0 P_+(k)$ where $P_\pm$ are the projectors onto the occupied and unoccupied bands. The energy difference between the valence and conduction bands can hence be made large without changing the eigenstates of the system.

\begin{figure}[tbp]
\includegraphics[width=4.3in]{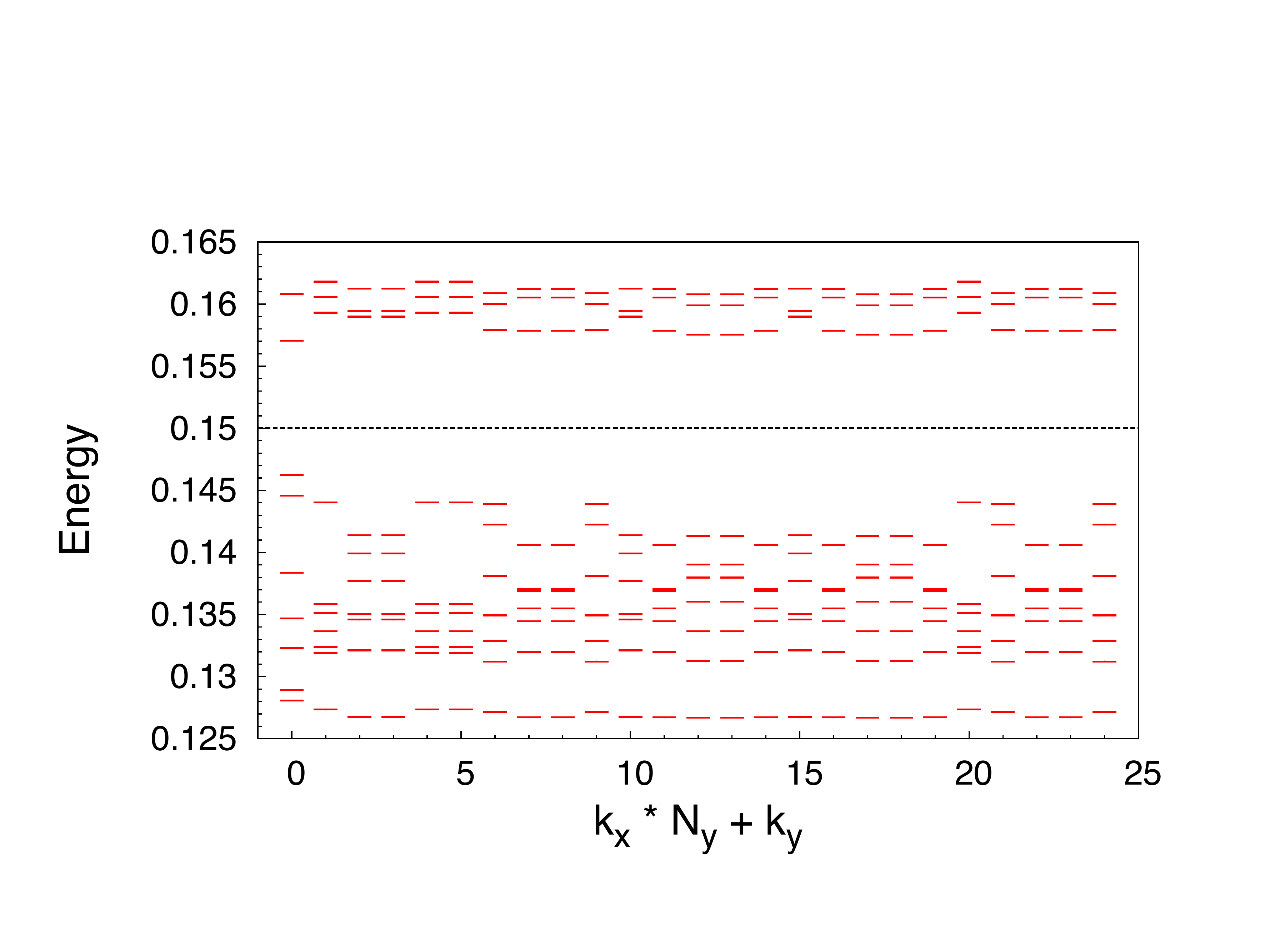}
\caption{Low-lying spectrum of the three-body interaction for $N_e=12$ electrons and $N_x=N_y=5$ }\label{Moore-ReadFCIquasihole1}
\end{figure}

 We now fractionally fill the flat valence band of this insulator and add interactions.  We diagonalize the interaction Hamiltonian directly in the filled band, neglecting the conduction band. This is similar to the Lowest-Landau Level (LLL) projection in the usual Fractional Quantum Hall Effect. Several interactions we use are:  
\beq
H_{\text{two-body}} = \sum_{<AB>} n_A n_B
\eneq where $A,B$ are nearest neighbor $A, B$ sites. In the current section we also use other many-body interactions such as three-body:
\beq
H_{\text{three-body}} = \sum_{<AB_1>, <A,B_2>} n_A n_{B_1} n_{B_2} +  n_B n_{A_1} n_{A_2}  
\eneq where $B_1, B_2$ are the two nearest neighbors of the site $A$ and $A_1, A_2$ are the two nearest neighbors of the site $B$.  The first of the $3$-kinds of  four-body interactions used in our exact diagonalization reads (see Fig[\ref{checkerboardlatticeanddifferentinteractions}]):


\beq
H_{\text{four-body 1}} = \sum_{<A_1, B_1>, <A_1 ,B_2>, <<A_1, A_2>>, <<B_1, B_2>>} n_{A_1} n_{A_2} n_{B_1} n_{B_2}  
\eneq 

where $B_1, B_2$ are the two nearest neighbors of the sites $A_1$, and also of the site $A_2$. The other two four-body interactions used which we denote $H_{\text{four-body 2}}$, $H_{\text{four-body 3}}$ can be easily understood from  Fig[\ref{checkerboardlatticeanddifferentinteractions}]. All the numerical calculations are performed with $t_2=(2-\sqrt{2})/2 t_1$ as discussed in \cite{sun-PhysRevLett.106.236803}. The total translation operators in the $x,y$ directions commute with both the single and many-body Hamiltonian and hence the eigenstates are indexed by total momentum quantum numbers $(K_x, K_y)$ which are the sum of the momentum quantum numbers of each of the $N$ particles modulo $(N_x, N_y)$. The basis states are $\prod_{i=1}^N \gamma^\dagger_{-, \vec{k_1}} \ldots \gamma^\dagger_{-, \vec{k_N}} \ket{0}$ (we work in the "LLL", and the $\gamma^\dagger_{-,\vec{k}}$'s are the creation operators for a particle of momentum $\vec{k}$ in the valence band). When acting on the basis states, the $c_{\vec{k}, \alpha} = u_{-,\alpha, \vec{k} }\gamma_{-, \vec{k}}$ where $u_{-,\alpha, \vec{k} }$ is the $\alpha = A,B$ component of the eigenstate of the occupied band of $h_2(k)$ or $h_2^{FB}(k)$ (they have identical eigenstates). Diagonalizing directly in the valence band provides for large numerical efficiency, allowing us to go to very large system sizes. The inversion symmetry of the single particle problem is maintained at the interacting level, and the interacting spectrum has an exact  $(K_x, K_y) \rightarrow (-K_x, -K_y) $ symmetry which can and has been used as checkup.  The $H_{\text{two-body}} $ has been used to obtain the Abelian, filling $1/3$ FCI topological state.

\begin{figure}[tbp]
\includegraphics[width=4.3in]{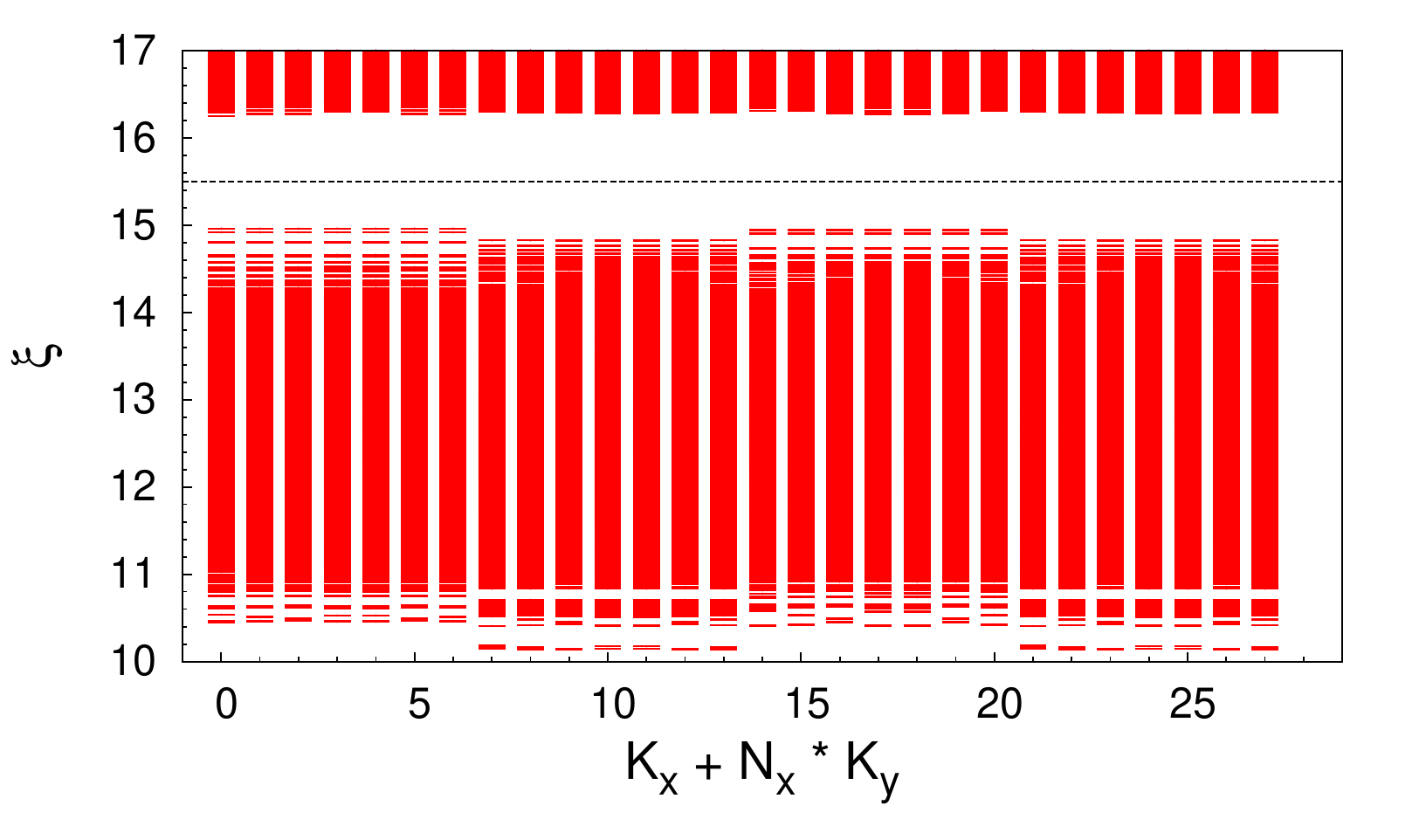}
\caption{Entanglement spectrum for the groundstates of the three-body interaction for $N_e=14$ electrons and $N_x=7, N_y=4$ and $N_A=6$ particles. The number of states per momentum sector  below the dotted line  is $8463$ in the even $k_y$ sectors, $8486$ in the odd $k_y$ sectors. This is the counting predicted by our FQH to FCI mapping. 
 }\label{Moore-ReadFCIEntSpectrum}
\end{figure}

We now use $H_{\text{three-body}} $ and, borrowing from FQH experience, we look for the MR Pfaffian at filling $\nu=1/2$. We diagonalize $H_{\text{three-body}} $  for up to $N=14$ particles on a $N_x \times N_y = 7 \times 4$ lattice, and present the spectrum in Fig[\ref{Moore-ReadFCIGroundstate}]. The spectrum is separated into a six-fold quasidegenerate groundstate (at momentum $(0,2)$ we have four states, separated into two exactly degenerate doublets) at momenta consistent with our FQH to FCI mapping (in this case we also have to use the center of mass degeneracy mismatch factor, which is $2$). A clear and detailed presentation of the resolution of states in momentum sectors is shown in  Fig[\ref{Moore-ReadFCIGroundstate}].  Both the number of groundstates and their momenta are consistent with the MR Pfaffian state, with a large gap separation to the continuum of states. We now add quasiholes to the system, by performing the diagonalization of the $3$-body Hamiltonian for $N_e=12$ electrons on a $N_x \times N_y=5 \times 5$ lattice. This represents the addition of one flux quantum to the system.  The spectrum is seen in Fig[\ref{Moore-ReadFCIquasihole1}], and we observe a low-energy manifold of states (below the dashed line) with $7$ states in each momentum sector separated by a clear energy gap from a high-energy continuum of states. This is indeed the correct counting for the non-Abelian quasiholes of the MR state based on our emergent symmetry principle.  We then close the analysis of the MR state by looking at the particle entanglement spectrum of the $N_e=14, N_x=7, N_y=4$ degenerate states when the number of particles not traced out of the system is $N_A=6$. By our theory, this spectrum should show the counting, per momentum sector, in a low energy manifold of the quasiholes of the MR state for $6$ particles in $28$ orbitals . The results are presented in Fig[\ref{Moore-ReadFCIEntSpectrum}], where we again see that a low-lying manifold of states is separated by  a gap form a manifold of spurious continuum. The number of states per momentum sector  below the dotted line in  Fig[\ref{Moore-ReadFCIEntSpectrum}] is $8463$ in the even $k_y$ sectors, $8486$ in the odd $k_y$ sectors, the correct numbers of non-Abelian quasiholes predicted by our FQH to FCI mapping.

\begin{figure}[tbp]
\includegraphics[width=3.5in]{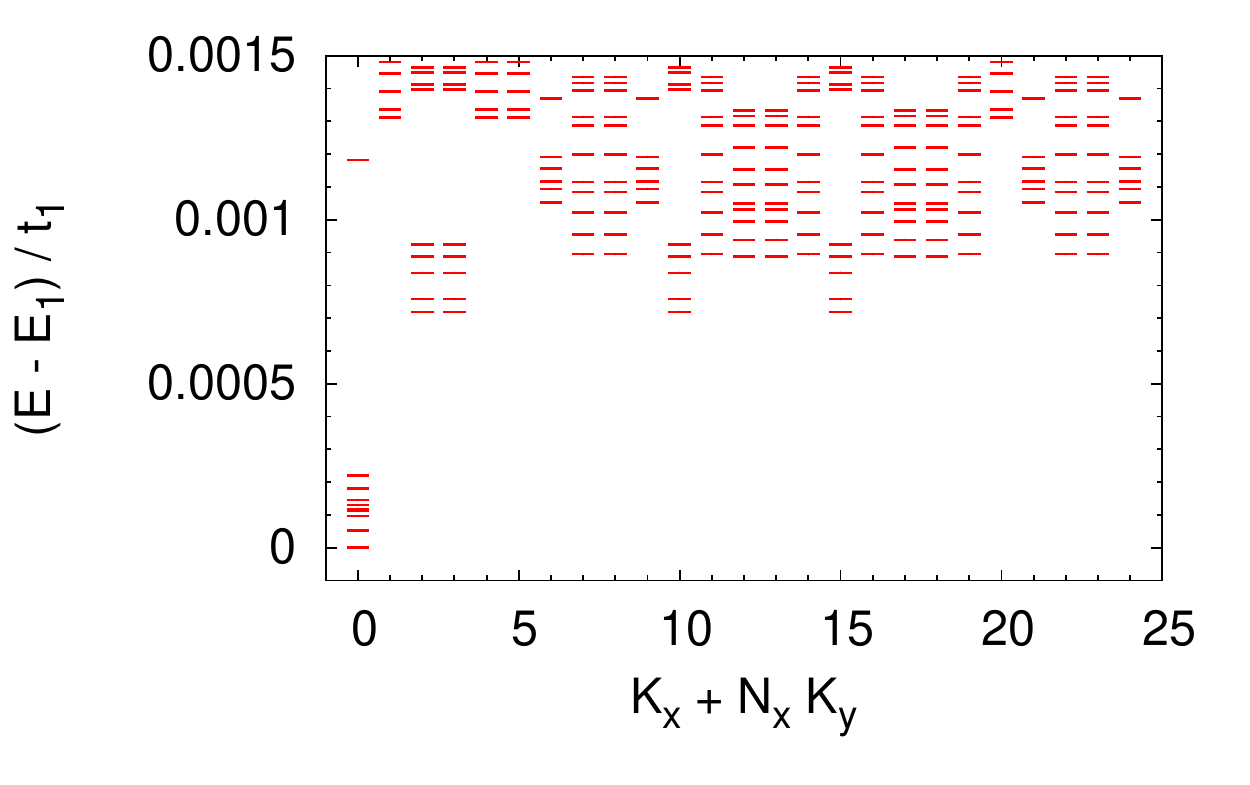}
\includegraphics[width=3.3in]{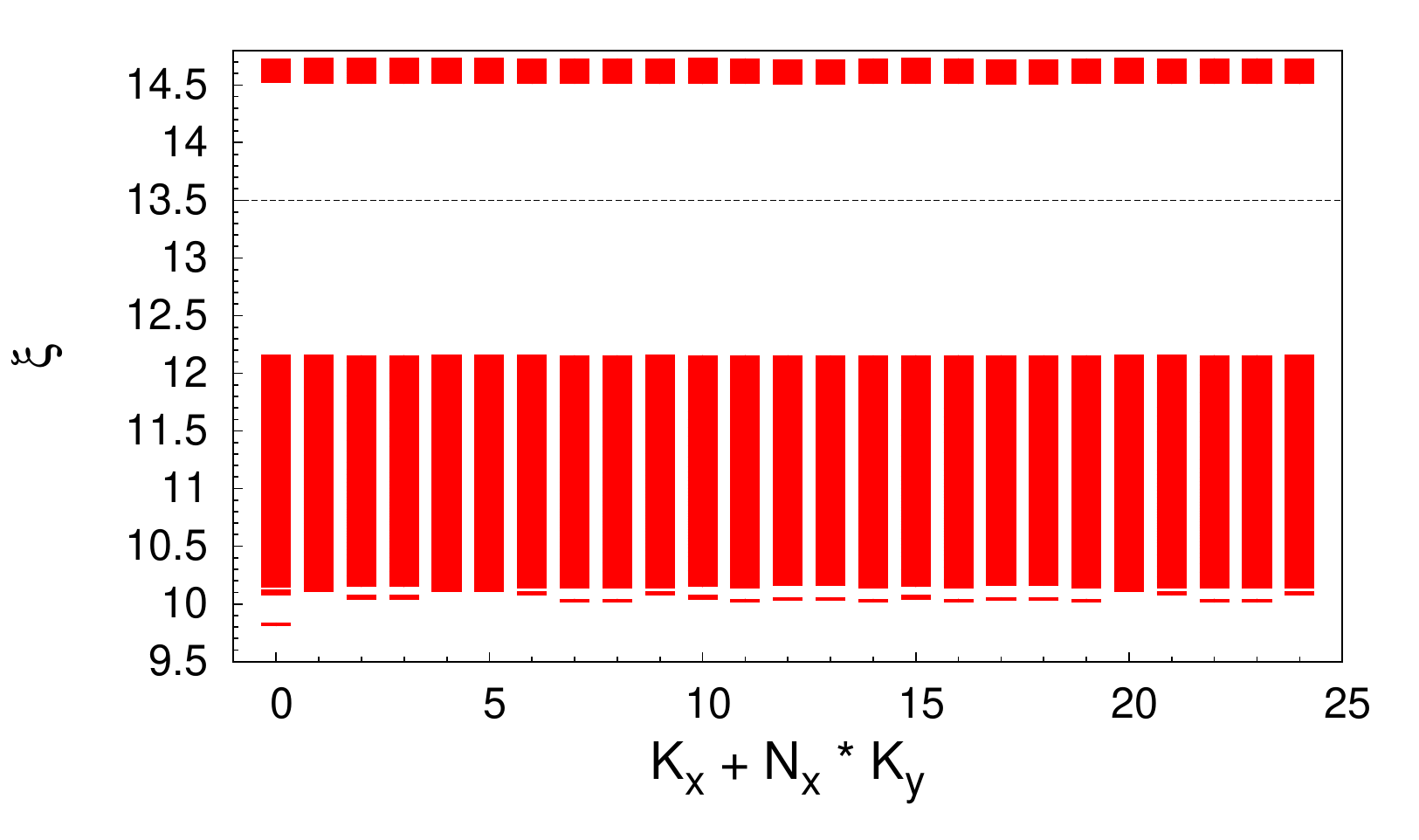}
\caption{{\it Left panel} : Low-lying spectrum of the four-body interaction for $N_e=15$ electrons and $N_x=N_y=5$. The groundstate manifold is made of the 10 low energy states in the $(k_x=0,k_y=0)$ sector. {\it Right panel} : Entanglement spectrum for the groundstates of the four-body interaction for $N_e=15$ electrons and $N_x=N_y=5$ and $N_A=5$ particles. The number of states per momentum sector  below the dotted line  is $2050$ except in the $(k_x=0,k_y=0)$ sector that has 2055 states. This is the correct counting predicted by our FQH to FCI mapping.}\label{Read-RezayiFCISpectrum}
\end{figure}

We conclude that we have presented sufficient proof that the FCI state at half-filling with three-body potential is of MR type in the FCI. We now move to the four-body interaction. The $\mathbb{Z}_3$ RR state occurs at filling factor $3/5$. We have performed the calculations up to $N=15$. Fig. \ref{Read-RezayiFCISpectrum} shows the energy spectrum for $N=15$ with a lattice size $N_x=N_y=5$. We see a tenfold low energy manifold at momenta consistent with our FQH to FCI mapping, clearly separated from the higher energy continuum. The particle entanglement spectrum of this groundstate manifold is given in Fig. \ref{Read-RezayiFCISpectrum}. While all $N_A$ sectors display a full entanglement gap, we have chosen the $N_A=5$ sector. This case exhibits an identical counting (2050 states) in all momentum sectors except in the $(k_x=0,k_y=0)$ sector where the counting is 2055. Once again, this result is in agreement with our FQH to FCI mapping.


\section{Conclusions}

In this manuscript, we have explained the emergent translational symmetry apparent in the low-energy manifold of states of a fractionally filled Chern insulator when the insulator exhibits a topologically ordered state. We have shown that the one-body GMP algebra obtained at long wavelength in \cite{Parameswaran-2011arXiv1106.4025P} maps the problem of a Chern insulator on a $N_x \times N_y$ lattice in the problem of a FQH state in the LLL with number of fluxes $N_s = N_x N_y$ piercing the torus and with a periodic background potential superimposed on the many-body potential already present in the system. We revisited the continuum FQH problem and presented a detailed derivation of the many-body relative translational symmetries in the system (after the center of mass motion is taken out) presented in a classic paper by Haldane \cite{haldane-PhysRevLett.55.2095}. We then showed that the BZ of the FCI is just a folding of the one for the FQH, and pointed out the direct connection between the center of mass degeneracies in the FCI and those in the FQH. This FQH-FCI map allows us to compute the number of states that arise per momentum sector of the FCI \emph{if the FCI is in the same topological phase as the FQH}. We showed how to obtain, in the FQH effect, the counting of quasihole states per $2D$ relative momentum sector for a large series of states exhibiting a generalized Pauli principle \cite{bernevig-08prl246802}. These states include the Laughlin, MR and RR series, as well as the Jack polynomials states, the Haffnian\cite{green-10thesis}, and many others, including extensions to spin-unpolarized states\cite{ardonne-PhysRevB.84.205134,estienne-2011arXiv1107.2534E}.  The counting construction makes use of  only combinatoric methods and involves counting partitions. We showed how to use the combinatoric counting rule and the FQH-FCI map to obtain the number of states per momentum sector in both the energy and the entanglement spectrum of the FCI at filling $1/3$. We then proceeded on a comprehensive numerical analysis of the spectra of Chern insulators with three and four-body interactions. We provided a proof of principle that Moore-Read states and Read-Reazayi $\mathbb{Z}_3$ non-Abelian states can be stabilized in Chern insulators and provided a numerical check of our FQH-FCI mapping.

\section{Acknowledgements} BAB wishes to thank Z. Papic, F.D.M. Haldane, T.L. Hughes, S. Sondhi, and S.A. Parameswaran for useful discussions. BAB was supported by Princeton Startup Funds, NSF CAREER DMR-095242, ONR - N00014-11-1-0635, Darpa - N66001-11-1-4110 and NSF-MRSEC DMR-0819860 at Princeton University. BAB thanks Technion, Israel, and Ecole Normale Superieure, Paris, for generous hosting during the stages of this work.

\section{Appendices}

\subsection{Appendix $1$: Details of the Girvin-MacDonald-Platzman Algebra}

In this appendix, we prove in detail that 
\beq
\sum_{n_2} u_{\alpha, k}^{n_1 \star} u_{\alpha, k+q}^{n_2} u_{\beta, k+q}^{n_2 \star} u_{\beta, k+ w +q}^{m_2} - u_{\beta, k}^{n_1 \star} u_{\beta, k+ w}^{n_2}  u_{\alpha, k+w}^{n_2 \star} u_{\alpha, k+w+q}^{m_2} \approx  \frac{i}{2} (q_i w_j - w_i q_j) F_{ij}^{n_1 m_2}
\eneq
 The expansion (double index means summation):
\beq
u_{\alpha, k+q}^n = u_{\alpha, k}^n + q_i \partial_i  u_{\alpha, k}^n  + \frac{1}{2} q_i q_j \partial_i \partial_j u_{\alpha, k}^n 
\eneq gives:
\begin{eqnarray}
& u_{\alpha, k}^{n_1 \star} (u_{\alpha, k}^{n_2} + q_i \partial_i  u_{\alpha, k}^{n_2}  + \frac{1}{2} q_i q_j \partial_i \partial_j u_{\alpha, k}^{n_2} )\cdot \nonumber \\  & (u_{\beta, k}^{n_2 \star} + q_i \partial_i  u_{\beta, k}^{n_2 \star}  + \frac{1}{2} q_i q_j \partial_i \partial_j u_{\beta, k}^{n_2 \star} )(u_{\beta, k}^{m_2 } + (q_i + w_i) \partial_i  u_{\beta, k}^{m_2 }  + \frac{1}{2}( q_i+ w_i )(q_j+ w_j) \partial_i \partial_j u_{\beta, k}^{m_2} ) - \nonumber \\ & - u_{\beta, k}^{n_1 \star} (u_{\beta, k}^{n_2} + w_i \partial_i  u_{\beta, k}^{n_2}  + \frac{1}{2} w_i w_j \partial_i \partial_j u_{\beta, k}^{n_2} )\cdot \nonumber \\  & (u_{\alpha, k}^{n_2 \star} + w_i \partial_i  u_{\alpha, k}^{n_2 \star}  + \frac{1}{2} w_i w_j \partial_i \partial_j u_{\alpha, k}^{n_2 \star} )(u_{\alpha, k}^{m_2 } + (q_i + w_i) \partial_i  u_{\alpha, k}^{m_2 }  + \frac{1}{2}( q_i+ w_i )(q_j+ w_j) \partial_i \partial_j u_{\alpha, k}^{m_2} ) = \nonumber \\ &=  (\delta_{n_1, n_2} + q_i u_{\alpha, k}^{n_1 \star}  \partial_i  u_{\alpha, k}^{n_2}  + \frac{1}{2} q_i q_j u_{\alpha, k}^{n_1 \star}  \partial_i \partial_j u_{\alpha, k}^{n_2} )\cdot  (\delta_{n_2,m_2} + (q_i + w_i)  u_{\beta, k}^{n_2 \star}  \partial_i  u_{\beta, k}^{m_2 } \nonumber \\ & +\frac{1}{2}( q_i+ w_i )(q_j+ w_j)   u_{\beta, k}^{n_2 \star} \partial_i \partial_j u_{\beta, k}^{m_2} +  q_i u_{\beta, k}^{m_2 }   \partial_i  u_{\beta, k}^{n_2 \star} + q_i(q_j+ w_j) (\partial_i  u_{\beta,k}^{n_2 \star} )(\partial_j u_{\beta,k}^{m_2}) + \frac{1}{2} q_i q_j u_{\beta, k}^{m_2 } \partial_i \partial_j u_{\beta, k}^{n_2 \star} ) - \nonumber \\ & - 
(\delta_{n_1, n_2} + w_i u_{\beta, k}^{n_1 \star}  \partial_i  u_{\beta, k}^{n_2}  + \frac{1}{2} w_i w_j u_{\beta, k}^{n_1 \star}  \partial_i \partial_j u_{\beta, k}^{n_2} )\cdot  (\delta_{n_2,m_2} + (q_i + w_i)  u_{\alpha, k}^{n_2 \star}  \partial_i  u_{\alpha, k}^{m_2 } \nonumber \\ & +\frac{1}{2}( q_i+ w_i )(q_j+ w_j)   u_{\alpha, k}^{n_2 \star} \partial_i \partial_j u_{\alpha, k}^{m_2} +  w_i u_{\alpha, k}^{m_2 }   \partial_i  u_{\alpha, k}^{n_2 \star} + w_i(q_j+ w_j) (\partial_i  u_{\alpha,k}^{n_2 \star} )(\partial_j u_{\alpha,k}^{m_2}) + \frac{1}{2} w_i w_j u_{\alpha, k}^{m_2 } \partial_i \partial_j u_{\alpha, k}^{n_2 \star} ) = \nonumber \\ &=  \delta_{n_1, m_2}  +  (q_i + w_i)  u_{\beta, k}^{n_1 \star}  \partial_i  u_{\beta, k}^{m_2 } +\frac{1}{2}( q_i+ w_i )(q_j+ w_j)   u_{\beta, k}^{n_1 \star} \partial_i \partial_j u_{\beta, k}^{m_2} + q_i u_{\beta, k}^{m_2 }   \partial_i  u_{\beta, k}^{n_1 \star} +  q_i(q_j+ w_j) (\partial_i  u_{\beta,k}^{n_1 \star} )(\partial_j u_{\beta,k}^{m_2}) +\nonumber \\ & +  \frac{1}{2} q_i q_j u_{\beta, k}^{m_2 } \partial_i \partial_j u_{\beta, k}^{n_1 \star} + q_i u_{\alpha, k}^{n_1 \star}  \partial_i  u_{\alpha, k}^{m_2 } + q_j (q_i + w_i)  u_{\alpha, k}^{n_1 \star}  (\partial_j  u_{\alpha, k}^{n_2})   u_{\beta, k}^{n_2 \star}  \partial_i  u_{\beta, k}^{m_2 }  +  q_j  q_i u_{\alpha, k}^{n_1 \star}  (\partial_j  u_{\alpha, k}^{n_2}  ) u_{\beta, k}^{m_2 }   \partial_i  u_{\beta, k}^{n_2 \star}  +\nonumber \\  & +  \frac{1}{2} q_i q_j u_{\alpha, k}^{n_1 \star}  \partial_i \partial_j u_{\alpha, k}^{m_2} -  (q\leftrightarrow w, \alpha \leftrightarrow \beta) 
\end{eqnarray}
The zeroth order term in q cancels. We now group the first and second order terms. The first order terms are:
\begin{eqnarray}
& (q_i + w_i)  u_{\beta, k}^{n_1 \star}  \partial_i  u_{\beta, k}^{m_2 } +  q_i (u_{\beta, k}^{m_2 }   \partial_i  u_{\beta, k}^{n_1 \star} + u_{\alpha, k}^{n_1 \star}  \partial_i  u_{\alpha, k}^{m_2 }  ) - (q\leftrightarrow w, \alpha \leftrightarrow \beta)  = \nonumber  \\ &(q_i + w_i)  u_{\beta, k}^{n_1 \star}  \partial_i  u_{\beta, k}^{m_2 } +  q_i \partial_i (u_{\beta, k}^{m_2 }  u_{\beta, k}^{n_1 \star}) - (q\leftrightarrow w, \alpha \leftrightarrow \beta)  = \nonumber  \\ &  (q_i + w_i)  u_{\beta, k}^{n_1 \star}  \partial_i  u_{\beta, k}^{m_2 }  - (q\leftrightarrow w, \alpha \leftrightarrow \beta)  =0
\end{eqnarray} (summation over $\alpha, \beta$ was implied. 
Hence the first order term is zero. 

The second order term is (remember double index summation):
\begin{eqnarray}
&\frac{1}{2}( q_i+ w_i )(q_j+ w_j)   u_{\beta, k}^{n_1 \star} \partial_i \partial_j u_{\beta, k}^{m_2} +  q_i(q_j+ w_j) (\partial_i  u_{\beta,k}^{n_1 \star} )(\partial_j u_{\beta,k}^{m_2}) +\nonumber \\ & +  \frac{1}{2} q_i q_j u_{\beta, k}^{m_2 } \partial_i \partial_j u_{\beta, k}^{n_1 \star} +q_j (q_i + w_i)  u_{\alpha, k}^{n_1 \star}  (\partial_j  u_{\alpha, k}^{n_2})   u_{\beta, k}^{n_2 \star}  \partial_i  u_{\beta, k}^{m_2 }  +  q_j  q_i u_{\alpha, k}^{n_1 \star}  (\partial_j  u_{\alpha, k}^{n_2}  ) u_{\beta, k}^{m_2 }   \partial_i  u_{\beta, k}^{n_2 \star}  + \frac{1}{2} q_i q_j u_{\alpha, k}^{n_1 \star}  \partial_i \partial_j u_{\alpha, k}^{m_2} - \nonumber \\  & - (q\leftrightarrow w, \alpha \leftrightarrow \beta)  = \nonumber \\ & =(q_i(q_j+ w_j)- w_i(q_j+w_j))  (\partial_i  u_{\beta,k}^{n_1 \star} )(\partial_j u_{\beta,k}^{m_2})   + \frac{1}{2} (q_i q_j - w_i w_j) u_{\beta, k}^{m_2 } \partial_i \partial_j u_{\beta, k}^{n_1 \star}  + \nonumber \\ & 
+ q_j (q_i + w_i)  u_{\alpha, k}^{n_1 \star}  (\partial_j  u_{\alpha, k}^{n_2})   u_{\beta, k}^{n_2 \star}  \partial_i  u_{\beta, k}^{m_2 } - w_j (q_i + w_i)  u_{\beta, k}^{n_1 \star}  (\partial_j  u_{\beta, k}^{n_2})   u_{\alpha, k}^{n_2 \star}  \partial_i  u_{\alpha, k}^{m_2 }  +  \nonumber\\ & 
+  q_j  q_i u_{\alpha, k}^{n_1 \star}  (\partial_j  u_{\alpha, k}^{n_2}  ) u_{\beta, k}^{m_2 }   \partial_i  u_{\beta, k}^{n_2 \star}  -   w_j  w_i u_{\beta, k}^{n_1 \star}  (\partial_j  u_{\beta, k}^{n_2}  ) u_{\alpha, k}^{m_2 }   \partial_i  u_{\alpha, k}^{n_2 \star}   + 
\frac{1}{2}( q_i q_j - w_i w_j)  u_{\alpha, k}^{n_1 \star}  \partial_i \partial_j u_{\alpha, k}^{m_2}   = \nonumber \\ & =   (q_i w_j- w_i q_j)  (  (\partial_i  u_{\beta,k}^{n_1 \star} )(\partial_j u_{\beta,k}^{m_2})  -   u_{\alpha, k}^{n_1 \star}  (\partial_j  u_{\alpha, k}^{n_2})   u_{\beta, k}^{n_2 \star}  \partial_i  u_{\beta, k}^{m_2 }  )+
\nonumber \\ & + \frac{1}{2} (q_i q_j - w_i w_j)( u_{\beta, k}^{m_2 } \partial_i \partial_j u_{\beta, k}^{n_1 \star}+   u_{\alpha, k}^{n_1 \star}  \partial_i \partial_j u_{\alpha, k}^{m_2}    +  2 (\partial_i  u_{\beta,k}^{n_1 \star} )(\partial_j u_{\beta,k}^{m_2})) + \nonumber \\ & + q_i q_j(u_{\alpha, k}^{n_1 \star}  (\partial_j  u_{\alpha, k}^{n_2}  ) u_{\beta, k}^{m_2 }   \partial_i  u_{\beta, k}^{n_2 \star}  +  u_{\alpha, k}^{n_1 \star}  (\partial_j  u_{\alpha, k}^{n_2})   u_{\beta, k}^{n_2 \star}  \partial_i  u_{\beta, k}^{m_2 }) - w_i w_j(u_{\beta, k}^{n_1 \star}  (\partial_j  u_{\beta, k}^{n_2}  ) u_{\alpha, k}^{m_2 }   \partial_i  u_{\alpha, k}^{n_2 \star} + u_{\beta, k}^{n_1 \star}  (\partial_j  u_{\beta, k}^{n_2})   u_{\alpha, k}^{n_2 \star}  \partial_i  u_{\alpha, k}^{m_2 } )\label{bigequation}
\end{eqnarray}
We now observe several simplifications. The last line of the above equation vanishes due to the identity:
\beq
u_{\alpha, k}^{n_1 \star}  (\partial_j  u_{\alpha, k}^{n_2}  ) u_{\beta, k}^{m_2 }   \partial_i  u_{\beta, k}^{n_2 \star}  +  u_{\alpha, k}^{n_1 \star}  (\partial_j  u_{\alpha, k}^{n_2})   u_{\beta, k}^{n_2 \star}  \partial_i  u_{\beta, k}^{m_2 } = u_{\alpha, k}^{n_1 \star}  (\partial_j  u_{\alpha, k}^{n_2}  )  \partial_i ( u_{\beta, k}^{m_2 }    u_{\beta, k}^{n_2 \star} )=0
\eneq 
The second but last line of Eq[\ref{bigequation}] also cancels because of the identity:
\beq
A_{ij}(u_{\beta, k}^{m_2 } \partial_i \partial_j u_{\beta, k}^{n_1 \star}+   u_{\alpha, k}^{n_1 \star}  \partial_i \partial_j u_{\alpha, k}^{m_2}    +  2 (\partial_i  u_{\beta,k}^{n_1 \star} )(\partial_j u_{\beta,k}^{m_2})) = A_{ij} \partial_i \partial_j ( u_{\beta, k}^{m_2 }    u_{\beta, k}^{n_2 \star} )=0
\eneq when $A_{ij}$ is any symmetric matrix. We are left with
\begin{eqnarray}
& \sum_{n_2} u_{\alpha, k}^{n_1 \star} u_{\alpha, k+q}^{n_2} u_{\beta, k+q}^{n_2 \star} u_{\beta, k+ w +q}^{m_2} - u_{\beta, k}^{n_1 \star} u_{\beta, k+ w}^{n_2}  u_{\alpha, k+w}^{n_2 \star} u_{\alpha, k+w+q}^{m_2} \approx  \nonumber \\ & (q_i w_j- w_i q_j)  (  (\partial_i  u_{\beta,k}^{n_1 \star} )(\partial_j u_{\beta,k}^{m_2})  -   u_{\alpha, k}^{n_1 \star}  (\partial_j  u_{\alpha, k}^{n_2})   u_{\beta, k}^{n_2 \star}  \partial_i  u_{\beta, k}^{m_2 } ) =\nonumber \\ & =\frac{1}{2}(q_i w_j- w_i q_j)  (  (\partial_i  u_{\beta,k}^{n_1 \star} )(\partial_j u_{\beta,k}^{m_2})   -   u_{\alpha, k}^{n_1 \star}  (\partial_j  u_{\alpha, k}^{n_2})   u_{\beta, k}^{n_2 \star}  \partial_i  u_{\beta, k}^{m_2 } -(i\leftrightarrow j)  )  
\end{eqnarray}
By denoting the non-Abelian Berry potential 

\beq
A_j^{n_1, n_2}=i u_{\beta k}^{n_1 \star} \partial_j u_\beta^{m_2}
\eneq
we find that
\beq
  (\partial_i  u_{\beta,k}^{n_1 \star} )(\partial_j u_{\beta,k}^{m_2})   -   u_{\alpha, k}^{n_1 \star}  (\partial_j  u_{\alpha, k}^{n_2})   u_{\beta, k}^{n_2 \star}  \partial_i  u_{\beta, k}^{m_2 } -(i\leftrightarrow j)   =- i F_{ij}^{n_1, n_2}
  \eneq
where
$F_{ij}^{n_1, n_2} = \partial_i A_{j}^{n_1,n_2} - \partial_j A_{i}^{n_1,n_2}  - i [A_i, A_j]^{n_1,n_2}$.

\subsection{Appendix $2$: Relation Between the Density Algebras and Chern Number}

The projected densities in the  commutator algebra in Eq[\ref{nonabelianmatrix}] are required to not commute if the topological insulator is to have a non-zero Chern number. Taking the trace of the commutator in Eq[\ref{nonabelianmatrix}] with the $\rho_{-q-w} =\frac{1}{\sqrt{N_s}}  \sum_{k} u_{\alpha k-q -w}^{m_1} u_{\alpha k}^{n_2\star}  \gamma_{k}^{n_2\dagger} \ket{0} \bra{0} \gamma_{k-q-w}^{m_1}$ we obtain, to second order in $q,w$ (in the long-wavelength limit):
\beq
[\rho_q, \rho_w] \rho_{-q-w}  =-\frac{i}{2} (q_i w_j - w_i q_j) \frac{1}{N_s^{3/2}} \sum_{k, n_1 m_2} F_{ij}^{n_1, m_2} (k) \gamma_{k}^{n_1 \dagger} \ket{0}{\bra{0}} \gamma_{k}^{m_2}
\eneq where we have used the fact that $ u_{\alpha k-q -w}^{m_1} u_{\alpha k}^{n_2\star}=\delta_{m_1, n_2}$ to zeroth order in $q,w$:
\beq
[\rho_{q_x \hat{x}}, \rho_{w_y \hat{y}}] \rho_{-q_x \hat{x}-w_y\hat{y}}  =\frac{i}{2} q_x w_y \frac{1}{N_s^{3/2}} \sum_{k, n_1 m_2} F_{xy}^{n_1, m_2} (k) \gamma_{k}^{n_1 \dagger} \ket{0}{\bra{0}} \gamma_{k}^{m_2}
\eneq We now take the trace of  the above to obtain:
\beq
\frac{2}{i} \frac{\sqrt{N_s}}{q_x w_y}Tr( [\rho_{q_x \hat{x}}, \rho_{w_y \hat{y}}] \rho_{-q_x \hat{x}-w_y\hat{y}}) = \frac{1}{N_s} \sum_{k, m} F_{xy}^{m,m}(k) = C
\eneq with $q_x = \frac{2 \pi}{N_x}$, $q_y= \frac{2 \pi}{N_y}$ the lowest value for the momenta one can obtain on the lattice. Hence the Chern number can be expressed as the long-wavelength limit of the commutator of the projected density operators in the lower sub-band.

\subsection{Appendix $3$: Relation Between the Position Operators and the Chern Number }
An alternative way of presenting this result is by using the position operator in the $X$ direction:
\beq
X=\frac{1}{\sqrt{N_s}}\sum_{j, \alpha}  e^{i \Delta q j } \ket{j, \alpha} \bra{j, \alpha} = \frac{1}{\sqrt{N_s}}\sum_{k} c_{k \alpha} \ket{0}\bra{0} c_{k+ \Delta q \alpha}
\eneq where by $\Delta k j$ we mean $\Delta \vec{k} \cdot \vec{j}$.  The projected density operator in the lowest band is:
\beq
P X P = \frac{1}{\sqrt{N_s}} \sum_{k} u_{k \alpha}^{n_1\star} u_{k+ \Delta q}^{n_2} \gamma_{n_1 k}^\dagger \ket{0} \bra{0} \gamma_{k + \Delta q}^{n_2}= \rho_{\Delta q}
\eneq Hence the projected density operator for $\Delta q= 2 \pi/N_x$ equals the projected position operator. There is an alternate expression for the projected position and density operators. They are translational operators in $k$-space: 

\beq
P X P \ket{k, n}= \rho_{\Delta q} \ket{k, n}= u^{n_1\star}_{k  - \Delta q, \alpha} u^n_{k \alpha} \ket{k - \Delta q, n_1} \approx  e^{-i\int_{k}^{k + \Delta q } A^{n_1, n}(k)} \ket{k - \Delta q, n_1}
\eneq For one occupied band, the density or position operators just translate the band at different momentum, whereas for more than one occupied bands, the translation is accompanied by a rotation. We now see that:
\beq
\frac{2}{i} \frac{N_s^{3/2 }}{(2\pi)^2}Tr( [PXP, PYP] X^{-1} Y^{-1}) = \frac{1}{N_s} \sum_{k, m} F_{xy}^{m,m}(k) = C
\eneq

\subsection{Appendix $4$: Chern Number of a Landau Level}

 We outline how a simple calculation for the Chern number of lowest Landau Level. The wavefunctions for the landau levels in the continuum do not have two momenta $k_x, k_y$, and hence the momentum expression for the Chern number cannot be used. We obtain the Chern number as:
\beq
C=-\frac{\hbar^2}{l^2} \lim_{\Delta_x, \Delta_y \rightarrow 0} \frac{1}{\Delta_x \Delta_y}
\text{Tr}[ [PXP, PYP] X^{-1}Y^{-1}]\eneq where in the LLL we have the expressions
\beq
PXP =\frac{1}{\sqrt{V}} P e^{i \Delta_x x} P = P e^{i\Delta_x (z+ \bar{z})/2)} P, \;\;\;\;\; PYP =\frac{1}{\sqrt{V}} P e^{i \Delta_y y} P = P e^{\Delta_y (z- \bar{z})/2)} P
\eneq LLL wavefunctions $\psi(z)$ have the property that $\Pi \psi(z) =0$ where $\Pi =\Pi_x+ i \Pi_y$, $\Pi_x, \Pi_y$ being the canonical momenta in the presence of a magnetic field. We then have:
\beq
Pe^{za+ \bar{z} b} P= P e^{i \frac{l^2}{h} (\Pi a- \Pi^\dagger b) } e^{i \frac{l^2}{h} (K^\dagger b- Ka) } P \eneq where we have used $z=(i l^2/h) (\Pi -K)$, $K= K_x + i K_y$, and the fact that $\Pi$ commutes with $K$.  We now need to disentangle the exponent containing $\Pi, \Pi^\dagger$ to obtain: $e^{i \frac{l^2}{h} (\Pi a- \Pi^\dagger b) }  = e^{-i \frac{l^2}{h}  \Pi^\dagger b } e^{i \frac{l^2}{h} \Pi a }  e^{ab l^2/h^2}$.  Since $e^{i \frac{l^2}{h} \Pi a } P = P$, $P e^{-i \frac{l^2}{h}  \Pi^\dagger b } = P$, we have 
\beq
Pe^{za+ \bar{z} b} P= e^{ab l^2/h^2} Pe^{i \frac{l^2}{h} (K^\dagger b- Ka) } P
\eneq. This allows us the computation of the terms involved in the Chern number:
\beq
[PXP, PYP] = \frac{1}{V} e^{-\frac{\Delta_x^2+ \Delta_y^2}{4} \frac{l^2}{\hbar^2}}[ Pe^{i\frac{l^2}{\hbar} \Delta_x  K_y}P, P e^{-i\frac{l^2}{\hbar} \Delta_y  K_x}P]
\eneq Moreover, again after projection to the LLL, and after some algebra, we find: 
\begin{eqnarray}
&\text{Tr}[ [PXP, PYP] X^{-1}Y^{-1}] =  \frac{ 1}{V} e^{-\frac{l^2}{\hbar^2}\frac{\Delta_x^2+ \Delta_y^2 }{2}} \times \nonumber \\ &  \text{Tr} [ e^{-\frac{l^2}{\hbar^2}\frac{\Delta_x  \Delta_y }{2}}
e^{-i \frac{l^2}{\hbar}   \Delta_x K_y}   Pe^{i\frac{l^2}{\hbar} \Delta_x  K_y} P e^{-i\frac{l^2}{\hbar} \Delta_y  K_x}P  e^{i \frac{l^2}{\hbar} \Delta_y K_x }   -  e^{\frac{l^2}{\hbar^2}\frac{\Delta_x  \Delta_y }{2}}
e^{i \frac{l^2}{\hbar}   \Delta_y K_x}   Pe^{-i\frac{l^2}{\hbar} \Delta_y  K_x} P e^{i\frac{l^2}{\hbar} \Delta_x  K_y}P  e^{-i \frac{l^2}{\hbar} \Delta_x K_y }]
\end{eqnarray} We now observe that, if we write down $P =\int d^2r \ket{\psi(r)} \bra{\psi(r)}$ then the quantity $e^{-i \frac{l^2}{\hbar}   \Delta_x K_y}   Pe^{i\frac{l^2}{\hbar} \Delta_x  K_y} $ is again a projection operator $P$ due to the fact that $e^{-i \frac{l^2}{\hbar}   \Delta_x K_y} $ are translation operators. We find that: $Tr[e^{-i \frac{l^2}{\hbar}   \Delta_x K_y}   Pe^{i\frac{l^2}{\hbar} \Delta_x  K_y} P e^{-i\frac{l^2}{\hbar} \Delta_y  K_x}P  e^{i \frac{l^2}{\hbar} \Delta_y K_x }]=V$. We hence obtain:
\beq
C= -\frac{\hbar^2}{l^2} \lim_{\Delta_x, \Delta_y \rightarrow 0} \frac{1}{\Delta_x \Delta_y} e^{-\frac{l^2}{\hbar^2}\frac{\Delta_x^2+ \Delta_y^2 }{2}}( e^{-\frac{l^2}{\hbar^2}\frac{\Delta_x  \Delta_y }{2}}- e^{\frac{l^2}{\hbar^2}\frac{\Delta_x  \Delta_y }{2}}) = 1
\eneq

\subsection{Appendix $5$: Center of Mass Degeneracy Mismatch}

The center of mass degeneracy in the FQH is $q = N_s/GCD(N_s, N_e)$ (with $N_S= N_x N_y$) while in the FCI is $q_x q_y= (N_x/GCD(N_x, N_e)) (N_y/GCD(N_y, N_e))$. Their ratio is
\beq
\frac{q}{q_x q_y} = \frac{GCD(N_x, N_e) \cdot GCD(N_y, N_e)}{ GCD(N_x N_y, N_e)}
\eneq We can easily prove that this is an integer by applying the decomposition of a number in primes $p_i$: $N_x = \prod_i p_i^{\alpha_i}$, $N_y = \prod_i p_i^{\beta_i}$, $N_e = \prod_i p_i^{\theta_i}$ where $\alpha_i, \beta_i, \theta_i$ are all integer powers. The ratio then becomes:
\beq
\frac{q}{q_x q_y} =\prod_i p_i^{\min(\alpha_i, \theta_i) + \min(\beta_i, \theta_i) - \min(\alpha_i+ \beta_i, \theta_i)}
\eneq We can now analyze the different cases to prove that all powers of $p_i$ are positive in the above expression. Since our analysis is valid for \emph{any} $i$, we drop the index. We have the cases 1) $\alpha \le \theta,\beta \le \theta, \alpha + \beta \ge \theta$ in which case $\min(\alpha, \theta) + \min(\beta, \theta) - \min(\alpha+ \beta, \theta) = \alpha + \beta -\theta \ge 0$; 2) $\alpha \le \theta, \beta \le \theta, \alpha+\beta \le \theta$ in which case $\min(\alpha, \theta) + \min(\beta, \theta) - \min(\alpha+ \beta, \theta) = 0$; 3) $\alpha \ge \theta,  \beta \le \theta, \alpha+\beta \ge \theta$ in which case $\min(\alpha, \theta) + \min(\beta, \theta) - \min(\alpha+ \beta, \theta) =\beta$; 4) $\alpha \ge \theta, \beta \le \theta, \alpha+\beta \le \theta$ in which case $\min(\alpha, \theta) + \min(\beta, \theta) - \min(\alpha+ \beta, \theta) =\theta-\alpha\ge 0$; 5) $\alpha \ge \theta, \beta \ge \theta, \alpha+\beta \ge\theta$in which case $\min(\alpha, \theta) + \min(\beta, \theta) - \min(\alpha+ \beta, \theta) =\theta$. We see that in all cases possible, $\min(\alpha, \theta) + \min(\beta, \theta) - \min(\alpha+ \beta, \theta) $ is a positive integer which makes $q$ divisible by $q_x q_y$. 

\bibliography{Emergent.bib}

\end{document}